\documentclass[journal, 12pt,onecolumn, draftclsnofoot]{IEEEtran}

\usepackage{ragged2e}
\usepackage{floatrow}
\usepackage{multirow}
\usepackage{amsthm}
\usepackage[utf8]{inputenc}
\usepackage[english]{babel}
\theoremstyle{remark}
\newtheorem{theorem}{Theorem}
\newtheorem{lemma}{Lemma}

\newtheorem{remark}{Remark}

\newtheorem{corollary}{Corollary}
\usepackage{multicol}
\usepackage{epsfig}
\usepackage{epstopdf}
\usepackage{float}
\usepackage{perpage}
\MakeSorted{figure}
\MakeSorted{table}
\usepackage{array}
\usepackage{graphicx}

\usepackage{stfloats}
\usepackage{amsfonts}
\usepackage{amssymb}
\usepackage{soul}

\usepackage{enumitem}							
\usepackage[english]{babel}
\usepackage[utf8]{inputenc}
\usepackage[linesnumbered,ruled,vlined]{algorithm2e}
\usepackage{upgreek}
\usepackage{bm} 
\usepackage{hyperref}
\usepackage{float}
\floatstyle{plaintop}
\restylefloat{table}
\hypersetup{linktocpage} % links to only the page number
\hypersetup{
colorlinks,
citecolor=black,
filecolor=black,
linkcolor=black,
urlcolor=black
}
\usepackage{adjustbox}
\usepackage{pifont}% http://ctan.org/pkg/pifont
\newcommand{\cmark}{\ding{51}}%
\newcommand{\xmark}{\ding{55}}%
\usepackage{makecell}



\usepackage{cite}
\DeclareMathAlphabet\mathbfcal{OMS}{cmsy}{b}{n}
\ifCLASSINFOpdf
\else
\fi
\usepackage[cmex10]{amsmath}

\usepackage{array}
\usepackage{fixltx2e}

\usepackage[linesnumbered,ruled,vlined]{algorithm2e}

\usepackage[font=footnotesize,labelfont=bf, figurename=Fig.]{caption} 
\usepackage{subcaption}
\usepackage{fix-cm}
\usepackage{comment}

\usepackage{mathtools}
\usepackage{pifont}

\newcommand{\Amat}{\mathbf{A}}
\newcommand{\Bmat}{\mathbf{B}}
\newcommand{\Cmat}{\mathbf{C}}
\newcommand{\Dmat}{\mathbf{D}}

\newcommand{\Imat}{\mathbf{I}}
\newcommand{\Jmat}{\mathbf{J}}

\newcommand{\Pmat}{\mathbf{P}}
\newcommand{\Qmat}{\mathbf{Q}}
\newcommand{\Rmat}{\mathbf{R}}
\newcommand{\Smat}{\mathbf{S}}

\newcommand{\Wmat}{\mathbf{W}}
\newcommand{\Xmat}{\mathbf{X}}

\newcommand{\av}{\mathbf{a}}

\newcommand{\fv}{\mathbf{f}}

\newcommand{\hv}{\mathbf{h}}
\newcommand{\nv}{\mathbf{n}}
\newcommand{\pv}{\mathbf{p}}

\newcommand{\uv}{\mathbf{u}}
\newcommand{\vv}{\mathbf{v}}

\newcommand{\xv}{\mathbf{x}}
\newcommand{\yv}{\mathbf{y}}
\newcommand{\zv}{\mathbf{z}}

\newcommand{\E}{\mathbb{E}}

\newcommand\eqa{\stackrel{\mathclap{\normalfont(a)}}{=}}

\begin{document}
\bstctlcite{IEEEexample:BSTcontrol}
	\title{Hardware-Impaired Rician-Faded Cell-Free Massive MIMO Systems With Channel Aging}
%\title{Cell-Free Rician-Faded  Massive MIMO Systems:  Analysis And Optimization With Channel Aging And Hardware Impairments}
	\author{Venkatesh Tentu, Dheeraj N Amudala, Anish Chattopadhyay and Rohit Budhiraja \thanks{The authors are with the Dept. of Electrical Engineering, IIT Kanpur, India. email: \{tentu, dheeraja, anishchat20,  rohitbr\}@iitk.ac.in. A part of this work is accepted for presentation in IEEE International Conference on Communications 2023. 
}}
\maketitle
\vspace{-2.1cm}
\begin{abstract}
\vspace{-.2cm}
We study the impact of channel aging on the uplink of a cell-free (CF) massive multiple-input multiple-output (mMIMO) system by considering i) spatially-correlated Rician-faded channels; ii) hardware impairments at the access points and user equipments (UEs); and iii) two-layer large-scale fading decoding (LSFD).  We first derive a closed-form spectral efficiency (SE) expression for this system, and later propose two novel optimization techniques to optimize the non-convex SE metric by exploiting the minorization-maximization (MM) method. The first one requires a numerical optimization solver, and has a high computation complexity. The second one with closed-form transmit  power updates,  has a trivial computation complexity. We numerically show that  i) the two-layer LSFD scheme effectively mitigates the interference due to channel aging for both low- and high-velocity UEs; and ii) increasing the number of AP antennas does not mitigate the SE deterioration due to channel aging. We numerically characterize the optimal pilot length required to maximize the SE for various UE speeds.  We also numerically show that the proposed closed-form MM optimization yields the same SE as that of the first technique, which requires numerical solver, and that too with a much reduced time-complexity.
%for LSFD and single layer decoding.
%This can help a system designer in deciding their suitable value depending on the UE operating conditions.
 %------------------- 
%\colr{This technique is eminently suitable for high-velocity UEs which are shown to require frequent power updates.} \colr{We also numerically characterize the hardware impairment levels which can provide a high SE for high- and low-velocity UEs. This can help a system designer in deciding their suitable value depending on the UE operating conditions.}  
 %------------------- 
 %We numerically show that the effect of random phase-shifts on the SE i) increases with non-ideal hardware; and ii) reduces with increasing UE mobility. 
 %When UEs experience pilot contamination, the channel aging does not affect the estimation error. %We investigate that LSFD can mitigate the effect of channel aging for low/high velocities and inter-user interference for low velocity. 
 %-------------------
 %\colb{We numerically show that the proposed dynamic ADC architecture reduces the SE loss due to hardware impairments.}  We also show that LSFD can effectively mitigate  the detrimental effects of i) channel aging for both low and high UE velocities; and ii) IUI for low-velocity UEs but not for high-velocity UEs.
\end{abstract}
\vspace{-.8cm}
\begin{IEEEkeywords}\vspace{-0.1in}
	Cell-free,  channel aging, hardware impairments, minorization-maximization (MM), Rician fading.
\end{IEEEkeywords}
\vspace{-.8cm}
\section{Introduction} \vspace{-.3cm}
Cell-free (CF) massive multiple-input multiple output (mMIMO) is being investigated as a key technology for beyond fifth-generation wireless systems due to its  high spectral efficiency (SE) and improved coverage~\cite{HQNgo2017}. A CF mMIMO system consists of a large number of access points (APs) which are randomly deployed over a large coverage area,  and  are connected to a central processing unit (CPU) via high-speed fronthaul links. The APs cooperate via these fronthaul links to mitigate the inter-user interference (IUI) caused due to multiple user equipments (UEs) being served on the same time-frequency resource~\cite{HQNgo2017,EMIL_CORR_RAYL_20}.
Initial CF mMIMO works commonly considered  single layer decoding (SLD) schemes, wherein APs  individually combine their respective receive signals using the locally-estimated channel information, while the CPU handle only data  detection~\cite{HQNgo2017,EMIL_CORR_RAYL_20}. 
{Ngo \textit{et al.} in \cite{HQNgo2017} investigated the SE gains provided by a CF mMIMO system with SLD, over its conventional small-cell counterpart. The authors in~\cite{EMIL_CORR_RAYL_20} proposed scalable combiners and precoders for a spatially-correlated Rayleigh-faded CF mMIMO system with SLD, and showed that they outperform conventional maximum ratio processing.}  
We know that fifth generation (5G) cellular networks have recently been deployed. The channel models considered to evaluate different technologies for 5G systems e.g., mMIMO, contain line of sight (LoS) components, along with the non LoS (NLoS) one\cite{GUE_channel_3gp_Rici}. This makes the channel \textit{Rician distributed}.

Further, the local SLD techniques, despite having a low implementation complexity, fail to effectively suppress the {IUI}. This can be improved by performing the second level of combining at the CPU, referred to as the large-scale fading decoding  (LSFD)~\cite{Emil2019_rician_phase,Ozlem_CFwpt_21,JZhang2021}. {The key difference between LSFD and SLD receivers is that the former requires additional statistical channel parameters such as large-scale fading coefficients at the CPU. These large-scale fading parameters remain constant over multiple coherence intervals, and thus can be easily made available at the CPU~\cite{Emil2019_rician_phase,Ozlem_CFwpt_21,JZhang2021}.} 
%------------------------------
%The LSFD depends only on the large-scale fading coefficients of the channels between the APs and the UEs, which remain constant over multiple coherence intervals, and thus can be easily made available at the CPU~\cite{Emil2019_rician_phase,Ozlem_CFwpt_21,JZhang2021}.
%------------------------------
Ozdogan \textit{et al.} in \cite{Emil2019_rician_phase} showed the improved SE gains obtained due to LSFD for a CF mMIMO system with spatially-correlated Rician-faded channels. {Demir \textit{et al.} in \cite{Ozlem_CFwpt_21} maximized the minimum SE of a wireless-powered CF mMIMO system by considering LSFD and the spatially-correlated Rician-faded channels.} Zhang \textit{et al.} in \cite{JZhang2021} proposed  local zero forcing combiners for an uncorrelated CF mMIMO system, and then integrated these combiners with LSFD, and investigated their SE.

Majority of CF mMIMO works, including the aforementioned ones in \cite{HQNgo2017,EMIL_CORR_RAYL_20,Emil2019_rician_phase,Ozlem_CFwpt_21,JZhang2021}, considered a block fading channel model, wherein the channel remains constant over a coherence interval.  The  5G cellular systems are designed for a UE speed of up to $500$ km/h, whose channel  then continuously varies with time \cite{g5G_book}. 
The channel estimated by the APs, consequently,  become outdated with time~\cite{kong_aging_15,Anastasios_aging_17}. This channel aging phenomenon can drastically degrade the gains accrued by  CF mMIMO technology. 
The authors in~\cite{Anastasios_aging_17,SBL_2019,YUAN_ML_AGING_20,Papazaf2017multicell} and~\cite{chopra_AGING_21,zheng_aging_20,Zheng_aging_correlated_21} recently investigated the 
effect of channel aging on cellular  and CF mMIMO systems, respectively. The authors in \cite{Anastasios_aging_17} derived asymptotic power scaling laws which showed the detrimental effect of channel aging on a single-cell correlated Rayleigh-faded mMIMO system. 
%---------------------------------------------------
% //Jianpeng \textit{et. al} in \cite{SBL_2019} proposed a machine-learning-based channel predictor for a single-cell mMIMO system with channel aging, and investigated the SE performance over its conventional channel estimator counterpart.
%---------------------------------------------------YUAN_ML_AGING_20
The authors in \cite{SBL_2019,YUAN_ML_AGING_20} proposed a machine-learning-based channel predictor for a single-cell mMIMO system with channel aging, which could be used for both uplink and downlink systems.  Papazafeiropoulos \textit{et al.} in \cite{Papazaf2017multicell} analyzed the outage probability of a multi-cell mMIMO system and showed that its preferred to have massive number of antennas under channel aging conditions.
%Papazafeiropoulos \textit{et al.} in \cite{Papazaf2017multicell} analyzed the outage probability of a multi-cell mMIMO system with channel aging.
%Recently, authors investigated the effect of channel aging in CF mMIMO~\cite{chopra_AGING_21,Zheng_aging_correlated_21}
Chopra \textit{et. al} in \cite{chopra_AGING_21} showed that the impact of channel aging is higher on a CF mMIMO system than on a  cellular mMIMO system. This work, however, considered uncorrelated Rayleigh-faded channels and local SLD scheme at the APs. 
Zheng \textit{et. al} in \cite{zheng_aging_20} and \cite{Zheng_aging_correlated_21} analyzed the effect of channel aging in a CF mMIMO system with LSFD, for uncorrelated and correlated Rayleigh fading channels, respectively. {The authors in \cite{Zheng_aging_correlated_21} also proved that the CF mMIMO system is more robust to channel aging than a small-cell system.}

The above CF mMIMO works~\cite{Emil2019_rician_phase,Ozlem_CFwpt_21,JZhang2021,chopra_AGING_21,zheng_aging_20,Zheng_aging_correlated_21} assumed high-quality radio frequency (RF) transceivers and high-resolution analog-to-digital converters (ADCs)/digital-to-analog converters (DACs) at the APs and UEs. The RF  transceiver chips used to design 5G cellular system, and its evolved CF mMIMO systems  have inherent hardware distortion, which is commonly characterized using the error vector magnitude (EVM) metric. It is usually specified in the device data sheet \cite{adrv_ds}. {The effect of hardware distortion caused by the low-cost hardware can be suppressed by using the calibration schemes and the compensation algorithms, but the residual impairments still remains \cite{Masoumi_CF_HW_20,Papazaf2021CF_HW,CF_UAV_22,ZhengZZZA20,ElhoushyH20}. The
impact of these residual impairments on CF mMIMO systems should be further studied.}
%---------------------------------------------------
% The low-cost hardware induces hardware distortion  the non-linear amplifications in the low-noise amplifier (LNA) and quantization errors in the ADC/DAC
%--------------------------------------------------- 
%Such components are particularly susceptible to HWIs such as in-phase/quadrature-phase (I/Q)-imbalance [14], oscillator
%phase noise (PN) [15], [16], and high power amplifier non-linearities. Even if calibration schemes and compensation algorithms are utilized at the transmitter and receiver, respectively, a certain amount of distortions, known as residual HWIs,
%remains and can be categorized into additive and multiplicative distortions. The model of additive HWIs includes additive Gaussian noises at both the transmitter and receiver, expressing the aggregate effect of many impairments, and has been grounded based on its analytical tractability and experimental validation.
%---------------------------------------------------
%Masoumi \textit{et al.} in~\cite{Masoumi_CF_HW_20} derived a closed-form SE expression for hardware-impaired CF mMIMO system with uncorrelated Rayleigh-faded channels.
%------------------------------------------
{Masoumi \textit{et al.} in~\cite{Masoumi_CF_HW_20} analyzed the effect of hardware-impairments in a CF mMIMO system with uncorrelated Rayleigh-faded channels.} 
 The authors in \cite{Papazaf2021CF_HW} derived an approximate SE expression for a spatially-correlated Rayleigh-faded CF mMIMO system with hardware impairments. Tentu \textit{et. al} in \cite{CF_UAV_22} investigated the SE of an unmanned aerial vehicle enabled CF mMIMO system with RF impairments and spatially-correlated Rician channels. All these works considered only local SLD schemes, and also ignored channel aging in their analysis. 
{Zhang \textit{et al.} in \cite{ZhengZZZA20} analyzed the SE of a hardware-impaired CF mMIMO system with LSFD, and  showed that the hardware distortion non-trivially impacts the LSFD performance.} This work, however, ignored channel aging. The authors in \cite{ElhoushyH20} investigated the joint impact of hardware impairments and channel aging for a CF mMIMO system, but only for uncorrelated Rayleigh channels,  and that too without LSFD.
%------------------------------------------
%Masoumi \textit{et al.} in~\cite{Masoumi_CF_HW_20}  derived the uplink SE of a hardware-impaired uncorrelated Rayleigh-faded  CF mMIMO system with limited-capacity fronthaul link between each AP and CPU.  Elhousy \textit{et. al} in~\cite{ElhoushyH20} compared the SE of a hardware-impaired spatially-uncorrelated Rayleigh-faded CF mMIMO system with its small-cell counterpart, and showed that small-cell system performance degrades significantly.
%------------------------------------------

Practical CF mMIMO systems, to reduce the energy consumption and implementation costs, also employ low-resolution ADC/DACs~\cite{zhang2019_Low_ADC, zhang_CF_ADC_rician_22, hu2019_tcom}.
The authors in~\cite{zhang2019_Low_ADC, zhang_CF_ADC_rician_22, hu2019_tcom} studied the effect of low-resolution ADCs in CF mMIMO systems. Zhang \textit{et. al} in \cite{zhang2019_Low_ADC} analyzed the SE of spatially-uncorrelated Rayleigh-faded CF mMIMO system with low-resolution ADC/DACs, and showed its improved  energy efficiency over high-resolution ADC/DACs. 
Zhang \textit{et. al} in \cite{zhang_CF_ADC_rician_22} investigated the uplink and downlink SE of an uncorrelated Rician-faded CF mMIMO system with low-resolution ADCs at the APs. Hu \textit{et. al} in \cite{hu2019_tcom} investigated the asymptotic SE of an uncorrelated Rayleigh-faded CF mMIMO system, and showed that the SE is mainly limited by the UE ADC resolution, when low-resolution ADCs are used at both APs and UEs.
%------------------------------------
%\colb{To further reduce the cost and energy efficiency of CF mMIMO system, the authors in \cite{Zhang_CF_HW_18,Masoumi_CF_HW_20,CF_UAV_22} assumed that the APs and UE \colr{additionally} employ low-cost RF transceivers. The low-cost RF transceivers, however, experience power amplifier non-linearities and phase noise, which degrade the benefits accrued from CF mMIMO technology~\cite{Zhang_CF_HW_18,ZhengZZZA20,Masoumi_CF_HW_20,ElhoushyH20,Zhang_CF_access_19,Elhoushy_CF_survey_21}.}
% Sven \textit{et. al} in \cite{jcobsson_ADCs_17} showed that  in a mMIMO system, ADCs with only a few bits of resolution are required to achieve an SE close to the one obtained using high-resolution ADCs. 
%showed that the \colr{asymptotic SE approximation for individual UE} (\colb{not clear}) \colb{Rayleigh uncorrelated} is mainly limited by its ADC resolution, when low-resolution ADCs are used at both APs and UEs. 
%------------------------------------
Low resolution ADC/DACs  improve the energy efficiency by reducing the power consumption, but also cause non-negligible SE loss due to coarse quantization. To alleviate this problem, the authors in~\cite{zhang_21_ADC_DAC,zhang_20_mixed_ADC_DAC} considered a mixed-ADC architecture, wherein one fraction of AP antennas has a high-resolution ADCs, while the other has low-resolution. Zhang \textit{et. al} in \cite{zhang_21_ADC_DAC} and \cite{zhang_20_mixed_ADC_DAC} investigated the SE of a CF mMIMO system with mixed-ADC architecture at the APs by considering spatially-uncorrelated Rayleigh- and Rician-fading respectively, and showed its superiority over its low-resolution counterpart.

 Recently, the authors in \cite{xiong_CF_21_dynamic_ADC,Verenzuela2021} proposed a dynamic ADC architecture and derived closed-form SE expressions for CF mMIMO with spatially-correlated Rayleigh channels. The dynamic ADC architecture offers the flexibility to tune the ADC resolution of each AP antenna. {This provides system designers with extra degrees-of-freedom for system design and optimization.} References~\cite{zhang2019_Low_ADC, zhang_CF_ADC_rician_22, hu2019_tcom,zhang_21_ADC_DAC,zhang_20_mixed_ADC_DAC,xiong_CF_21_dynamic_ADC,Verenzuela2021} considered a local SLD scheme and investigated the effect of low-/mixed-/dynamic-resolution ADCs alone. These works \textit{ignored RF impairments and channel aging}. 
{Also, the existing works~\cite{zhang_CF_ADC_rician_22,CF_UAV_22,zhang_20_mixed_ADC_DAC} modeled the Rician fading channel with a static line-of-sight phase. A small change in the UE location induces significant phase-shift in the LoS path. The authors in~\cite{Emil2019_rician_phase,Ozlem_CFwpt_21} modeled these phase-shifts as a uniformly distributed random variable, and derived closed-form SE expression for a CF mMIMO system, \textit{but without considering channel-aging, and hardware impairments}.}
 It is crucial to analyze the joint effect of RF impairments and ADC/DAC quantization in the presence of channel aging and Rician phase-shifts, which is a crucial gap in the existing CF mMIMO literature. \textit{The current work fills this gap, by analyzing a CF mMIMO system with two-layer LSFD, channel aging, Rician phase-shifts, low-cost RF chains and dynamic ADC architecture.}

We next summarize in Table~\ref{tab:literature} the relevant CF mMIMO literature focusing on LSFD, channel aging, RF and ADC hardware impairments and spatially-correlated Rician faded channels with phase-shifts. We infer from Table~\ref{tab:literature} that the existing correlated Rician-faded CF mMIMO literature has not yet investigated the: i) SE with channel aging and two-stage LSFD; ii) RF impairments and dynamic ADC architecture; and iii) sum SE optimization. Additionally, for Rayleigh channels, the CF mMIMO LSFD literature, has not investigated the i) SE with channel aging and non-ideal hardware and; ii) sum SE optimization.
%-
%
%----------------------------------------
\begin{table}[t]
	\caption{Summary of CF mMIMO literature focusing on LSFD and Channel aging.}\vspace{-6pt}
	\label{tab:literature}\footnotesize
	\begin{tabular}{|c|c|c|c|c|c|c|c|c|}	
		\hline
		Ref. & LSFD & Channel aging & \makecell{Rayleigh/\\[-5pt] Rician} & Correlation & \makecell{RF\\[-5pt]impairments}& \makecell{ADC\\[-5pt]architecture} & Optimization   \\ \hline\hline
		\cite{Emil2019_rician_phase} &\cmark &\xmark & Rician &\xmark &\xmark & ideal & \xmark\\ \hline
		\cite{chopra_AGING_21} &\xmark &\cmark & Rayleigh &\xmark &\xmark &ideal &\xmark \\ \hline
		\cite{Zheng_aging_correlated_21} &\cmark &\cmark &Rayleigh &\cmark &\xmark &ideal &\xmark \\ \hline
		\cite{ZhengZZZA20} &\cmark &\xmark &Rayleigh & \xmark&\cmark &ideal &\xmark \\ \hline 
		\cite{ElhoushyH20} &\xmark &\cmark &Rayleigh &\xmark &\cmark & ideal &max-min SE \\ \hline 
		\cite{zhang_CF_ADC_rician_22} &\xmark &\xmark &Rician &\xmark &\xmark &low-resolution &weighted max-min SE \\ \hline
		\cite{zhang_20_mixed_ADC_DAC} &\xmark &\xmark &Rician &\xmark &\xmark & mixed-resolution &\xmark\\ \hline
		\cite{xiong_CF_21_dynamic_ADC} &\xmark &\xmark &Rayleigh &\cmark &\xmark &dynamic-resolution &\xmark \\ \hline
		\textbf{Pr.} &\cmark &\cmark &\textbf{Rician} & \cmark &\cmark &\textbf{dynamic-resolution} & \textbf{sum-SE} \\ \hline
	\end{tabular}\vspace{-25pt}
\end{table}
The \textbf{main contributions} of the current work address these~gaps as follows: 
\begin{itemize}
\item We consider a spatially-correlated  Rician-faded  CF mMIMO system with channels aging, and investigate the impact of dynamic ADC architecture and low-cost hardware-impaired RF chains. We also consider the two-layer LSFD, and derive a closed-form SE expression by addressing the derivation difficulties caused by the combined modelling of channel aging, RF impairments, dynamic ADC architecture and spatially-correlated Rician channels.
 
\item We use the derived closed-form SE expression to maximize the non-convex SE metric by optimizing the UE transmit powers and the LSFD coefficients, which is the second contribution of this work. We propose two optimal power allocation schemes using minorization-maximization (MM) technique, which iteratively maximizes the convex surrogate of the non-convex objective~\cite{sun_majorization_17_TSP}. We propose a novel convex surrogate function, and analytically show that it satisfies the desirable surrogacy properties~\cite{sun_majorization_17_TSP}. This approach, however, has a high computation complexity as it requires off-the-shelf optimization solvers~\cite{sun_majorization_17_TSP}. We next design a low-complexity practically-implementable power allocation technique by using the Lagrangian dual transform technique~\cite{shen_QT_TSP_18}. This enables us in designing the closed-form transmit power update, which has a trivial computational complexity. It is extremely useful for designing optimal power for practical CF mMIMO systems with channel aging.  

\item  We show that LSFD can effectively mitigate  the detrimental effects of i) channel aging for both low and high UE velocities; and ii) IUI for low-velocity UEs but not for high-velocity UEs.  We also show that the increased number of AP antennas cannot  mitigate the SE degradation due to channel aging. 
\end{itemize}
The rest of the paper is organized as follows. Section~\ref{Sys_model_section}  discusses the channel model, uplink channel estimation and data transmission, and LSFD for the proposed hardware-impaired CF mMIMO system with channel aging. Section~\ref{SE-analysis_sect} derives and analyzes the closed-form SE expression for the aforementioned system. Section~\ref{sect_SE_opti} proposes two optimal power allocation schemes to optimize SE. Section~\ref{simulation_section1} first numerically validates the efficacy of the derived closed-form SE. It then investigates the effect of channel aging and LSFD on the SE. It then compares the complexity of the proposed optimization techniques. Section~\ref{Conclusion7} finally concludes the paper.

\textit{Notations:} Bold-faced lower- and upper-case alphabet denote vectors and matrices, respectively. Superscripts $(\cdot)^{*}$, $(\cdot)^{T}$, $(\cdot)^{H}$ and $(\cdot)^{-1}$ denote the conjugate, transpose, conjugate transpose and inverse, respectively, and $\mathbb{E}\{\cdot\}$ is the expectation operator. Trace and diagonal of a matrix $\Xmat$ are denoted as $\text{tr}(\Xmat)$ and $\text{diag}(\Xmat)$, respectively. Also, $\|\cdot \|$ is the Euclidean 2-norm, $|\cdot|$ is the absolute value and $\text{real}\{\cdot\}$ is the real-part of the argument. The notation $\mathcal{CN}(\mathbf{0},\Imat)$ denotes a complex circular Gaussian random vector with zero mean and covariance matrix $\Imat$. \\
%We use $\mathbf{0}$ to denote the null vector, and $\Imat$ to denote the identity matrix.

%---------------------------------------------------------------------------
\vspace{-1.2cm}
\section{System Model} \label{Sys_model_section}
\begin{figure*}[htpb] %\setcounter{figure}{1} %{0.5\textwidth}
	\centering 
	%\begin{subfigure}[b]{.45\linewidth}%\setcounter{subfigure}{3}
	\includegraphics[scale=0.6]{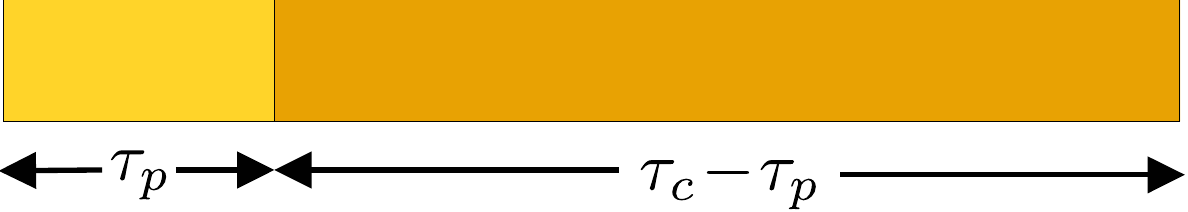} \vspace{-10pt} 
	\caption{\small Structure of resource block of length $\tau_c$ time instants.}
	\label{trasmission_frame1}
\end{figure*}

\vspace{-.1cm}
We consider the uplink of a CF mMIMO system with $M$ multi-antenna APs and $K$ single-antenna UEs. Each AP has $N$ antennas.  We assume, similar to \cite{Ozlem_CFwpt_21, hu2019_tcom}, that the APs are randomly distributed over a large geographical area, and are connected to a CPU via high-speed fronthaul links. To reduce the system hardware cost and power consumption, APs and UEs are equipped with low-cost hardware-impaired RF chains. Further, the APs have a dynamic-resolution ADC architecture, wherein each AP antenna can be connected to a different resolution ADC. This is unlike  \cite{zhang2019_Low_ADC,zhang_21_ADC_DAC,zhang_20_mixed_ADC_DAC}, which assume that the ADCs either have a low  or a mixed resolution.  The  UEs are designed using low-resolution DACs. 
In \textit{correlated} Rician-faded cell-free mMIMO systems, the channel does not harden. For high speed UEs, the channel will age. To investigate channel aging effect,  we consider, as shown in Fig. \ref{trasmission_frame1},  a resource block of length $\tau_c$ time instants.  
%%-----------------------------------------------------
\vspace{-.7cm}
%-----------------------------------------------------
The channel remains constant for a time instant, and varies across time instants in a correlated manner. This temporal correlation is modeled later using Jake's model~\cite{Zheng_aging_correlated_21}.  We also assume that each resource block is divided into uplink training and data transmission intervals of length $\tau_p$ and $(\tau_c-\tau_p)$ time instants, respectively.  We next model the UE-AP channel.

\vspace{-.8cm}
\subsection{Channel model} %\underline{\textbf{Channel model:}}
\vspace{-.2cm}
 The  channel from the $k$th UE to the $m$th AP at the $\lambda$th time instant is denoted as  $\hv_{mk}[\lambda]$. Due to dense AP deployment, the channel $\hv_{mk}[\lambda]$ contains  both LoS and NLoS paths. 
It, therefore, has a  Rician probability density function (pdf), and is modeled as follows \cite{Emil2019_rician_phase}:
\begin{align}
 \!\!\!   \hv_{mk}[\lambda] =  \bar{\hv}_{mk}e^{j\phi_{mk}^{\lambda}}\!+\! \Rmat_{mk}^{\frac{1}{2}}\tilde{\hv}_{mk}[\lambda]. \label{eq: h_mk[lambda]} %\\[-32pt] \notag
\end{align} 
Here $\bar{\hv}_{mk} = \sqrt{\frac{ \!K_{mk}\beta_{mk}}{\!K_{mk}+1}}\breve{\hv}_{mk}$, $ \Rmat_{mk} \!= \frac{\beta_{mk}}{\!K_{mk}\!+\!1}\breve{\Rmat}_{mk}$. The term $K_{mk}$ is the Rician factor, and $\beta_{mk}$ is  the large-scale fading coefficient. 
 The LoS term $ \breve{\hv}_{mk}$ is modeled  as $ \breve{\hv}_{mk}=[1, e^{j\psi_{mk}}, \cdots, e^{j(M-1)\psi_{mk}}]^T$, where $\psi_{mk}$ is the angle of arrival from the $k$th UE to the $m$th AP. The term, $\tilde{\hv}_{mk}[\lambda] $, with pdf $\mathcal{CN}(\mathbf{0},\mathbf{I}_{N})$ models the small-scale NLoS fading. The term $\breve{\Rmat}_{mk}$ is the spatial correlation matrix. The LoS phase-shift  $\phi_{mk}^{\lambda}$ at the $\lambda$th instant is uniformly distributed between $[-\pi, \pi]$. 
 The long-term channel statistics i.e., $ \bar{\hv}_{mk}$, $\beta_{mk}$ and $\Rmat_{mk}$, remain constant over a resource block and, similar to~\cite{Emil2019_rician_phase}, are assumed to be perfectly known at the AP.  Most of the  Rician CF works e.g.,~\cite{Jin_CF_EXP_CORR_20,Femenias_CF_CORR_Rici_20,CF_UAV_22}, assume that the  LoS phase-shift $\phi_{mk}^{\lambda}$ is  static. A slight change in UEs position, however, can radically modify the phase. It is, thus, crucial to consider a dynamic phase-shift $\phi_{mk}^{\lambda}$ in the Rician channel to practically model it \cite{Emil2019_rician_phase}.   We accordingly, similar to~\cite{Emil2019_rician_phase}, assume that $\phi_{mk}^{\lambda}$ varies as frequently as small-scale fading, and that the AP is unaware of it. As an AP does not have the prior knowledge of $\phi_{mk}^{\lambda}$, it estimates the channel $\hv_{mk}$, without its~knowledge.
 %A small change in the UE location may create a change in phase-shift and small-scale fading. \colr{existing works- static phase}
%The LoS phase-shift $\phi_{mk}^{\lambda}$, similar to small-scale fading component, varies at each time instant \cite{Emil2019_rician_phase}.

As discussed earlier, due to UEs mobility, the AP-UE channel in a {resource block varies across time instants, which
causes channel aging~\cite{Zheng_aging_correlated_21}. To analyze its effect, we model the channel $\hv_{mk}[n]$ at the $n$th time instant $1\leq n \leq \tau_c$},  as a combination of the channel $\hv_{mk}[\lambda]$ at the time instant $\lambda$,  and the innovation component as follows \cite{zheng_aging_20,Zheng_aging_correlated_21}:\vspace{-3pt} 
\begin{align}
    \hv_{mk}[n] =\rho_{k}[\lambda-n]\hv_{mk}[\lambda] + \sqrt{1-\rho_{k}^{2}[\lambda- n]} \Big( \bar{\hv}_{mk}e^{j\phi_{mk}^{n}} + \fv_{mk}[n] \Big). \label{eq_channel_h[n]}\\[-32pt]\notag
\end{align}
%$\overline{\rho}_{k}[\lambda- n]=\sqrt{1-\rho_{k}^{2}[\lambda- n]}$
The temporal correlation $\rho_{k}[\lambda-n]$, based on the Jake's model~\cite{Zheng_aging_correlated_21}, is given as $\rho_{k}[\lambda-n]= J_{0}(2\pi f_{d,i}T_s (\lambda-n))$.  Here $J_{0}(\cdot)$ is the zeroth-order Bessel function, and $T_s$ is the sampling time. The term  $f_{d,i}=(v_{k}f_{c})/c$ is the Doppler spread, with $v_{k}$, $f_{c}$ and $c$ being the user velocity, carrier frequency and the velocity of light, respectively. The innovation component $\fv_{mk}[n]$ is independent of the channel at the time instant $\lambda$ i.e., ${\hv}_{mk}[\lambda]$, and has a pdf $\mathcal{CN}(\mathbf{0},\Rmat_{mk})$~\cite{Zheng_aging_correlated_21}.
\vspace{-0.6cm}
\subsection{Uplink training}
\vspace{-0.2cm}
%\colr{$1\leq n \leq \tau_p$, $n=t_i$} $\phi_{k}[t_k]\in \{0,1\}$. 
Recall that the uplink training phase consists of $\tau_p$ time instants. The $k$th UE, similar to~\cite{Zheng_aging_correlated_21}, transmits its pilot signal $\sqrt{\tilde{p}_{k}}\phi_{k}[t_k]$ at the time instant $t_k\subset \{1, \dots, \tau_p\}$. Here  $|\phi_{k}[t_k]|^2=1$. 
We assume that the uplink training duration $\tau_p< K$. The number of UEs transmitting pilot at a particular time instant is more than one, which causes pilot contamination~\cite{Emil2019_rician_phase}. The set of UEs that transmit pilots at the time instant $t_k$ is denoted as  $\mathcal{P}_k$.   The $k$th UE feeds its pilot signal to the low-resolution DAC, which distorts it.  This distortion is commonly analyzed using Bussgang model~\cite{demir_bussgang_21}. The distorted DAC output, based on the Bussgang model, is:\vspace{-5pt}
\begin{align}
s_{\text{DAC},k}^{p}[t_k]=Q(\sqrt{\tilde{p}_{k}}\phi_{k}[t_k])=\alpha_{d,k}\sqrt{\tilde{p}_{k}}\phi_{k}[t_k]+\upsilon^{p}_{\text{DAC},k}[t_k].\\[-32pt]\notag
\end{align}
%------------------------------------------------------
% \colr{Need to change the flow, First discuss ADC/DAC impairments, the equation and then the Bussgang model in detail. How Bussgang model is used for modelling the quantization noise, using distortion factor and quantization noise. We then discuss the RF impairments, the equation, and then its modelling.} 
 %------------------------------------------------------
% \colr{We assume that all the UEs are equipped with low-cost RF chains and Dynamic-resolution ADC/DAC architecture, which are prone to hardware impairments~\cite{dey_21_ADC_DAC}. We model these RF impairments using EVM model which adds an independent distortion term to the transmit/receive signal~\cite{emil_HW_15}. The ADC/DAC impairments are modeled using Bussgang model~\cite{demir_bussgang_21}. The pilot signal of $k$th UE is fed to a low-resolution DAC which distorts it as follows}
%------------------------------------------------------
%\begin{align}
%s_{\text{DAC},k}[t_k]=Q(\sqrt{p_{p_{k}}}\phi_{k}[t_k])=\alpha_{d,k}\sqrt{p_{p_{k}}}\phi_{k}[t_k]+\upsilon_{\text{DAC},k}[t_k].
%\end{align}
%------------------------------------------------------
Here $\alpha_{d,k} = 1\!-\iota_{d,k}$ with $\iota_{d,k}$ being the DAC distortion factor. The term $\upsilon^{p}_{\text{DAC},k}[t_k]$ is  zero-mean DAC quantization noise, which is uncorrelated with the input pilot signal, and has a variance $\alpha_{d,k}(1\!-\!\alpha_{d,k})\tilde{p}_{k}\mathbb{E}\big\{|\phi_{k}[t_k]|^2\big\}$ \cite{Emil2019_rician_phase}. The $k$th UE DAC output signal $s_{\text{DAC},k}^{p}[t_k]$ is  fed to its hardware- impaired RF chain, which adds a distortion term 
$\xi_{\text{RF},k}^{p}[t_k]$
\cite{emil_HW_15}.  The effective uplink pilot signal is, therefore,  given as $s_{\text{RF},k}^{p}[t_k] =  s_{\text{DAC},k}^{p}[t_k] +\xi_{\text{RF},k}^{p}[t_k]$. The distortion $ \xi_{\text{RF},k}^{p}[t_k]$ is independent of the input signal, and has a pdf $ \mathcal{CN}\big(0,\kappa_{t,k}^{2}\big(\mathbb{E}\{s^{p}_{\text{DAC},k}[t_k](s_{\text{DAC},k}^{p}[t_k])^{H}\}\big)\big)$~\cite{emil_HW_15}. The term $\kappa_{t,k}$ in the pdf models the UE transmit EVM~\cite{emil_HW_15}.
At the $t_k$ time instant, the $m$th AP receives at its antennas,  the sum of  pilot signals transmitted by UEs in the set $\mathcal{P}_k$ i.e.,\vspace{-5pt}  
\begin{align}
\yv_{m}^p[t_k]=\sum\limits_{i\in\mathcal{P}_{k}}\hv_{mi}[t_k]s_{\text{RF},i}^{p}[t_k].\label{eq_AP_rx_signal_without_HW}\\[-33pt]\notag
\end{align}
 
 To reduce the system cost, the  APs are designed using hardware-impaired RF chains and a dynamic-ADC architecture, which enables us to vary the resolution of each ADC from $1$ to a maximum of $D$ bits. The $m$th AP feeds the above pilots signal received at its antenna to the hardware-impaired RF chains, whose distorted output, based on the EVM model, is {\cite{emil_HW_15}}:\vspace{-5pt}  
\begin{align}
\yv_{\text{RF},m}^p[t_k] =  \yv_{m}^p[t_k]  +\boldsymbol{\eta}_{\text{RF},m}^{p}[t_k]+\zv_{m}^{p}[t_k].\label{eq_RX_signal_AP_RF_HW1}\\[-32pt]\notag
\end{align}
The term $\boldsymbol{\eta}_{\text{RF},m}^{p}[t_k]$, with pdf $\mathcal{CN}(\boldsymbol{0},\kappa_{r,m}^{2}\Wmat_{m}[t_k])$, models the RF hardware distortion. The term $\kappa_{r,m}^{2}$ is the AP receiver EVM and $\Wmat_{m}[t_k] = \text{diag}\big(\E\big\{\yv_{m}^p[t_k](\yv_{m}^p[t_k])^{H}\big| \hv_{mk}[t_k]\big\}\big)$. Note that the receive hardware impairment is proportional to the power of UEs transmit signals~\cite{emil_HW_15}. The vector $\zv_{m}[t_k]$, with pdf  $\mathcal{CN}(\boldsymbol{0},\mathbf{I}_N)$, is the additive white Gaussian noise (AWGN) at the $m$th AP.  The RF chain output is then fed to the dynamic-resolution ADCs, which distort it by adding quantization noise. The distorted ADCs output, based on the Bussgang model~\cite{demir_bussgang_21}, is given as\vspace{-4pt} 
\begin{align}
\yv_{\text{ADC},m}^{p}[t_k]= \Amat_{m}\yv_{\text{RF},m}^p[t_k]+{\nv^p_{\text{ADC},m}}[t_k]=\Amat_{m} \big(\yv_{m}^{p}[t_k]  +\boldsymbol{\eta}^p_{\text{RF},m}[t_k]+\zv^p_{m}[t_k] \big) +{\nv^p_{\text{ADC,m}}}[t_k].\label{eq_RX_signal_AP_RF_HW_ADC}\\[-31pt]\notag
\end{align}
 The matrix $\Amat_{m}=\text{diag}(1-\iota_{m,n},\cdots,1-\iota_{m,n})$, where $\iota_{m,n}$ is  the ADC distortion factor~\cite{demir_bussgang_21}. The zero-mean additive ADC quantization noise ${\nv^p_{\text{ADC},m}}[t_k]$ is uncorrelated with $\yv_{\text{RF},m}^{p}[t_k]$. It has a  covariance of $\Bmat_{m}\text{diag}\big(\mathbb{E}\big \{\yv_{\text{RF},m}^{p}[t_k](\yv_{\text{RF},m}^{p}[t_k])^{H}\big|\hv_{mk}[t_k]\big \}\big)$ with $\Bmat_{m}=\Amat_{m}(\Imat_{N}-\Amat_{m})$~\cite{demir_bussgang_21}. The existing CF mMIMO literature \cite{Zhang_CF_HW_18,ZhengZZZA20,Masoumi_CF_HW_20,ElhoushyH20,hu2019_tcom,zhang2019_Low_ADC,zhang_CF_ADC_rician_22}, except \cite{xiong_CF_21_dynamic_ADC}, has not investigated the dynamic-resolution ADC architecture with different diagonal elements of $\Amat_{m}$ i.e., $\iota_{m,p} \neq \iota_{m,q}$ for $p \neq q$. 
 It significantly complicates the SE analysis and derivation, when  compared with \cite{Zhang_CF_HW_18,ZhengZZZA20,Masoumi_CF_HW_20,ElhoushyH20,hu2019_tcom,zhang2019_Low_ADC,zhang_CF_ADC_rician_22}, 
which considers either low- or mixed-ADC architecture at the APs. Xiong \textit{et. al} in \cite{xiong_CF_21_dynamic_ADC} considered a CF mMIMO system with dynamic-ADC architecture, but with correlated Rayleigh fading channels and ideal RF chains at the APs and UEs. \textit{The current work, in contrast, studies the dynamic ADC/DAC architecture for a CF mMIMO system with i) low-cost RF chains; ii) spatially-correlated Rician channel with phase-shifts; and iii) channel aging.} The proposed architecture is generic, and reduces to its low-resolution and mixed-resolution counterparts by choosing $\Amat_m = (1-\iota_{m})\mathbf{I}_N$  and $\Amat_m= \text{blkdiag}\{(1-\iota_{m})\mathbf{I}_{\gamma},\mathbf{I}_{N-\gamma}\}$ with $0 \leq \gamma \leq N$, respectively.
% \begin{remark}
% 	{In practice, the elements of distortion vector $\boldsymbol{\eta}^p_{\text{RF},m}[t_k]$ are correlated. The authors  in \cite{emil_HW_impact_19} showed that this correlation can be neglected  in a mMIMO system, without significantly affecting the performance. We, therefore, similar to~\cite{Zhang_CF_HW_18,ZhengZZZA20,Masoumi_CF_HW_20,ElhoushyH20},
% 		assume that the elements of $\boldsymbol{\eta}^p_{\text{RF},m}[t_k]$ are uncorrelated, for a tractable SE analysis.}
% 	\end{remark}
% 
%  \colr{need to bring out dynamic-resolution difficulties. i.e., How the dynamic-resolution makes the matrix diagonal terms different and how these make the SE analysis difficult.}

%\colr{need to add footnote about \colb{uncorrelated hardware distortion}. see sdey paper} - for RF or ADC?
%-------------------------------------------------------------------------------
%$\Smat^{m}[t_k]=diag(\E\{\boldsymbol{y}_{RF}^{m}[t_k](\boldsymbol{y}_{RF}^{m}[t_k])^{H}|h_{mk}[t_k]\})$ for k=1,...K.
%-------------------------------------------------------------------------------

We next substitute expression of $\yv_{m}^{p}[t_k]$ from \eqref{eq_AP_rx_signal_without_HW} in \eqref{eq_RX_signal_AP_RF_HW_ADC}, and re-express $\yv_{\text{ADC},m}^{p}[t_k]$ as \vspace{+0.1pt}
\begin{align}
   \yv_{\text{ADC},m}^{p}[t_k]&= \!\sum_{i\in \mathcal{P}_k} \! \Amat_{m}\big(\hv_{mi}[t_k]\big(\!{\alpha_{d,i}\sqrt{\tilde{p}_{i}}}+\upsilon_{\text{DAC},i}^{p}[t_k]\!+\!\xi_{\text{RF},i}^{p}[t_k]\big)\!+\!\boldsymbol{\eta}_{\text{RF},m}^{p}[t_k]\!+\zv_{m}^{p}[t_k]\big)\!+\nv_{\text{ADC},m}^{p}[t_k]. \notag %\label{eq_Rx_pilot_ADC}
\end{align} 
%-------------------------------------------------------------------------------
% \begin{align}
%   & \colb{\yv_{\text{ADC},m}[t_k]}= \!\sum_{i\in \mathcal{P}_k} \! \Amat_{m}\bigg\{\!\hv_{mi}[t_k]\Big(\!\big(1\!-\!\rho_{d,i}\big)\sqrt{p_{i}}+n_{\text{DAC},i}[t_k]\!+\!\eta_{tu_{i}}[t_k]\!\Big)\!+\!\boldsymbol{\eta}_{r,\text{AP}}^{m}[t_k]\!+\zv_{m}[t_k]\bigg\}\!+\nv_{\text{ADC},m}[t_k] \nonumber\\
%     & \qquad =\!\sum_{i\in \mathcal{P}_k}\!\!\Amat_{m}\Big(\!\overline{\hv}_{mk}\Big(\! \rho_{k}[\lambda \!-\!t_k]e^{j\phi_{mk}^{\lambda}} \!+\!\overline{\rho_{k}}[\lambda \!-\!t_k]e^{j\phi_{mk}^{t_k}}\! \Big)\!+\!\rho_{k}[\lambda \!-\! t_k]\Rmat_{mk}^{1/2}\tilde{\hv}_{mk}[\lambda]\! +\!\overline{\rho_{k}}[\lambda \!-\! t_k]\fv_{mk}[t_k] \Big) \nonumber\\
%     & \qquad \qquad \Big(\!(1\!-\!\rho_{d,i})\sqrt{p_{i}} \! +\!\upsilon_{\text{DAC},i}[t_k]\!+\!
%         \eta_{tu_{i}}[t_k]\! \Big)\!+\!\Amat_{m}\boldsymbol{\eta}_{r,\text{AP}}^{m}[t_k]\!+\!\Amat_{m}\zv_{m}[t_k]\!+\boldsymbol{n}_{\text{ADC}}^{m}[t_k] \label{eq:y_m[t_k]}
% \end{align}
%-------------------------------------------------------------------------------
 Recall that the channel between two different time instants ages, and is consequently  correlated. The received signal $\yv_{\text{ADC},m}^{p}[t_k]$ can be exploited while estimating channel at any other time instant also. The channel estimate quality will, however, deteriorate with increasing time difference between the pilot transmission ($1<n<\tau_p$) and the considered channel realization ($\tau_p+1<n<\tau_c$). We, therefore, without loss of generality, estimate the channel at the time instant $\lambda = \tau_p +1$, and use these estimates to obtain the channels at all other time instants ($n >\lambda$). To estimate the channel at the $\lambda$th time instant, we express the received pilot signal $\yv_{\text{ADC},m}^{p}[t_k]$ in terms of the channel at the time instant $\lambda$, using \eqref{eq_channel_h[n]}  as follows\vspace{-4pt}
\begin{align}
   \yv_{\text{ADC},m}^{p}[t_k]&  =\!\sum_{i\in \mathcal{P}_k}\!\!\Amat_{m}\left(\! \rho_{k}[\lambda \!-\! t_k]\hv_{mk}[\lambda]\! +\!\overline{\rho}_{k}[\lambda \!-\!t_k]\big(\bar{\hv}_{mk} e^{j\phi_{mk}^{t_k}} +\fv_{mk}[t_k]\big) \right) \Big(\!{\alpha_{d,i}\sqrt{\tilde{p}_{i}}} \! +\!\upsilon^p_{\text{DAC},i}[t_k]\!
    \nonumber\\[-6pt]
    &  \qquad + \xi^p_{\text{RF},i}[t_k] \Big)\!  +\Amat_{m}\boldsymbol{\eta}_{\text{RF},m}^p[t_k]+\Amat_{m}\zv_{m}^p[t_k]+\nv_{\text{ADC},m}^p[t_k]. \label{eq:y_m[t_k]} %\\[-31pt]\notag
\end{align}
Here $\overline{\rho}_{k}[\lambda \!-\!t_k] = \sqrt{1-\rho_{k}^{2}[\lambda \!-\!t_k]}$.
%\colr{explain the equation.}\\
 Using $\yv_{\text{ADC},m}^{p}[t_k]$, the channel $\hv_{mk}[\lambda]$  is estimated in the following Theorem, which is proved in Appendix~\ref{appendix_LMMSE}.
 %The access points (APs) are not aware of the random-phases shifts and the linear MMSE estimator is called phase-unaware estimator. 
\begin{theorem}\label{ch_estimate_theorem}
For a hardware-impaired CF mMIMO system with spatially-correlated Rician fading and phase-shifts, the linear minimum mean square error (LMMSE) estimate of $\hv_{mk}[\lambda]$ is given~as\vspace{-4pt}
\begin{align}
    \hat{\hv}_{mk}[\lambda]=\sqrt{\tilde{p}_{k}} \alpha_{d,k}\rho_{k}[\lambda-t_k]\bar{\Rmat}_{mk}\Amat_{m}\boldsymbol{\Psi}_{mk} \yv_{\text{ADC},m}^{p}[t_k]\;, \text{where } \overline{\mathbf{R}}_{mk}=\big( \bar{\hv}_{mk}\bar{\hv}_{mk}^{H}+\Rmat_{mk}\big).  \label{eq: LMMSE} \\[-32pt]\notag
\end{align}
The matrix $\boldsymbol{\Psi}_{mk} = \!\Big(\sum_{i\in \mathcal{P}_k}\!\Amat_{m} \alpha_{d,i}(1+\kappa_{t,i}^{2})\tilde{p}_{i}\mathbf{\overline{R}}_{mk}\Amat_{m}^H \!
    +\!\left(\mathbf{B}_{a}^{m}+ \kappa^{2}_{r,m}\Amat_{m}\right)\Jmat_k + \sigma^{2}\Amat_{m}\! \Big)^{\!-1}\!\!$ with
$\Jmat_k=\sum_{j\in\mathcal{P}_{k}}(1+\kappa_{t,j}^{2})\alpha_{d,j}\tilde{p}_{j}\text{diag}\big(\bar{\Rmat}_{mj}\big)$.
\end{theorem}
The LMMSE estimation error $\tilde{\hv}_{mk}[\lambda]\!=\!\hv_{mk}[\lambda]\!-\hat{\hv}_{mk}[\lambda]$ has a zero mean, and covariance matrix $\Cmat_{mk} \!=\! \bar{\Rmat}_{mk}\!-\! \alpha_{d,k}^2\tilde{p}_{k}\rho_{k}^2[\lambda\!-\!t_k]\bar{\Rmat}_{mk}\Amat_{m} \Amat_{m}^H \bar{\Rmat}_{mk}^H $. The estimate $\hat{\hv}_{mk}[\lambda]$ is uncorrelated with~the~error $\tilde{\hv}_{mk}[\lambda]$. 
The channel estimator, derived in \cite{Zheng_aging_correlated_21}, by assuming Rayleigh fading and ideal RF chains and ADC/DAC, cannot be used herein. This is because the hardware-impaired RF chains and dynamic/low-resolution ADC/DAC architecture change the structure and computation of $\boldsymbol{\Psi}_{mk}$.\vspace{-4pt} 
%\colr{This is due to the hardware-impaired RF chains and dynamic/low-resolution ADC/DAC architecture, we need to compute $\boldsymbol{\Psi}_{mk}$, which \cite{Zheng_aging_correlated_21} need~not.}

%----------------------------------------------
\begin{corollary} \label{ph_aware_estimator}
If the $m$-th AP knows the LoS phase-shift $\phi_{mk}^{\lambda}$, the MMSE estimate of the UE $k$ channel, $\hv_{mk}[\lambda]$, is given as\vspace{-5pt}
\begin{align}
    \hat{\hv}_{mk}[\lambda]= \bar{\hv}_{mk}e^{j\phi_{mk}^{\lambda}} + \sqrt{\tilde{p}_{k}}\alpha_{d,k}\rho_{k}[\lambda-t_k]\Rmat_{mk}\Amat_{m}\widetilde{\boldsymbol{\Psi}}_{mk} \big(\yv_{\text{ADC},m}^{p}[t_k]- \overline{\yv}_{\text{ADC},m}^{p}[t_k] \big)\;, \text{where }  \label{eq:LMMSE_p_aware}
\end{align}
$\overline{\yv}_{\text{ADC},m}^{p}[t_k] = \sum\limits_{i\in \mathcal{P}_k} \sqrt{\tilde{p}_{i}} \Bar{\hv}_{mk}\Big(\rho_{k}[\lambda-t_k] e^{j\phi_{mk}^{\lambda}} + \overline{\rho}_{k}[\lambda-t_k] e^{j\phi_{mk}^{t_k}}\Big)$.
The matrix $\widetilde{\boldsymbol{\Psi}}_{mk}$ can be obtained from ${\boldsymbol{\Psi}}_{mk}$ in \eqref{eq: LMMSE} by replacing $\bar{\Rmat}_{mk}$ with $\Rmat_{mk}$. {The analytical proof is not included here due to lack of space. It can, however, be derived on the lines similar to Appendix~\ref{appendix_LMMSE}.}
\end{corollary}
%----------------------------------------------

\begin{remark}
{We note that the LMMSE channel estimation method does not fully utilize the temporal correlation among different time sample. Its major advantage, however,  is its closed-form solution which crucially helps us in deriving a closed form SE expression in Section~\ref{SE-analysis_sect}, which is a function only of the long-term channel statistics. This closed-form SE expression is then crucially used to derive  a low-complexity solution for its optimization, which we do in Section~\ref{sect_SE_opti}. In contrast, the \textit{iterative} Sparse Bayesian Learning (SBL) method in~\cite{SBL_2019} exploits the channel temporal correlation, but does not provide a closed-form solution. This will radically complicate the SE closed form derivation and its optimization. Further,  the SBL method proposed in~\cite{SBL_2019} cannot be trivially  extended to our CF mMIMO system with RF impairments, ADC/DAC quantization noise and Rician phase-shifts. Reference~\cite{SBL_2019} did not consider these impairments. A closed-form SBL method for the system considered herein, which can help in deriving closed form SE expression, however, is an interesting direction of future work.}
\end{remark}
\vspace{-.8cm}
%\underline{\textbf{Uplink Transmission:}}
\subsection{Uplink Transmission}
Let $s_{k}[n]$, with $\mathbb{E}\{|s_{k}[n]|^2\}=1$, be the information symbol which the $k$th UE wants to transmit at the $n$th  time instant. The symbol $s_{k}[n]$, after scaling with power control coefficient $\sqrt{p_k}$, is fed to the low-resolution DAC. Its distorted output, based on the Bussgang model~\cite{demir_bussgang_21}, is given as $    s_{\text{DAC},k}[n] \!= \!\alpha_{d,k}\sqrt{p_k}s_{k}[n]\!+\!\upsilon_{\text{DAC},k}[n]
$. The term $\upsilon_{\text{DAC},k}[n]$~is the DAC quantization noise~\cite{demir_bussgang_21}, and has a zero mean and  variance $\!(1\!-\alpha_{d,k})\alpha_{d,k}p_{k}$. It is uncorrelated with the information signal $s_{k}[n]$. The DAC output $s_{\text{DAC},k}[n]$ is fed to the low-cost hardware-impaired RF chain, whose distorted output, based on the EVM model~\cite{emil_HW_15}, is given  as\vspace{-7pt}
%The distorted RF signal is modeled using EVM, which adds an independent distortion term $\xi_{\text{RF},k}[n]$ to its input signal~\cite{emil_HW_15}.
%------------------------------------------------------
\begin{align}
    s_{\text{RF},k}[n]= s_{\text{DAC},k}[n] + \xi_{\text{RF},k}[n]  =\alpha_{d,k}\sqrt{p_k}s_{k}[n]+\upsilon_{\text{DAC},k}[n]+ \xi_{\text{RF},k}[n]. \\[-32pt]
\nonumber
\end{align}
 The distortion term $\xi_{\text{RF},k}[n]$ is the transmit RF hardware impairment of the $k$th UE. It has a pdf $\mathcal{CN}(0,\kappa_{t,k}^{2}(\mathbb{E}\{s_{\text{DAC},k}[n](s_{\text{DAC},k}[n])^{H}\}))$,  with $\kappa_{t,k}$ being its transmit EVM~\cite{Emil2019_rician_phase}. 

The $m$th AP receives the following sum signal at its antennas: $\yv_m[n] = \sum\limits_{k=1}^{K}\hv_{mk}[n]s_{\text{RF},k}[n]$.  The AP feeds this receive signal to its hardware-impaired RF chains, which distorts it as\cite{emil_HW_15}:\vspace{-7pt}
\begin{align} \label{eq_y_m_RF_HW}
    \yv_{\text{RF},m}[n]=  \yv_{m}[n]  +\boldsymbol{\eta}_{\text{RF},m}[n]+{\zv_{m}[n]}. \\[-32pt]
\nonumber
\end{align}
Here $\boldsymbol{\eta}_{\text{RF},m}[n]$, with pdf $\mathcal{CN}(\mathbf{0},\kappa_{r,m}^{2}\Wmat^{m}[n])$, is the receiver hardware distortion, and $\Wmat^{m}[n] =\text{diag}(\mathbb{E}\{\yv_{m}[n]\yv_{m}[n]^{H}| \hv_{mk}[n]\})$. The term $\zv_{m}[n]$, with pdf $ \mathcal{CN}(\boldsymbol{0},\mathbf{I}_N)$, is the AWGN at the $m$th AP. The RF chain output is then fed to the dynamic-resolution ADCs, whose quantized and

\begin{figure*}[htpb] %\setcounter{figure}{1} %{0.5\textwidth}
	\centering 
	%\begin{subfigure}[b]{.45\linewidth}%\setcounter{subfigure}{3}
	\includegraphics[scale=0.7]{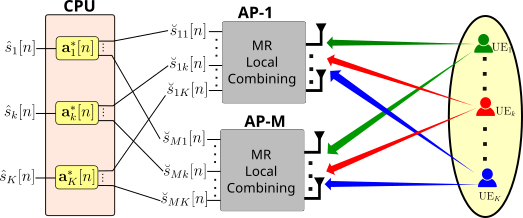} \vspace{-10pt} 
	\caption{\small LSFD architecture for CF mMIMO system.}
	\label{LSFD1}
\end{figure*}

\hspace{-10pt} noisy output, based on the Bussgang model~\cite{demir_bussgang_21}, is given as follows:%\vspace{+0.1pt}
\begin{equation}
 \yv_{\text{ADC},m}[n]= \Amat_{m}\yv_{\text{RF},m}[n] + \nv_{\text{ADC},m}[n]= \Amat_{m} \Big(\sum_{k=1}^{K}\hv_{mk}[n]s_{\text{RF},k}[n]  +\boldsymbol{\eta}_{\text{RF},m}[n]+\zv_{m}[n]\Big) + \nv_{\text{ADC},m}[n].\notag  %\\[-33pt] \nonumber
\end{equation}%\vspace{-0.1pt}
The matrix $\Amat_m = \text{diag}\{\iota_{m,1},\cdots,\iota_{m,N}\}$, with $\iota_{m,i}$ being the ADC distortion factor for the $i$th antenna of the $m$th AP. The vector $\nv_{\text{ADC},m}$ is the quantization noise, which is uncorrelated with the input signal $\yv_{\text{RF},m}[n]$. It has a zero mean and covariance $\boldsymbol{\Theta}_{m} = \Bmat_{m}\Smat^{m}[n]$, where $\Bmat_{m}=\Amat_{m} (\Imat_N - \Amat_{m})$ and $\Smat^{m}[n]=\text{diag}(\E\{{\yv_{\text{RF},m}[n]}{\yv^{H}_{\text{RF},m}[n]}|\hv_{mk}[n]\})$~\cite{demir_bussgang_21}. \newline  
\textbf{Principles of two-layer decoding:} %\underline{\textbf{Two-layer decoding:}}
The CF system considered herein,  as shown in Fig.~\ref{LSFD1},  employs two-layer decoding to mitigate IUI. In the first-stage, each AP combines its received signal by using local channel estimates, which only partially mitigate the IUI. To mitigate the residual IUI,  all APs sends their locally-combined received signal to the CPU, which  performs the second-layer LSFD. The CPU computes the LSFD weights based on the large-scale fading coefficients, which remains constant for $100$s of coherence intervals~\cite{Ozlem_CFwpt_21,Emil_CF_BOOK_20}. These coefficients, can thus, be easily computed and stored at the CPU~\cite{Ozlem_CFwpt_21}. We now explain the two-step decoding in detail.
 %%-----------------------------------------------------
 %\vspace{-.7cm}
%-----------------------------------------------------

%---------------------------------------------------------------------

The $m$th AP first uses channel estimate $\Hat{\hv}_{mk}[\lambda]$ to combine the distorted received signal as\vspace{-4pt} 
\begin{align}
\Breve{s}_{km}[n] = \Hat{\hv}^{H}_{mk}[\lambda]{\yv_{\text{ADC},m}[n]}. \\[-32pt]
\nonumber
\end{align} 
It sends the combined signal $\Breve{s}_{km}[n]$ to the CPU which performs second-layer LSFD as  $\Hat{s}_{k}[n]= \sum_{m=1}^{M}a^{*}_{mk}[n]\Breve{s}_{km}[n]$. Here $a_{mk}$ is the complex LSFD coefficient of the $m$th AP and the $k$th UE link. {This reduces the IUI by weighing the received signals from all APs. We note that $a^{*}_{mk}=1\; \forall m,k$, corresponds to the conventional matched filtering based SLD considered in~\cite{HQNgo2017}.} 

We now re-express the signal $\Hat{s}_{k}[n]$ as\vspace{-4pt}
\begin{align}\label{eq_signal_at_CPU}
   & \Hat{s}_{k}[n]=\underbrace{ \sum_{m=1}^{M}a_{mk}^{*}[n]\alpha_{d,k}\sqrt{p_{k}}\Hat{\hv}_{mk}^{H}[\lambda]\Amat_{m}\hv_{mk}[n]s_{k}[n]}_{\text{Desired Signal, }\widetilde{\text{DS}}_{k,n}} + \sum_{i\neq k}^{K}\underbrace{\sum_{m=1}^{M}a_{mk}^{*}[n]\alpha_{d,k}\sqrt{p_{i}} \Hat{\hv}_{mk}^{H}[\lambda]\Amat_{m}\hv_{mi}[n]s_{i}[n]}_{\text{Inter-user interference, } {\text{IUI}}_{ki,n}} \nonumber\\[-2pt]
    &\;\;+\underbrace{\sum_{m=1}^{M}a_{mk}^{*}[n]\Hat{\hv}_{mk}^{H}[\lambda]\Amat_{m}\Big(\sum_{i=1}^{K}\hv_{mi}[n]\upsilon_{\text{DAC},i}[n]\Big)}_{\text{UE DAC impairment, } {\text{DAC}}_{k,n}} +\underbrace{\sum_{m=1}^{M}a_{mk}^{*}[n]\Hat{\hv}_{mk}^{H}[\lambda]\Amat_{m}\Big(\sum_{i=1}^{K}\hv_{mi}[n]\xi_{\text{RF},i}[n]\Big)}_{\text{UE RF impairment, } {\text{TRF}}_{k,n}} \nonumber\\[-2pt]
    &\;\;+\underbrace{\sum_{m=1}^{M}a_{mk}^{*}[n]\Hat{\hv}_{mk}^{H}[\lambda]\Amat_{m}\boldsymbol{\eta_{\text{RF},m}}[n]}_{\text{AP RF impairment, } {\text{RRF}}_{k,n}}  +\underbrace{\sum_{m=1}^{M}a_{mk}^{*}[n]\Hat{\hv}_{mk}^{H}[\lambda]\nv_{\text{ADC},m}[n]}_{\text{AP ADC impairment, } {\text{ADC}}_{k,n}} +\underbrace{\sum_{m=1}^{M}a_{mk}^{*}[n]\Hat{\hv}_{mk}^{H}[\lambda]\Amat_{m}{\zv}_{m}[n]}_{\text{AWGN, } {\text{NS}}_{k,n}}. \notag \\[-15pt]
%\nonumber 
\end{align}

The first term $\widetilde{\text{DS}}_{k,n}$ denotes the desired signal, the second term $\sum_{i \neq k}^{K}{\text{IUI}}_{ki,n}$ is due to the signal transmitted by the UEs $i \neq k$ to APs. The third and  fourth terms denote the distortion due to the low-resolution DAC and the RF hardware at the UE, respectively. The fifth and the sixth terms denote the impairments due to the dynamic-resolution ADCs and the low-cost RF hardware at the AP. The last term  denotes the AWGN at the AP. 
We see that, in contrast to \cite[Eq. (15)]{Zheng_aging_correlated_21}, the low-cost RF and dynamic-resolution ADC/DAC adds four extra terms ${\text{DAC}}_{k,n}$, ${\text{TRF}}_{k,n}$, ${\text{RRF}}_{k,n}$ and ${\text{ADC}}_{k,n}$. These impairments also non-trivially modify the other terms in \eqref{eq_signal_at_CPU}, and require novel mathematical results to calculate the closed-form SE expression.
%The presence of hardware impairments requires calculating new terms in the SE closed-form expressions. Simplification of these terms require novel mathematical results which  
%\colr{we can compare this equation with the existing rayleigh work and show that our system is more generic.} \\
\vspace{-0.2cm}
\section{Spectral efficiency Analysis} \label{SE-analysis_sect} %\vspace{-0.2cm}
{We now derive a closed-form SE expression for the CF mMIMO system with i) channel aging; ii) RF and dynamic ADC/DAC architecture; and iii) spatially-correlated Rician channel with phase-shifts. The closed-form SE expression, derived using use-and-then-forget (UatF) technique \cite{Zheng_aging_correlated_21,Emil_CF_BOOK_20},  is valid for a finite number of antennas, and requires only long-term channel statistics.   The UatF technique re-expresses \eqref{eq_signal_at_CPU} by decomposing the desired signal $\widetilde{\text{DS}}_{k,n}$ therein  as follows:  
\begin{align}\label{eq_signal_BU11}
 & \Hat{s}_{k}[n]=\underbrace{  \sum\limits_{m=1}^{M}a_{mk}^{*}[n]\alpha_{d,k}\rho_{k}[n-\lambda]\sqrt{p_{k}}\E\big\{\hat{\hv}_{mk}^{H}[\lambda]\Amat_{m}\hv_{mk}[\lambda] \big\}   }_{\text{Desired signal, }{\text{DS}}_{k,n}}\!s_{k}[n] + {\text{IUI}}_{ki,n}  + {\text{DAC}}_{k,n} \\[-4pt]
  &\!\! + \underbrace{ \sum\limits_{m=1}^{M}a_{mk}^{*}[n]\alpha_{d,k}\rho_{k}[n-\lambda]\sqrt{p_{k}}\Big(\hat{\hv}_{mk}^{H}[\lambda]\Amat_{m}\hv_{mk}[\lambda] - \E\big\{\hat{\hv}_{mk}^{H}[\lambda]\Amat_{m}\hv_{mk}[\lambda] \big\} \Big) }_{\text{Beamforming uncertainty, } {\text{BU}}_{k,n}}\!s_{k}[n]  \nonumber\\[-3pt]
  & \!\!+\underbrace{ \sum\limits_{m=1}^{M}\!a_{mk}^{*}[n]\alpha_{d,k}\overline{\rho}_{k}[n\!-\!\lambda]\sqrt{p_{k}}\hat{\hv}_{mk}^{H}[\lambda]\Amat_{m}\bar{\fv}_{mk}[t_k]}_{ \text{Channel aging, } {\text{CA}}_{k,n}}\!s_{k}[n] 
 + {\text{TRF}}_{k,n} +  {\text{RRF}}_{k,n} + {\text{ADC}}_{k,n} + {\text{NS}}_{k,n}.\notag
\end{align}
We see that the desired signal term $\widetilde{\text{DS}}_{k,n}$ is now decomposed as $\widetilde{\text{DS}}_{k,n}={\text{DS}}_{k,n}+{\text{BU}}_{k,n}+{\text{CA}}_{k,n}$. The modified desired signal term ${\text{DS}}_{k,n}$, which is used for data detection, requires only long term channel information \cite{Zheng_aging_correlated_21,Emil_CF_BOOK_20}. The term ${\text{BU}}_{k,n}$ is the beamforming uncertainty \cite{Zheng_aging_correlated_21,Emil_CF_BOOK_20}.  The term ${\text{CA}}_{k,n}$, which denotes channel aging, is because the channel $\hv_{mk}[n]$ at the $n$th time instant is now expressed using \eqref{eq_channel_h[n]}  as a combination of the channel $\hv_{mk}[\lambda]$ at the time instant $\lambda$,  and its innovation component $\bar{\fv}_{mk}[t_k] = \bar{\hv}_{mk} e^{j\phi_{mk}^{t_k}} +\fv_{mk}[t_k]$. In \eqref{eq_signal_BU11}, the sum of various terms, except ${\text{DS}}_{k,n}$ can be treated as effective noise.
%The sum of BUI, CA, IUI, RF and ADC/DAC impairments at AP and UE, and AWGN are treated as the effective noise~\cite{Emil_CF_BOOK_20}.
 Using central limit theorem, the sum can be approximated as  a worst-case Gaussian noise~\cite{Emil_CF_BOOK_20}. 
%----------------------------------------------------------------------
With this assumption, a  SE lower bound for a hardware-impaired spatially-correlated Rician-faded CF mMIMO system with channel aging for a given LSFD weights is given as: $\text{SE}_{k,n}= \text{log}_{2}(1+ \overline{\text{SINR}}_{k,n})$, where \vspace{+0.1pt}
\begin{equation}
\!\!\!\!\overline{\text{SINR}}_{k,n} =  \frac{\overline{\text{DS}}_{k,n}}{\overline{\text{BU}}_{k,n}\!+\overline{\text{CA}}_{k,n} \!+\! \sum\limits_{i\neq k}^{K}\!\overline{\text{IUI}}_{ki,n} \!+ \!\overline{\text{DAC}}_{k,n} \!+\!\overline{\text{TRF}}_{k,n}  \!+\overline{\text{RRF}}_{k,n} \!+\overline{\text{ADC}}_{k,n} \!+\overline{\text{NS}}_{k,n}}.   \label{eq_SE_expectations} 
\end{equation}\vspace{-2pt}
%----------------------------------------------------------------------
%\begin{equation}
%    \text{SINR}_{k,n} =  \frac{\overline{\text{DS}}_{k,n}}{\begin{Bmatrix}\overline{\text{BU}}_{k,n}+\overline{\text{CA}}_{k,n} + \sum_{i\neq k}^{K}\text{UI}_{k,i,n} + \text{DACtx}_{k,n}    \\
%    +\text{ HItx}_{k,n}  +\text{ HIrx}_{k,n} +\text{ADCrx}_{k,n}+\overline{\text{NS}}_{k,n}\end{Bmatrix}}.   \label{eq_SE_expectations}
%\end{equation}
%----------------------------------------------------------------------
The term  $\overline{\text{DS}}_{k,n}$ is the desired signal power, $\overline{\text{BU}}_{k,n}$ is the beamforming uncertainty power, $\sum\limits_{i\neq k}^{K}\overline{\text{IUI}}_{ki,n}$ is the interference power,  $\overline{\text{DAC}}_{k,n}$ is the UE DAC impairment power,  $\overline{\text{TRF}}_{k,n}$ is the UE RF impairment power,  $\overline{\text{RRF}}_{k,n}$ is the AP RF impairment power,  $\overline{\text{ADC}}_{k,n}$ is the AP ADC impairment noise and $\overline{\text{NS}}_{k,n}$ is the AWGN variance.
{These terms are mathematically defined in Table~\ref{table_DS_IUI_HW_ADC}, where $ \bar{a}_{mk}^{\rho *}[n] = a_{mk}^{*}[n]\alpha_{d,k}\rho_{k}[n_{\lambda}]$ and $ \bar{a}_{mk}^{\bar{\rho} *}[n] = a_{mk}^{*}[n]\alpha_{d,k}\bar{\rho}_{k}[n_{\lambda}]$.}
\begin{table*}[]   
 %\footnotesize
 \centering 
 \begin{adjustbox}{width=\columnwidth,center} %\footnotesize %%\smallsize
    \caption{Simulated expressions for the desired signal and interference terms.}  \vspace{-0.05in}
   \label{table_DS_IUI_HW_ADC}
  \begin{tabular}{|c|c|} 
 \hline
$\overline{\mbox{DS}}_{k,n}\!=\!\Big| \sum\limits_{m=1}^{M}\bar{a}_{mk}^{\rho *}[n]\sqrt{p_{k}}\E\big\{\hat{\hv}_{mk}^{H}[\lambda]\Amat_{m}\hv_{mk}[\lambda] \big\} \Big|^{2} \!$ &   $ \overline{\mbox{CA}}_{k,n}\!=\!\E\Big\{\Big| \sum\limits_{m=1}^{M}\bar{a}_{mk}^{\bar{\rho} *}[n]\sqrt{p_{k}}\hat{\hv}_{mk}^{H}[\lambda]\Amat_{m}\bar{\fv}_{mk}[\lambda]  \Big|^{2} \Big\}$ \\ 
$\overline{\mbox{IUI}}_{ki,n}\!=\!\!\E\Big\{ \Big| \sum\limits_{m=1}^{M}a_{mk}^{*}[n]\alpha_{d,i}\sqrt{p_{i}} \hat{\hv}_{mk}^{H}[\lambda]\Amat_{m}\hv_{mi}[n] \Big|^{2} \Big\}$ & $\overline{\text{RRF}}_{k,n}\!=\!\E\Big\{\Big| \sum\limits_{m=1}^{M}a_{mk}^{*}[n]\hat{\hv}_{mk}^{H}[\lambda]\Amat_{m}\boldsymbol{\eta}_{r,m}^{\text{AP}}[n] \Big|^{2} \Big\}$    \\
 $\overline{\text{TRF}}_{k,n}\!=\!\E\Big\{\Big|  \sum_{m=1}^{M}a_{mk}^{*}[n]\hat{\hv}_{mk}^{H}[\lambda]\Amat_{m}\hv_{mi}[n]\xi_{\text{RF},i}[n]  \Big|^{2} \Big\}\!$ &  $\overline{\text{ADC}}_{k,n}=\E\Big\{ \Big| \sum\limits_{m=1}^{M}a_{mk}^{*}[n]\hat{\hv}_{mk}^{H}[\lambda]\nv_{\text{ADC},m}[n] \Big|^{2} \Big\}$ \\  
 $\overline{\mbox{DAC}}_{k,n}=\E\Big\{\Big|  \sum\limits_{m=1}^{M}a_{mk}^{*}[n]\hat{\hv}_{mk}^{H}[\lambda]\Amat_{m}\hv_{mi}[n]\upsilon_{\text{DAC},i}[n]  \Big|^{2} \Big\}$ & $\overline{\text{NS}}_{k,n}\!\!= \E\Big\{ \Big|\sum\limits_{m=1}^{M}a_{mk}^{*}[n]\hat{\hv}_{mk}^{H}[\lambda]\Amat_{m}\zv_{m}[n] \Big|^{2}\Big\}$  \\ 
 \hline 
        \multicolumn{2}{|c|}{$\overline{\mbox{BU}}_{k,n}\!=\!\E\Big\{\Big| \sum\limits_{m=1}^{M} \bar{a}_{mk}^{\rho *}[n] \sqrt{p_{k}}\Big(\hat{\hv}_{mk}^{H}[\lambda]\Amat_{m}\hv_{mk}[\lambda] - \E\big\{\hat{\hv}_{mk}^{H}[\lambda]\Amat_{m}\hv_{mk}[\lambda] \big\} \Big) \Big|^{2}  \Big\}\!$.} \\
 \hline
  \end{tabular}
  \end{adjustbox}\vspace{-1cm}
  \vspace{-1pt}
\end{table*}
The expectations therein need to be computed to derive the closed-form SE expression.
%---------------------------------------------------
%The presence of i) channel aging; ii) spatially-correlated Rician channel with phase-shifts; iii) RF and ADC/DAC impairments at the APs and UEs; and v) LSFD, significantly complicates their computation, when compared with \cite{Zheng_aging_correlated_21}, which considered \textit{Rayleigh-faded} CF channel-aging system, and \textit{that too with ideal hardware}.
%---------------------------------------------------
Simplifying these terms requires multiple novel mathematical results, which are proposed in the following lemma. Its proof is relegated to  Appendix~\ref{appen_Lemma}. We note that these results are a non-trivial extension of~\cite{Zheng_aging_correlated_21}.  \vspace{-3pt} 
%---------------------------------------------------
%-------------------------------------------------------
{\begin{lemma}
	\label{lemma_comp_term}
	 Consider two correlated random vectors $\hv_{mk}[\tilde{n}] =\rho_{k}[\lambda-\tilde{n}]\hv_{mk}[\lambda] + \bar{\rho}_{k}[\lambda- \tilde{n}] \Big( \bar{\hv}_{mk}e^{j\phi_{mk}^{\tilde{n}}} + \fv_{mk}[\tilde{n}] \Big)$ where $\tilde{n} \in \{n,t_k\}$ and
$\hv_{mi}[\lambda] =\bar{\hv}_{mk}e^{j\phi_{mk}^{\lambda}}+ \Rmat_{mk}^{\frac{1}{2}}\tilde{\hv}_{mk}[\lambda]$. The vectors $\fv_{mk}[\tilde{n}], \tilde{\hv}_{mk}[t_k]$ and $\tilde{\hv}_{mk}[n]$ are distributed as $\mathcal{CN}(\mathbf{0}, \Rmat_{mk})$, and with the deterministic diagonal matrices $\Amat_{m},\Pmat_{mk} \in \mathbb{R}^{N \times N}$, then the following results in Table~\ref{Lemma_Closed_form1} hold. Here $ \varepsilon(\rho_i)=\big( \rho_{i}^{2}[\lambda-t_k]\Bar{\rho}_{i}^{2}[n-\lambda]+1 \big)$.\vspace{-10pt}
\begin{table*}[tbph] 
  \centering 
  \begin{adjustbox}{width=\columnwidth,center} %\footnotesize %%\smallsize 
    \caption{Closed-form expressions obtained in Lemma~\ref{lemma_comp_term}.\vspace{-7pt}}
   \label{Lemma_Closed_form1}
  \begin{tabular}{|c|c|} 
 \hline
\makecell{i)}  &  \makecell{ $\E\left\{\hv_{mi}^{H}[t_k]\Amat_{a}^{m}\Pmat_{mk} \text{diag}(\hv_{mi}[n]\hv_{mi}^{H}[n])\Pmat_{mk}^{H}\Amat_{a}^{m}\hv_{mi}[t_k] \right\}
  =    \varepsilon(\rho_i) \text{Tr}\left( \bar{\Rmat}_{mi} \Amat_{a}^{m}\Pmat_{mk} \text{diag}\big( \bar{\Rmat}_{mi} \big)\Pmat_{mk}^{H}\Amat_{a}^{m} \right) $ \\ $+\rho_{i}^{2}[\lambda\!-\!t_k] \Big( 2\text{real}\!\left\{ \text{Tr}\!\left( \bar{\hv}_{mi}\bar{\hv}_{mi}^{H}\Amat_{m}\Pmat_{\!mk}\text{diag}\left( \Pmat_{\!mk}^{H}\Amat_{m}\Rmat_{mi} \right)\! \right)\right\} \!+\!\text{Tr}\left( \Rmat_{mi}\Amat_{m}\Pmat_{\!mk}\text{diag}\left( \Pmat_{\!mk}^{H}\Amat_{m}\Rmat_{mi} \right)\right)\! \Big)$   }  \\   
 \hline 
 \makecell{ii)} & \makecell{$ \E\{ \hv_{mi}^{H}[\lambda]\Amat_{m}\Pmat_{mk}\hv_{mi}[\lambda]\hv_{mi}^{H}[\lambda]\Pmat_{mk}^{H}\Amat_{m}\hv_{mi}[\lambda] \} $ 
 \\ 
 $ = \text{tr}\left( \bar{\Rmat}_{mi} \Amat_{m}\Pmat_{\!mk}\bar{\Rmat}_{mi} \Pmat_{\!mk}^{H}\Amat_{m}\right) \!+ \left| \text{tr} \left( \Rmat_{mi}\Amat_{m}\Pmat_{\!mk}\right) \right|^{2} \!+2\text{real} \big\{ \bar{\hv}_{mi}^{H}\Amat_{m}\Pmat_{\!mk}{\bar{\hv}}_{mi}\text{tr} \left( \Rmat_{mi}\Pmat_{\!mk}^{H}\Amat_{m} \right) \! \big\} $ 
  } \\
 \hline 
  \end{tabular}
  \end{adjustbox}\vspace{-1cm}
\end{table*} 
%%----------------------------------------------- 
%-------------------------------------------------------
%\begin{align}
%&\E\left\{\hv_{mi}^{H}[t_k]\Amat_{a}^{m}\Pmat_{mk} \text{diag}(\hv_{mi}[n]\hv_{mi}^{H}[n])\Pmat_{mk}^{H}\Amat_{a}^{m}\hv_{mi}[t_k] \right\}   \nonumber\\
%&   = \rho_{i}^{2}[\lambda\!-\!t_k] \Big( 2\text{real}\!\left\{ \text{Tr}\!\left( \bar{\hv}_{mi}\bar{\hv}_{mi}^{H}\Amat_{m}\Pmat_{\!mk}\text{diag}\left( \Pmat_{\!mk}^{H}\Amat_{m}\Rmat_{mi} \right)\! \right)\right\} \!+\!\text{Tr}\left( \Rmat_{mi}\Amat_{m}\Pmat_{\!mk}\text{diag}\left( \Pmat_{\!mk}^{H}\Amat_{m}\Rmat_{mi} \right)\right)\! \Big) \nonumber\\
%&\; + \big(  \rho_{i}^{2}[\lambda-t_k]\Bar{\rho}_{i}^{2}[n-\lambda]+1 \big) \text{Tr}\left( \bar{\Rmat}_{mi} \Amat_{a}^{m}\Pmat_{mk} \text{diag}\big( \bar{\Rmat}_{mi} \big)\Pmat_{mk}^{H}\Amat_{a}^{m} \right), \label{eq_h_h_mk_aging} 
%\end{align}
%and 
%\begin{align}
%&\E\{ \hv_{mi}^{H}[\lambda]\Amat_{m}\Pmat_{mk}\hv_{mi}[\lambda]\hv_{mi}^{H}[\lambda]\Pmat_{mk}^{H}\Amat_{m}\hv_{mi}[\lambda] \} \nonumber\\
%& = \text{tr}\left( \bar{\Rmat}_{mi} \Amat_{m}\Pmat_{\!mk}\bar{\Rmat}_{mi} \Pmat_{\!mk}^{H}\Amat_{m}\right) \!+ \left| \text{tr} \left( \Rmat_{mi}\Amat_{m}\Pmat_{\!mk}\right) \right|^{2} \!+2\text{real} \big\{ \bar{\hv}_{mi}^{H}\Amat_{m}\Pmat_{\!mk}{\bar{\hv}}_{mi}\text{tr} \left( \Rmat_{mi}\Pmat_{\!mk}^{H}\Amat_{m} \right) \! \big\}.  \label{eq_subpart_gamma_1_exp1}
%\end{align}
%-------------------------------------------------------
\end{lemma}}

%---------------------------------------------------
%\colr{compare it with existing rayeligh work and then discuss, it is just not enough to mention that these terms gets significantly complicated.}
%---------------------------------------------------

We next derive the closed-form  SE in the following theorem, which depends on the 
long-term channel statistics. This theorem extensively uses Lemma~\ref{lemma_comp_term}.

\begin{theorem} \label{theorem_lower_bound}
The closed-form SE expression for a CF mMIMO system with an arbitrary number of antennas, whose spatially-correlated Rician channel experience aging,  has RF and ADC/DAC impairments both at the APs and UEs, and employs two-stage LSFD,  is given as $\text{SE}_{sum}=\frac{1}{\tau_c}\sum\limits_{n=\lambda}^{\tau_c}\sum\limits_{k=1}^K\log_2(1+\overline{\text{SINR}}_{k,n})$ where \vspace{-6pt}
\begin{align}
	 &\!\!\!\!\overline{\text{SINR}}_{k,n}\!=\!\frac{\alpha_{d,k}^{2}p_{k}|\av_{k}^{H}[n]\boldsymbol{\delta}_{k,n}|^{2}}{\begin{Bmatrix} \! 
	 		\av_{k}^{H}[n]\Big( \Bmat_{k,n} \!+\boldsymbol{\Lambda}_{k,n} \!+\sum\limits_{i\neq k}^{K} \alpha_{d,i}^{2}p_{i}\Cmat_{ki,n}\! + \sum\limits_{i=1}^{K} \big(1\!-\alpha_{d,i} \!+\!  \kappa_{t,i}^{2} \big)\alpha_{d,i}p_{i}\Cmat_{ki,n} \\[-3pt] 
	 		+\sum\limits_{i=1}^{K} \kappa^2_{r,m}{\Dmat}_{ki,n} + \sum\limits_{i=1}^{K}\overline{\Dmat}_{ki,n} + \text{diag}\big(\sigma^{2}\text{tr}\big(\overline{\boldsymbol{\Gamma}}_{mk}\Bmat_{m} \big) \big) + \sigma^2 \Qmat_{k} \Big) \av_{k}[n]
	 \!\!\end{Bmatrix}} =\frac{\Delta_{k,n}}{\Omega_{k,n}}.\label{eq_SINR_k}
\end{align}%\vspace{-2pt}
%\begin{align}
%    .
%\end{align}
Here $\boldsymbol{\delta}_{k,n}=[\delta_{k,n}^{1},\hdots,\delta_{k,n}^{M}]^{\text{T}} \;\in \mathbb{C}^{M\times1}$, 
$\Bmat_{k,n}=\text{diag}([b_{k,n}^{1},\hdots, b_{k,n}^{M}]) \;\in \mathbb{C}^{M\times M}$.
The terms $\delta_{k,n}^{m}$, $b_{k,n}^{m}$, $\boldsymbol{\Lambda}_{k,n},\Cmat_{ki,n}$,  $\Dmat_{ki,n}, \overline{\Dmat}_{ki,n}, \overline{\boldsymbol{\Gamma}}_{mk}$ and $\Qmat_{k} $ are derived in Appendix~\ref{appen_SE_terms}. 
%------------------------------------------------------------
%\begin{proof}
%Refer to Appendix~\ref{}
%\end{proof}
\end{theorem}
%------------------------------------------------------------
%------------------------------------------------------------
{We see from \eqref{eq_SINR_k} that the $\overline{\text{SINR}}_{k,n}$ expression is a generalized Rayleigh quotient with respect to $\av_{k}[n]$. We now state a Lemma from~\cite{Emil_CF_BOOK_20}, which is then used to optimize the  LSFD coefficients.}
\begin{lemma} \label{lemma_rayeligh_quoti}
For a fixed vector $\av \in \mathbb{C}^{N \times 1}$ and positive definite matrix $\Bmat \in \mathbb{C}^{N \times N}$, it holds that 
$\underset{\vv}{\text{Max}} \; \frac{|\vv^H \av|^2}{\vv^H \Bmat \vv} = \av^H \Bmat^{-1} \av$,
where the maximum is attained at $\vv = \Bmat^{-1}\av$.
\end{lemma}
The optimal LSFD coefficients, using Lemma~\ref{lemma_rayeligh_quoti}, are given as \vspace{-4pt}
%------------------------------------------------------------
%\begin{align} \label{eq_LSFD_coefficients}
%\colr{\av_{k}^{*}[n] =  \Bigg( \cdots \Bigg)^{-1} t_k}
%\end{align}
%------------------------------------------------------------
\begin{align}
    \av_{k}^{*}[n] =& \Big( \Bmat_{k,n} +\boldsymbol{\Lambda}_{k,n} +\sum\limits_{i\neq k}^{K} \alpha_{d,i}^{2}p_{i}\Cmat_{ki,n} + \sum\limits_{i=1}^{K} \big(1\!-\alpha_{d,i} \!+\!  \kappa_{t,i}^{2} \big)\alpha_{d,i}p_{i}\Cmat_{ki,n} \notag \\[-3pt] 
   &\;\; +\sum\limits_{i=1}^{K} \kappa^2_{r,m}{\Dmat}_{ki,n} + \sum\limits_{i=1}^{K}\overline{\Dmat}_{ki,n} + \text{diag}\big(\sigma^{2}\text{tr}\big(\overline{\boldsymbol{\Gamma}}_{mk}\Bmat_{m} \big) \big) + \sigma^2 \Qmat_{k}\Big)^{-1} \boldsymbol{\delta}_{k,n}.\label{LSFD_eqq}  \\[-33pt] \notag
\end{align}
%------------------------------------------------------------
%The conventional matched filter is obtained by setting LSFD coefficients as $\av_{k}[n] \!= \![1/L,\! \cdots\!, 1/L ]^T\!$.
%------------------------------------------------------------
{We next provide intuitive insights using the closed-form SE expression in \eqref{eq_SINR_k}.}
%------------------------------------------------------------
%\colr{can give first sub-heading as Theoretical validation and the second one as Impact of channel aging on different interferences} \subsubsection*{Impact of channel aging on different interferences}
%------------------------------------------------------------\subsubsection*{Theoretical validation}
\begin{corollary} 
{We first simplify the SE expression in \eqref{eq_SINR_k} for ideal hardware, and show that it matches with the existing ones in~\cite{HQNgo2017,Zheng_aging_correlated_21}. This not only validates our results, but also shows that the current work subsumes the analysis in~\cite{HQNgo2017,Zheng_aging_correlated_21}.} By setting i) Rician factor $K_{mk}= 0$; ii) $\Rmat_{mk} = \beta_{mk}\Imat_{N}$; and iii) ideal RF hardware and high ADC/DAC resolution, we~have\vspace{+0.1pt} %the $\overline{\text{SINR}}_{kn}$ expression is 
\begin{align}\label{eq_SINR_k_rayl_ideal_HW}
		 \!\!\overline{\text{SINR}}_{k,n} =
		\frac{p_{k}\rho_{k}^{2}[n-\lambda]|\sum\limits_{m=1}^{M}a^{*}_{mk}[n] N \overline{\gamma}_{mk}  |^{2}}
		{\begin{Bmatrix} \sum\limits_{i=1}^{K}\sum\limits_{m=1}^{M}\!\big|a_{mk}^{*}[n]\big|^{2}p_{i}N\overline{\gamma}_{mk}\beta_{mi} + N^{2}\sum\limits_{i\in\mathcal{P}_{k}}p_{i}\rho_{i}^{2}[n-\lambda]\Big|\sum\limits_{m=1}^{M}\!a_{mk}^{*}[n]\sqrt{\overline{\gamma}_{mk}\overline{\gamma}_{mi}}\Big|^{2} 
			\\+ \sigma^{2}\sum\limits_{m=1}^{M}|a_{mk}^{*}[n]|^{2}N\overline{\gamma}_{mk}\end{Bmatrix}}.% \nonumber
	\end{align}
%	\begin{align}\label{eq_SINR_k_rayl_ideal_HW}
%	& \overline{\text{SINR}}_{k,n} \!=\\[-10pt]
%	&\frac{p_{k}\rho_{k}^{2}[n-\lambda]|\sum\limits_{m=1}^{M}a^{*}_{mk}[n] N \overline{\gamma}_{mk}  |^{2}}
%	{ \sum\limits_{i=1}^{K}\sum\limits_{m=1}^{M}\!\big|a_{mk}^{*}[n]\big|^{2}p_{i}N\overline{\gamma}_{mk}\beta_{mi} \!+\! N^{2}\!\sum\limits_{i\in\mathcal{P}_{k}}\!p_{i}\rho_{i}^{2}[n-\lambda]\Big|\!\sum\limits_{m=1}^{M}\!a_{mk}^{*}[n]\sqrt{\overline{\gamma}_{mk}\overline{\gamma}_{mi}}\Big|^{2} 
%		\!\!+\! \sigma^{2}\!\sum\limits_{m=1}^{M}\!|a_{mk}^{*}[n]|^{2}N\overline{\gamma}_{mk}}. \nonumber
%	\end{align}
Here $\overline{\gamma}_{mk} \!= \text{tr}(\overline{\boldsymbol{\Gamma}}_{mk}) $.	The above expression matches with~\cite[Eq. (21)]{Zheng_aging_correlated_21}. Further, by  assuming $\rho_k[n\!-\!\lambda] \!=1$, $N\!=1$ and $\av_{mk}[n] = 1/L$, the expression in \eqref{eq_SINR_k_rayl_ideal_HW} matches with~\cite[Eq. $(27)$]{HQNgo2017}. 
\end{corollary}
%-----------------------------------------------
%\underline{\textbf{Analysis of impact of channel aging:}} \subsubsection*{Impact of channel aging on different interferences}
\begin{corollary}\label{corollary_interference_order}  {We now investigate the effect of channel aging on the desired signal, interference, RF and ADC/DAC distortion power terms in \eqref{eq_SINR_k}.}
We first consider the term $\overline{\mbox{IUI}}_{ki,n}$, which is simplified in \eqref{eq_interference_final} in Appendix~\ref{appen_SE_terms}.
For $i\in\mathcal{P}_{k}$, the term $\overline{\mbox{IUI}}_{ki,n}$ can be re-written using \eqref{eq_interference_final} as\vspace{-4pt}
\begin{align} 
 \overline{\mbox{IUI}}_{ki,n} = & 
\;\alpha_{d,i}^2p_{i} \bigg( \sum\limits_{m=1}^{M} \Big( |a_{mk}^{*}[n]|^2 \Big( \rho_{i}^{2}[n\!-\! \lambda]\big(\varsigma_{kim,n}^{(1)} \! + \varsigma_{kim,n}^{(2)} \! + \varsigma_{kim,n}^{(3)} \big)\nonumber \\[-6pt]
& \quad +\overline{\rho}_{i}^{2}[n\!-\! \lambda]\text{tr}\left(\boldsymbol{\overline{\Gamma}}_{mk}\Amat_{m}\bar{\Rmat}_{mi}\Amat_{m}  \right) \!\Big)  
+  \sum\limits_{m{'}\neq m}^{M}a_{mk}^{*}[n]a_{m{'}k}^{*}[n] c_{kin,mm{'}}^{(\in)} \Big) \bigg). \label{IUI_insight}  \\[-33pt] \notag
  \end{align}
We reproduce $\varsigma^{(1)}_{kin,m}$ from \eqref{eq_var_sigma_1_final} from Appendix~\ref{appen_SE_terms} for the sake of drawing insights:\vspace{-4pt} 
  \begin{align}\label{varsigma}
   \varsigma^{(1)}_{kin,m} =& \sum_{j\in\mathcal{P}_{k}} \alpha_{d,j}^{2}\tilde{p}_{j} \text{tr}\left( \bar{\Rmat}_{mj} \Amat_{m}\Pmat_{mk}\bar{\Rmat}_{mi} \Pmat_{mk}^{H}\Amat_{m}\right) + \alpha_{d,i}^{2}\tilde{p}_{i}\rho_{i}^{2}[\lambda-t_k]\Big(  \left| \text{tr} \left( \Rmat_{mi}\Amat_{m}\Pmat_{mk}\right)    \right|^{2} \nonumber \\[-4pt]
    & +2\text{real} \big( \overline{\hv}_{mi}^{H}\Amat_{m}\Pmat_{mk} {{\overline{\hv}}_{mi}} \text{tr} ( \Rmat_{mi}\Pmat_{mk}^{H}\Amat_{m} )  \big) \Big). 
\end{align}
{We see from \eqref{varsigma} that $\varsigma^{(1)}_{kin,m}$ consists of $\rho_i^2[\lambda - t_k]$, which does not vary with the time instant $n$. Similarly, we see from \eqref{eq_var_sigma_2} and \eqref{eq_var_sigma-3_final} that the terms $\varsigma_{kin,m}^{(2)}$ and $\varsigma_{kin,m}^{(3)}$ are also independent of the time instant $n$. Further, the term $c_{kin,mm^{'}}^{(\in)}$ in the second summation of \eqref{IUI_insight}, contains $\rho_{i}^2 [n-\lambda]$. As a result, the interference term $\overline{\mbox{IUI}}_{ki,n}$ can be expressed as $ e_1 \rho_i^2 [n-\lambda] + e_2 (1-\rho_i^2[n-\lambda]) + e_3 \rho_i^2 [n-\lambda] \triangleq e_2 + e_4 \rho_i^2 [n-\lambda]$, where $e_1,\dots, e_4$ are positive constants. The term $\overline{\text{IUI}}_{ki,n}$, due to channel aging,  decreases with time instant $n$ as $\mathcal{O} \big(e_2 + e_4  \rho^2_{i}[n\!-\! \lambda]\big)$.}
%The first, second and third terms are of the order $e_1$, $\rho^2_{i}[\lambda- t_k]$ and $\rho^2_{i}[\lambda- t_k]$ respectively, with $e_1$ being a constant. The term $\varsigma^{(1)}_{k,i,n,m}$, therefore, decreases as $\mathcal{O}(e_1 + \rho^2_{i}[\lambda- t_k]))$. 
%Similarly, we see from \eqref{eq_var_sigma_2} and \eqref{eq_var_sigma-3_final} that the terms $\varsigma_{k,i,m,n}^{(2)}$ and $\varsigma_{k,i,m,n}^{(3)}$ decrease as $\mathcal{O}(e_2 + \rho^2_{i}[\lambda\!-\! t_k])$ and $\mathcal{O}(e_3 + \rho^2_{i}[\lambda\!-\! t_k])$ respectively, with $e_2$ and $e_3$ being the constants. We also see that the term $c_{kin,mm^{'}}^{(\in)}$ in \eqref{IUI_insight} decreases as $\mathcal{O}( \rho^2_{i}[n\!-\! \lambda] \rho^2_{i}[\lambda\!-\! t_k])$ with $n$. We, therefore, conclude that for $i\in\mathcal{P}_{k}$, $\overline{\mbox{IUI}}_{kin}$ decreases as $\mathcal{O} \big( \rho^2_{i}[n\!-\! \lambda] ( e_4 + \rho^2_{i}[\lambda\!-\! t_k])\big)$, with $e_4$ being a constant. %\colr{form one user, total K-1 UE interf}

Similarly, the desired signal power $\overline{\mbox{DS}}_{k,n}$ and beamforming uncertainty power $\overline{\mbox{BU}}_{k,n}$ in \eqref{eq_SINR_k} reduce as $\mathcal{O}(\rho_k^2[n-\lambda])$ and $\mathcal{O} \big( \rho^2_{k}[n\!-\! \lambda] ( e_1 + \rho^2_{k}[\lambda\!-\! t_k])\big)$, respectively. 
The UE transmit and AP receiver RF impairments and ADC/DAC quantization noise power reduce as $\mathcal{O}(\epsilon+\rho_k^2[n-\lambda])$. We see that as $n > \lambda$ increases, the channel aging factor $\rho_k[n-\lambda]$ severely reduces $\overline{\mbox{DS}}_{k,n}$ and $\overline{\mbox{BU}}_{k,n}$. This is also validated later in Section~\ref{simulation_section1}, where we analyze the power of desired signal, RF impairments, and ADC/DAC quantization noise for different UE velocities.
\end{corollary}

\begin{remark}
In CF mMIMO systems, each AP may have a small number of two to four antennas. The system designer, may not  connect each AP antenna to a different resolution ADC, as it will increase the system complexity. The current framework, however, also allows a designer to analytically evaluate the SE of such CF mMIMO system by varying the ADC resolution across different APs, and by keeping them same within an AP. This design is practical as each AP is designed as a separate subsystem. 	For CF mMIMO systems, if AP has greater than four antennas, the dynamic ADC architecture could be implemented within each AP itself. 
\end{remark}

\vspace{-0.5cm}
\section{SE Optimization of Hardware-Impaired CF System With Channel Aging } \label{sect_SE_opti}
We now optimize the SE  for a given time instant $n$. The SE optimization problem, by using the SE expression in Theorem~\ref{theorem_lower_bound},  can be cast as follows:\vspace{-4pt}
\begin{align} \label{eq_SE_max} 
    \textbf{P1:} \; \underset{\pv }{\text{Max}} \; R_{n}(\pv) = \sum_{k=1}^{K} \log_{2} \bigg( 1 + \frac{{\Delta}_{k,n}(\pv)}{{\Omega}_{k,n}(\pv)} \bigg), 
\; \text{subject to }  0\leq p_{k}\leq P_{max} \; \forall k.  %\\[-33pt] \notag
\end{align}
The term $P_{max}$ is the maximum allowed UE transmit power, and the vector $\pv = [p_1, \cdots, p_K]^{T}$.
%Here the vector $\pv = [p_1, \cdots, p_K]^{T}$. The term $P_{max}$ is the maximum UE transmit power.
For a given time-instant $n$, the objective of $\textbf{P1}$ is not a concave function. 
To optimize  $\textbf{P1}$, we now use MM technique, which first finds a surrogate function that locally approximates the objective function with their difference minimized at the current point, and then iteratively maximizes it.  {The surrogate functions should be designed to have the following desirable features:
i) convexity and smoothness; and ii) existence of a closed-form minimizer \cite{sun_majorization_17_TSP}.
The first property helps in easily optimizing the surrogate problem,  which can be easily shown to converge to a stationary point of the original non-convex problem \cite{sun_majorization_17_TSP}. The second property helps in designing a low-complexity optimization solution. We use MM framework to propose two solutions. The surrogate function in the first case only has the first property but does not have the second one. The first solution thus has a high complexity and requires a numerical optimization solver e.g., CVX~\cite{boyd_convex_book_04}. The second solution, builds upon the first one, and  combines the MM approach with Lagrangian dual transform from \cite{shen_QT_TSP_18} to design another surrogate function, which satisfies the  second property also. This leads to its highly reduced complexity.}  %The authors in \cite{Emil_CF_BOOK_20,Ozlem_CFwpt_21} have also optimized transmit power for CF mMIMO systems, but only with ideal hardware, and that too without considering channel aging. 

We now briefly explain the MM technique from \cite{sun_majorization_17_TSP}. Consider the following non-concave maximization problem with a non-concave objective $f(\xv)$: 
	\begin{align}
	{\text{Maximize}}\;\; f(\xv) \text { subject to } \xv \in \mathcal{X}, \text{ with $\mathcal{X}$ being a convex set}. \notag 
	\end{align}
		The MM technique solves it by first constructing a convex surrogate $g(\xv|\hat{\xv}^{(t)})$ of $f(\xv)$ at a feasible point $\hat{\xv}^{(t)}$~\cite{sun_majorization_17_TSP}. It then iteratively generates a sequence of feasible points $\hat{\xv}^{(t+1)}$ by maximizing $g(\cdot|\hat{\xv}^{(t)})$. 
		The surrogate function $g(\xv|\hat{\xv}^{(t)})$, along with two aforementioned desirable properties, should be continuous in $(\xv,\hat{\xv}^{(t)})$. It should also lower bound the objective function $f(\xv)$. }
The surrogate function has to additionally satisfy the following two technical conditions~\cite{sun_majorization_17_TSP}:\vspace{-4pt} 
\begin{align}
\text{C1}: g(\hat{\xv}^{(t)}|\hat{\xv}^{(t)}) = f(\hat{\xv}^{(t)}), \text{  and  }
\text{C2}: \nabla_{\xv}g(\xv|\hat{\xv}^{(t)})|_{\xv = \hat{\xv}^{(t)}}= \nabla_{\xv} f(\xv)|_{\xv = \hat{\xv}^{(t)}}. \\[-33pt] \notag
\end{align}
Each MM iteration generates a feasible point $\hat{\xv}^{(t)}$. The function value at each iteration $f(\hat{\xv}^{(t)})$ increases, and then finally converges to a stationary point of the original problem~\cite{sun_majorization_17_TSP}.

%\begin{remark}
% \colb{The surrogate functions should be designed to have the following desirable features:
%i) convexity and smoothness; and ii) existence of a closed-form minimizer \cite{sun_majorization_17_TSP}.
%The first property helps in easily optimizing the surrogate problem,  which can be easily shown to converge to a stationary point of the original non-convex problem \cite{sun_majorization_17_TSP}. The second property helps in designing a low-complexity optimization solution. We use MM framework to propose two solutions. The surrogate function in the first case only has the first property but does not have the second one. The first solution thus has a high complexity. The second solution, builds upon the first one, and  combines the MM approach with Lagrangian dual transform from \cite{shen_QT_TSP_18} to design another surrogate function, which satisfies the  second property also. This leads to its highly reduced complexity.}
%\end{remark}

%\underline{\textbf{Minorization-maximization approach:}}
\vspace{-0.4cm}
\subsection{Minorization-maximization approach}
 We begin by constructing a novel surrogate function for the non-concave objective $R_n(\pv)$. We achieve this aim by first proposing a lower bound on $R_n(\pv)$, and then show that it satisfies C1 and C2. This will make it a valid surrogate function. {We now state a Lemma, which is then used to construct a lower bound on $R_n(\pv)$}.
  \begin{lemma} \label{eq_sarrogate_fun}
 For $N_k(\xv):\mathbb{R}^n \rightarrow \mathbb{R}_{+}$, $D_k(\xv):\mathbb{R}^n \rightarrow \mathbb{R}_{++}$ and an increasing function $f(\cdot)$, we have
 \begin{align}
 \sum_{k=1}^{2K}f\Big(\frac{N_k(\xv)^2}{D_k(\xv)} \Big) \geq  \sum_{k=1}^{2K} f(2N_k(\xv)- D_k(\xv)). 
 \end{align}
 \end{lemma}\vspace{-2pt}
{By substituting $N_k(\xv) = y_{k,n} \sqrt{{\Delta}_{k,n}(\pv) }$, $D_k(\xv) = y_{k,n}^2 {\Omega}_{k,n}(\pv) $, and $f(\cdot) = \text{log}_2(1+\cdot)$ in Lemma~\ref{eq_sarrogate_fun}, the objective function $R_n(\pv)$ is lower-bounded as follows:} 
\begin{align} \label{eq_sarrogate}
R_{n}(\pv) \geq \sum_{k=1}^{2K} \text{log}_2 \Big(1+ 2y_{k,n} \sqrt{{\Delta}_{k,n}(\pv) }- y_{k,n}^2 {\Omega}_{k,n}(\pv) \Big) =  \widetilde{R}_n(\pv, \yv_n).
\end{align} 
Here $y_{k,n}$ is a function of the feasible point $\pv^{(t)}$ as $y_{k,n} = \sqrt{{\Delta}_{k,n}(\pv^{(t)}) } / {\Omega}_{k,n}(\pv^{(t)})$, and $\yv_n = [y_{1,n}, \cdots, y_{K,n}]^{T}$. 
%\colr{We also denote $\yv_n = [y_{1,n}, \cdots, y_{K,n}]^{T}$.}
%The vector $\yv_n = [y_{1,n}, \cdots, y_{K,n}]^{T}$ with $y_{k,n} = \sqrt{{\Delta}_{k,n}(\pv) } / {\Omega}_{k,n}(\pv)$. 
%\colr{$y_k$ should be $y_{k,n}$} 
We now show that $\widetilde{R}_n(\pv, \yv_n)$ in \eqref{eq_sarrogate} satisfies conditions $\text{C1}$ and $\text{C2}$, and is a valid surrogate function. 
{\begin{itemize}[leftmargin = *]
\item	Condition $\text{C1}$ can be proved by substituting the variable update of $y_{k,n}$ and $\pv = \pv^{(t)}$ in~$R_n(\pv, \yv_n)$. 
\item	Condition $\text{C2}$ can be verified by differentiating both $\widetilde{R}_n(\pv,\yv_n)$ and $R_n(\pv)$ at $\pv = \pv^{(t)}$. 
%$\nabla_{\pv}R_n(\pv)|_{\pv = \pv^{(t)}} = \nabla_{\pv}R_n(\pv,\yv_n)|_{\pv = \pv^{(t)}}$.
\end{itemize}
By replacing $R_n(\pv)$ with the surrogate function $\widetilde{R}_n(\pv, \yv_n)$, problem $\textbf{P1}$ is recast as follows:}
\begin{subequations}
\begin{align}
    \textbf{P2:} \;\;\;& \underset{\pv}{\text{Maximize}} \;\;\;\; \sum_{k=1}^{2K} \text{log}_2 \Big(1+ 2y_{k,n} \sqrt{{\Delta}_{k,n}(\pv) }- y_{k,n}^2 {\Omega}_{k,n}(\pv) \Big) \label{eq_SE_max_sarrog}\\[-4pt] 
    &\text{subject to }  0\leq p_{k}\leq P_{max} \; \forall k. \label{eq_constraints_sa} \\[-33pt] \notag
\end{align}
\end{subequations} 
We note that the ${\Delta}_{k,n}(\pv)$ and ${\Omega}_{k,n}(\pv)$ in the objective of  $\textbf{P2}$ are concave and convex in $\pv$, respectively. Problem $\textbf{P2}$ now becomes concave which can be solved using CVX~\cite{boyd_convex_book_04}. The procedure to calculate optimal power coefficients begins by first constructing the surrogate function $R_n(\pv,\yv_n)$, with $y_{k,n} = \sqrt{{\Delta}_{k,n}(\pv^{(t)}) } / {\Omega}_{k,n}(\pv^{(t)})$, and then by solving  $\textbf{P2}$ using CVX.
% The proposed procedure of solving $\textbf{P2}$ is summarized in Algorithm~\ref{algo_SE}.
%%-----------------------------------------------------------------------------------------
%\begin{algorithm}
%	\DontPrintSemicolon
%	\footnotesize %\scriptsize  %
%	\KwIn{i) Initialize $\{p_k \}_{k=1}^{K}$ with equal power allocation, set maximum number of  iteration  $I$. }\vspace{-3pt}
%	\KwOut{$p_{k}$ as the solutions.}\vspace{-3pt}
%	\For{$i \gets 1$ \textbf{to} $I$}{\vspace{-3pt}
%		%\For{$k \gets 1$ \textbf{to} $K$}{\vspace{-3pt}
%		%	\For{$i_i^a \gets 1$ \textbf{to} $I_i$}{\vspace{-3pt}
%				For a given feasible $\pv^{(i)}=(p^{i}_{1},\cdots,p_{k}^{i}, \cdots, p_{K}^{i})$, compute the auxiliary variables $y_{k,n} = \sqrt{\boldsymbol{\Delta}_{k,n}(\pv) } / \boldsymbol{\Omega}_{k,n}(\pv)$. \\[-1pt]
%				Solve \textbf{P2} for SE with respect to $k$th UE i.e., $p_{k}$. \\[-1pt]
%				Obtain $\pv^{*}$ from step $3$ and update $\pv^{(i+1)}=(p^{*}_{1},\cdots,p_{k}^{*}, \cdots, p_{K}^{{*}})$. \\[-2pt]
%				Do until convergence $(\| \pv^{(i+1)}- \pv^{(i)} \|^2 \leq \epsilon)$.}
%				%}\vspace{-8pt}
%		%Do until convergence $(\| \qv^{(i_o,i_{*}^L)}- \qv^{(i_o-1,i_{*}^L)} \|^2 \leq \epsilon)$.}\vspace{-8pt}
%	\Return{$p_{k}$ .}\;
%	\caption{SE optimization using QT}\label{algo_SE}
%\end{algorithm}\vspace{-10pt}
%-------------------------------------------------------------------------
%------------------------------------------------------------
%\newline \underline{\textbf{Closed-form MM approach:}}
\vspace{-0.5cm}
\subsection{Closed-form MM approach}\vspace{-5pt}
The MM-based SE optimization has a high complexity. We now propose a practically-implementable SE optimization algorithm by combining MM approach with the Lagrangian dual transform~\cite{shen_QT_TSP_18}, \textit{which provides an iterative closed-form solution for the UE transmit powers $\pv$.} To use this approach,  problem \textbf{P1} is equivalently expressed as\vspace{-4pt}
\begin{align}
     \textbf{P3:} \;  \underset{\pv,\gamma_{k,n} }{\text{Maximize}}\;\;   \sum_{k=1}^{K} \log_{2} \big( 1 + \gamma_{k,n} \big) \; \text{  subject to }  \gamma_{k,n} \leq \frac{{\Delta}_{k,n}(\pv)}{{\Omega}_{k,n}(\pv)}, \; 0\leq p_{k}\leq P_{max} \; \forall k.
\end{align}
{The epigraph variable $\gamma_{k,n}$ moves the ratio out of the logarithm~\cite{boyd_convex_book_04}. Problem \textbf{P3} can be decomposed into outer and inner optimizations over $p_k$ and $\gamma_{k,n}$, respectively. The inner optimization in $\gamma_{k,n}$ is convex. The strong, duality therefore, holds. Its equivalent Lagrangian function is~\cite{boyd_convex_book_04}:}\vspace{-4pt}
\begin{align} \label{eq_lagrangian}
    \mathcal{L}(\gamma_{k,n},\lambda_{k,n})&=\sum_{k=1}^{K} \log_{2} \big( 1 + \gamma_{k,n} \big) -\sum_{k=1}^{K}\lambda_{k,n}\left(\gamma_{k,n}-\frac{{\Delta}_{k,n}(\pv)}{{\Omega}_{k,n}(\pv)}  \right).
\end{align}
Here $\lambda_{k,n}$ is the Lagrangian dual variable. Let $(\gamma_{k,n}^{\ast},\lambda_{k,n}^{\ast})$ be the saddle point  which satisfies the first-order condition i.e., ${\partial \mathcal{L}(\gamma_{k,n},\lambda_{k,n}) }/{\partial \gamma_{k,n}} \big|_{\gamma_{k,n}^{*}}  =0$, then\vspace{-4pt}
%the optimal $\lambda_{k,n}^{*}$ that maximizes the Lagrangian is given by using the first order condition
%\begin{align}
%    \frac{\partial \text{L}(\gamma_{k,n},\lambda_{k,n}) }{\partial p_{k}} \Bigg|_{\gamma_{k,n}^{*}}&=0  \nonumber\\
%    \implies \lambda_{k,n}^{*}&= \frac{1}{(1+\gamma_{k,n}^{*}(p))\ln{2}}= \frac{\Omega_{k,n}(p)}{(\Delta_{k,n}(p)+\Omega_{k,n}(p))\ln{2}}
%\end{align}
\begin{align} \label{eq_lambda_optimal}
 \lambda_{k,n}^{*}= \frac{1}{(1+\gamma_{k,n}^{*})\ln{2}} \eqa \frac{\Omega_{k,n}(\pv)}{(\Delta_{k,n}(\pv)+\Omega_{k,n}(\pv))\ln{2}}.
\end{align}
{Equality $(a)$ is because $\gamma_{k,n}^{\ast} \!=\! {\Delta_{k,n}(\pv)}/{\Omega_{k,n}(\pv)}$, for a fixed $\pv$}. {Substituting \eqref{eq_lambda_optimal} in \eqref{eq_lagrangian}, the problem $\textbf{P3}$ is rewritten as:
\begin{align}
	\hspace{-5pt}\textbf{P4}:\;&\underset{\pv,\gamma_{k,n}}{\text{Maximize}}\quad  \mathcal{L}(\gamma_{k,n},\lambda_{k,n}^{*})= \sum_{k=1}^{K}\left[\! \log_{2}\left(1+\gamma_{k,n} \right)-\frac{1}{\ln{2}} \left(\gamma_{k,n} -{\frac{\Delta_{k,n}(\pv)(1+\gamma_{k,n})}{\Delta_{k,n}(\pv)+\Omega_{k,n}(\pv)}} \right)  \!\right] \label{eq_maxim_lagran}\\
	& \text{subject to }\quad 0 \leq p_k \leq P_{max}.
	  \nonumber
\end{align}}
%$\underset{\pv}{\text{Maximize}} \quad \underset{\gamma_{k,n}}{\text{Maximize}}\quad \mathcal{L}(\gamma_{k,n},\lambda_{k,n}^{\ast})$ subject to $0 \leq p_{k,n} \leq P_{max}$ 
%Using \eqref{eq_lambda_optimal}, we re-write \eqref{eq_lagrangian}~as\vspace{-4pt}
%\begin{align}
%  \underset{\gamma_{k,n} }{\text{Max }}  \mathcal{L}(\gamma_{k,n},\lambda_{k,n}^{*})=\underset{\gamma_{k,n} }{\text{Max}}  \sum_{k=1}^{K}\left( \log_{2}\left(1+\gamma_{k,n} \right)-\frac{1}{\ln{2}} \left(\gamma_{k,n} -{\frac{\Delta_{k,n}(\pv)(1+\gamma_{k,n})}{\Delta_{k,n}(\pv)+\Omega_{k,n}(\pv)}} \right)  \right). \label{eq_maxim_lagran} 
%\end{align}
%\colb{Objective is in fraction- non-convex - inorder to make concave we first apply QT from - We first state the lemma from [] - using Lemma, eq 40 can be written as}
{The objective function of problem \textbf{P4} contains non-convex fractional terms. We now aim to solve \textbf{P4} by proposing a surrogate function that lower bounds the objective. The relevant surrogate function is obtained by substituting $N_k(\xv) = y_{k,n}\sqrt{\Delta_{k,n}(\pv) (1+\gamma_{k,n})}$ and $D_k(\xv) = y_{k,n}^2 \big(\Delta_{k,n}(\pv)+\Omega_{k,n}(\pv)\big)$ in Lemma~\ref{eq_sarrogate_fun}, and  is given as 
\begin{align}
	{\overline{R}}_{n}(\pv,\gamma_{k,n},\yv_n) = \sum_{k=1}^{K} \log_{2}(1+\gamma_{k,n}) -&\frac{1}{\ln{2}}\Big( \gamma_{k,n} -\big( 2y_{k,n}\sqrt{\Delta_{k,n}(\pv)(1+\gamma_{k,n})}\Big. \nonumber\\[-5pt]
	 &\Big. -y_{k,n}^2( \Delta_{k,n}(\pv)+\Omega_{k,n}(\pv) ) \big)\Big).
\end{align}
Here $\yv_n = [y_{1,n},\cdots,y_{K,n}]^T$ with $y_{k,n}$ being a function of the feasible point $\big(\pv^{(t)},\gamma_{k,n}^{(t)}\big)$: $y_{k,n}^{(t)}=\frac{\sqrt{\Delta_{k,n}(\pv^{(t)})(1+\gamma_{k,n}^{(t)})}}{\Delta_{k,n}(\pv^{(t)})+\Omega_{k,n}(\pv^{(t)})}$. The surrogate function ${\overline{R}}_n(\pv,\gamma_{k,n},\yv_n)$ can be shown to follow conditions C1 and C2, on the lines similar to that of $\widetilde{R}_n(\pv,\yv_n)$. The resultant problem is, therefore, given as}
%To make the objective concave in $\pv$, we use Lemma~\ref{eq_sarrogate_fun} 
%{To make the objective concave in $\pv$, we approximate it with the surrogate function in Lemma~\ref{eq_sarrogate_fun}. The resultant problem is given as:}\vspace{-4pt}
%\cite[Corollary 2]{shen_QT_TSP_18}  to decouple the fractional terms as follows: \colr{use MM Lemma to decouple the fractional term}
%\colr{We can state the lemma from QT paper and then use it to recast the problem}
\begin{align}
    \textbf{P5:}\;\; \underset{ \pv,\gamma_{k,n} }{\text{Maximize}}\quad& {\overline{R}}_{n}(\pv,\gamma_{k,n},\yv_{n})\quad \text{subject to}\quad 0 \leq p_k \leq P_{max}.
\end{align} 
Problem $\textbf{P5}$, for a fixed $\pv$, is convex in $\gamma_{k,n}$ and its optimal value is given as $\gamma_{k,n}^\ast = \Delta_{k,n}(\pv)/\Omega_{k,n}(\pv)$. We now provide the closed-form  expression to calculate optimal power in the following lemma.
%-------------------------------------------------------------------------
%Here  $y_{k,n}^{t}$ is the auxiliary variable. 
%For a fixed $(p_k,\gamma_{k,n})$,  $y_{k,n}$ can be computed using first-order condition as 
%-------------------------------------------------------------------------    
\begin{lemma}
For a fixed $\gamma_{k,n}$, problem \textbf{P5} is concave in $p_{k}$, and the optimal value of $p_k$, obtained using the first-order optimality condition, is given  as %\vspace{-3pt}
\begin{align}
    p_{k} & =\min\left(P_{{max}}, \frac{ y_{k,n}^{2}\left( 1 + \gamma_{k,n} \right) \alpha_{d,k}^{2}|\av_{k}^{H}[n]\boldsymbol{\delta}_{k,n}|^{2}} {\left(\alpha_{d,k}^{2}y_{k,n}^{2}|\av_{k}^{H}[n]\boldsymbol{\delta}_{k,n}|^{2} + l_{k}^{d}   \right)^{2} }   \right). \label{eq: closed-form QT}
\end{align}
Here 
$l_{k}^{d} =  \sum\limits_{i=1}^{K}\av_{i}^{H}[n]\Big(y_{i,n}^{2}\alpha_{d,i}(1+\kappa_{t,i}^{2})\Cmat_{ik,n}  +\sum\limits_{i=1}^{K}y_{i,n}^{2}(\kappa^2_{r,m}\Dmat_{ik,n}+\overline{\Dmat}_{ik,n}) \Big)\av_{i}[n]  
+ \av_{k}^{H}[n] y_{k,n}^{2}(\Bmat_{k,n} +\boldsymbol{\Lambda}_{k,n} - \alpha_{d,k}^{2}\Cmat_{kk,n})\av_{k}[n] $.\vspace{-5pt} 
\begin{proof}
Refer to Appendix~\ref{appen_closed_form_qt_update}. 
\end{proof}
\end{lemma}
%-------------------------------------------------------------------------
%-------------------------------------------------------------------------
%\vspace{-0.4cm}
%$l_{k}^{d} =  \sum_{k^{'}=1}^{K}\av_{k^{'}}^{H}[n]\Big(y_{k^{'},n}^{2}(1-\rho_{d,k^{'}})(1+\kappa_{t}^{2})\Cmat_{k^{'},k,n} \Big) - \av_{k}^{H}[n] y_{k,n}^{2}\alpha_{d,k}^{2}\Cmat_{k,k,n} 
% +\sum_{k^{'}=1}^{K}\av_{k^{'}}^{H}[n]y_{k^{'},n}^{2}\\(\Dmat^{\text{RF}}_{k^{'},k,n}+  \Dmat^{\text{ADC}}_{k^{'},k,n})$. 
%\colr{modify the below algorithm according to your closed-form QT.}
%-------------------------------------------------------------------------
%-----------------------------------------------------------------------------------------
The procedure to calculate optimal power coefficients is summarized in Algorithm~\ref{algo_SE_closed-form_QT}. By iteratively constructing the function $\overline{R}_n(\pv,\gamma_{k,n},\yv_n)$, and by updating the transmit powers $p_k$ and the epigraph variable $\gamma_{k,n}$, the algorithm converges to a local optimum of $\textbf{P1}$~\cite{sun_majorization_17_TSP}. %{We numerically also validate this aspect later in Fig.~\ref{fig_convergence} of Section~\ref{simulation_section1}}.
%\underline{\textbf{Computational complexity:}}
%\colr{ which is same as  that of \cite{Farooq2021} and \cite{Farooq2022}.}
\begin{algorithm}
	\DontPrintSemicolon
	\footnotesize %\scriptsize  %
	\KwIn{i) Initialize $\{p_k \}_{k=1}^{K}$ with equal power allocation, set maximum number of  iteration $I$ and stopping tolerance $\epsilon$. }\vspace{-3pt}
	\KwOut{$\pv^{\ast}$}\vspace{-3pt}
	\For{$i \gets 1$ \textbf{to} $I$}{\vspace{-3pt}
		%\For{$a \gets 1$ \textbf{to} $L$}{\vspace{-3pt}
			%\For{$i_i^a \gets 1$ \textbf{to} $I_i$}{\vspace{-3pt}
				For a given feasible $\pv = \pv^{(i)}$, compute the updates of variables $y_{k,n}$ and $\gamma_{k,n}$. \\
				%	Solve \textbf{P5} for GEE with respect to $a$th AP i.e., $\qv_{a}$. \\[-1pt]
				Obtain $\pv^{(i+1)}$ from \eqref{eq: closed-form QT}. \\[-2pt]
				Do until convergence $(\| \pv^{(i+1)}- \pv^{(i)} \|^2 \leq \epsilon)$.} \vspace{-5pt}
			\Return{$\pv^\ast = \pv^{(i+1)}$ .}\;
			\caption{SE optimization using closed-form MM approach}\label{algo_SE_closed-form_QT}
		\end{algorithm}\vspace{-0.4cm}
%-------------------------------------------------------------------------
%-------------------------------------------------------------------------\\

\textbf{Computational complexity:}  The complexity of CVX-based approach is dominated by that of solving $\textbf{P2}$ in each iteration, which  has $K$ optimization variables and $2K$ linear constraints. This approach has a worst-case complexity of $\mathcal{O}((3K)^{3/2}K^2)$~\cite{boyd_convex_book_04}. The update of variable  $\yv_n$ has trivial complexity.  The Algorithm~\ref{algo_SE_closed-form_QT},  which computes optimal transmit power in a closed-form,  has a trivial complexity. {Also, the proposed closed-form MM approach in Algorithm~\ref{algo_SE_closed-form_QT} has the same complexity as that of \cite{Farooq2021} and \cite{Farooq2022}, when applied for the system models therein.} \vspace{-10pt}

\vspace{-.3cm}
\section{Simulation Results} \label{simulation_section1}\vspace{-4pt}
We now numerically validate the derived closed-form SE expression in \eqref{eq_SINR_k}, and investigate the effect of i) channel aging; ii) RF and ADC/DAC impairments; iii) Rician channels with~phase-shifts; and iv) LSFD. For these studies, we consider a CF mMIMO system, wherein $M$ APs and $K$ UEs are randomly distributed within a geographical area of $1 \times 1 \; \text{Km}^2$. We, similar to \cite{HQNgo2017,Zheng_aging_correlated_21}, assume that the coverage area is wrapped around the edges to avoid the boundary effect. We assume a system bandwidth of $B = 20$~MHz, and a resource block length of $\tau_c = 100$ instants. Each AP has a uniform linear antenna array, whose correlation matrix $\Rmat_{mk}$, is modelled using the Gaussian local scattering model with $\text{ASD} = 30^{\circ}$~\cite{Emil_CF_BOOK_20}. The large scale fading coefficients and the Rician factors corresponding to the UE-AP channels are modeled as follows~\cite[Table-5.1]{GUE_channel_3gp_Rici}:\vspace{-3pt}
%---------------------------------
%The RF hardware impairments of transmit UEs and receiver APs, are set as $\kappa_{t,k}$ and $\kappa_{r,m}$ \colr{HW $\kappa$}
%---------------------------------
% We model the large-scale fading coefficients and Rician factor as~\cite[Table-5.1]{GUE_channel_3gp_Rici}
\begin{align} \label{eq_GUE_ch12}
\beta_{mk} =  -30.9 -26\text{log}_{10}(d_{mk})+ \varrho_{mk} \; \; \text{and } K_{mk} = 13-0.03 d_{mk} [\text{dB}] .
\end{align}
%---------------------------------
%\begin{align} \label{eq_GUE_ch12}
%\beta_{mk} = \begin{cases} -30.9 -26\text{log}_{10}(d_{mk})+ \varrho_{mk}  &  P^{\text{LoS}}_{mk} \neq 0 \\[-4pt]
%-34.53 -38\text{log}_{10}(d_{mk})+ \varrho_{mk} &  P^{\text{LoS}}_{mk} = 0, \end{cases} \; \text{and } K_{mk} = 13-0.03 d_{mk} [\text{dB}] .
%\end{align}
%---------------------------------
Here $\varrho_{mk}$ is the correlated shadow fading, with $\sigma_{sh}$ being the standard deviation. The term $d_{mk}$ is the 2D distance from~the~$k$th UE to the $m$th AP. 
%\colr{The term $P^{\text{LoS}}_{mk}$ is the probability of the presence of the LoS link, which is calculated as $P^{\text{LoS}}_{mk} = \frac{K_{mk}}{1+K_{mk}}$.}
We set the transmit and receiver RF impairment levels as $\kappa_{r,m}=\kappa_r$ and  $\kappa_{t,k}=\kappa_t$. We assume that each UE is  equipped with a $b$-bit DAC, and the APs have the following dynamic ADC architecture: $[b_1,b_2,b_3,b_4]$ i.e., out of $M$ APs, each of the $25\%$ of the APs have $b_1$, $b_2$, $b_3$, and $b_4$ bit ADC resolution, respectively.
  The ADC/DAC distortion factor $\rho$ for $b$ bits is given in~\cite{zhang2019_Low_ADC}.  We set the noise variance $\sigma^2_m =-94$~dBm, $M=64$~APs, $N=4$ antennas per AP, pilot power $\tilde{p}_{k}=10$~dBm, the velocity of UEs $v_k=54$~Km/hr. These parameters remain fixed unless explicitly specified. 
% \colr{Mention about how you are setting i) the distortion factor and bussgang gain values, ii) the dynamic architecture that you are considering for most of the plots, iii) the velocity of users considered for most of the plots.} 
%\colb{$95\%$ likely SE- 3GPP doc} \\
%-------------------------------------------------------------------------
%\colr{compare individual terms - DS, IUI, TRF, RRF, ADC, DAC - for different configurations\\ for different speeds K/2, K/2 -- different HW ADCs -- perfect and imperfect CSI } \\ 
%
%\colb{K/2 high speed (share same pilots) -- K/2 low speed (same pilots) -- 6-bit DAC and 1-bit DAC}\\
%\colb{K/2 high speed  -- K/2 low speed (diif vel UEs share same pilots) -- 6-bit DAC and 1-bit DAC}
%MN product -- dynamic ADC assignment -- to a subset of APs -- and at particular AP each antenna will have different ADC resolution  \\
%-------------------------------------------------------------------------
%{\textbf{\underline{Validation of closed-form SE:}}} 
\subsubsection{Validation of closed-form SE}
\begin{figure*}[htbp]
	\centering\vspace{-5pt}
	\centering%\vspace{-10pt}
	\begin{subfigure}{.24\textwidth}
		\centering
		\includegraphics[width=\linewidth,height=\linewidth]{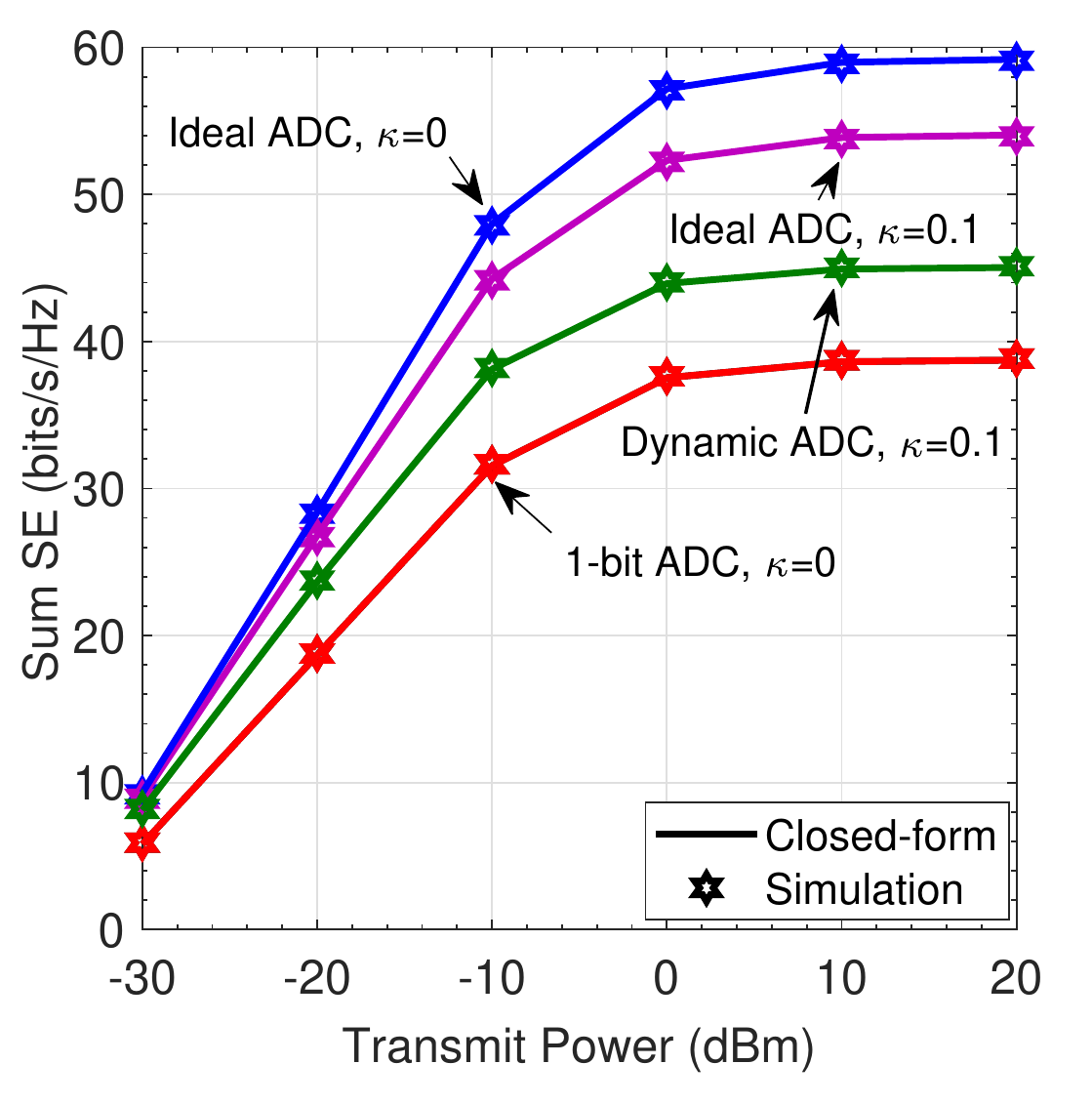}\vspace{-7pt}
		\caption{ \small} 
		\label{fig:SE_validation_ideal_dynamic_ADC}
	\end{subfigure}
	\begin{subfigure}{.24\textwidth}
		\centering
				\includegraphics[width=\linewidth,height=\linewidth]{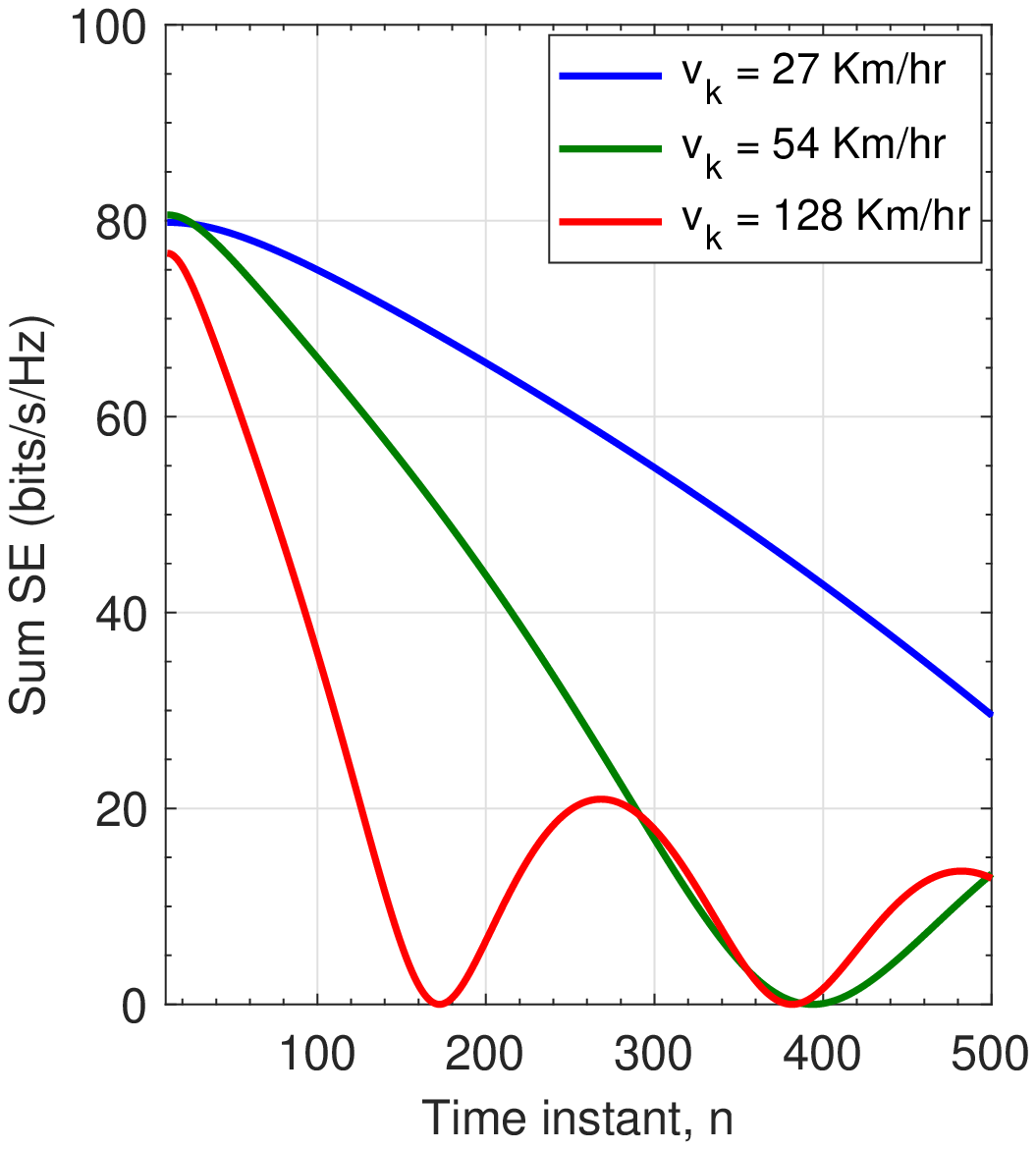}\vspace{-7pt}
		\caption{\small}
		\label{SE_vs_timeInstant}
	\end{subfigure}
	\begin{subfigure}{.24\textwidth}
	\centering
	\includegraphics[width=\linewidth,height=\linewidth]{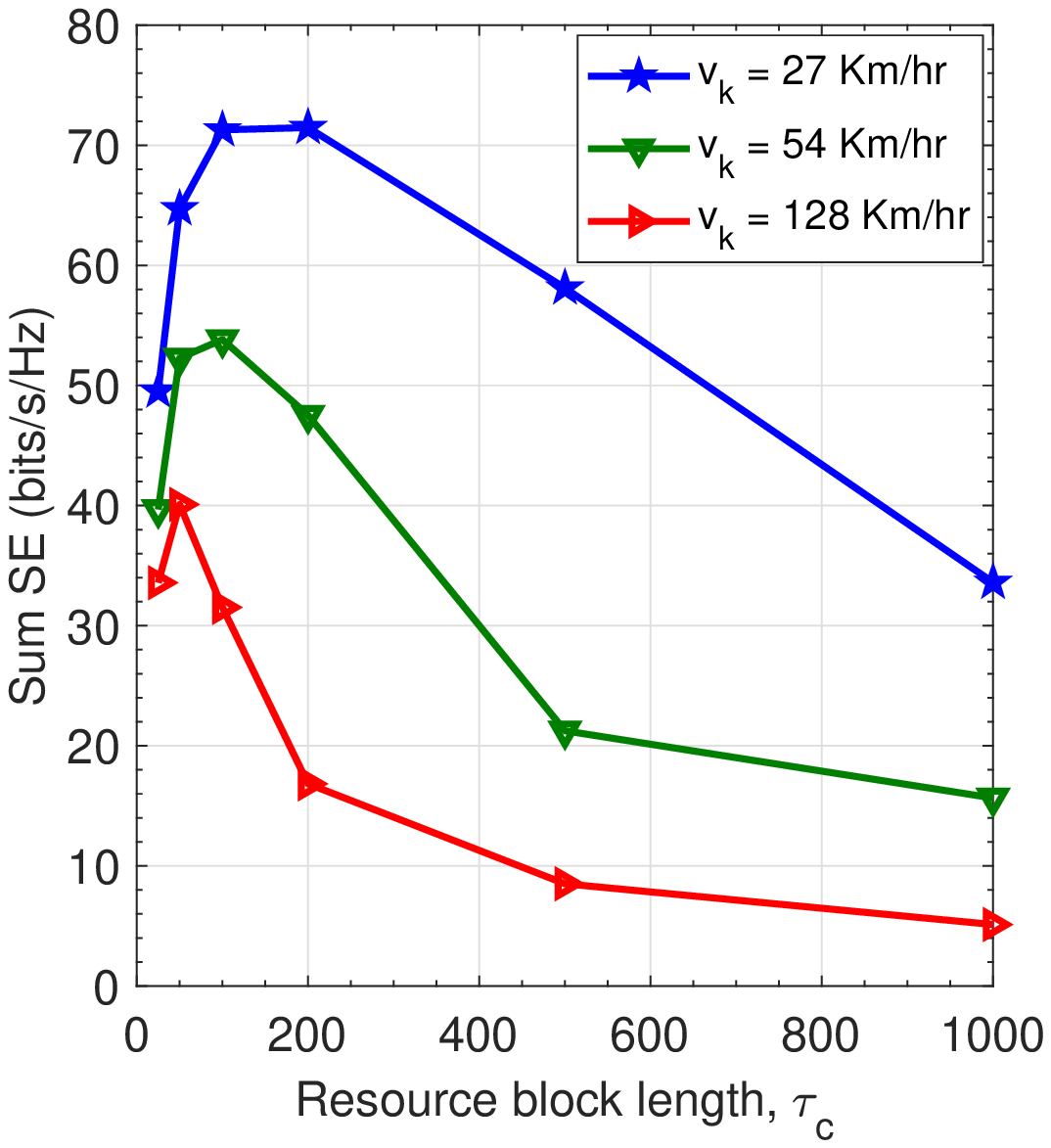}\vspace{-7pt}
	\caption{\small }
	\label{SE_vs_resourceBlock}  
\end{subfigure}
		\begin{subfigure}{.24\textwidth}
		\centering
		\includegraphics[width=\linewidth,height=\linewidth]{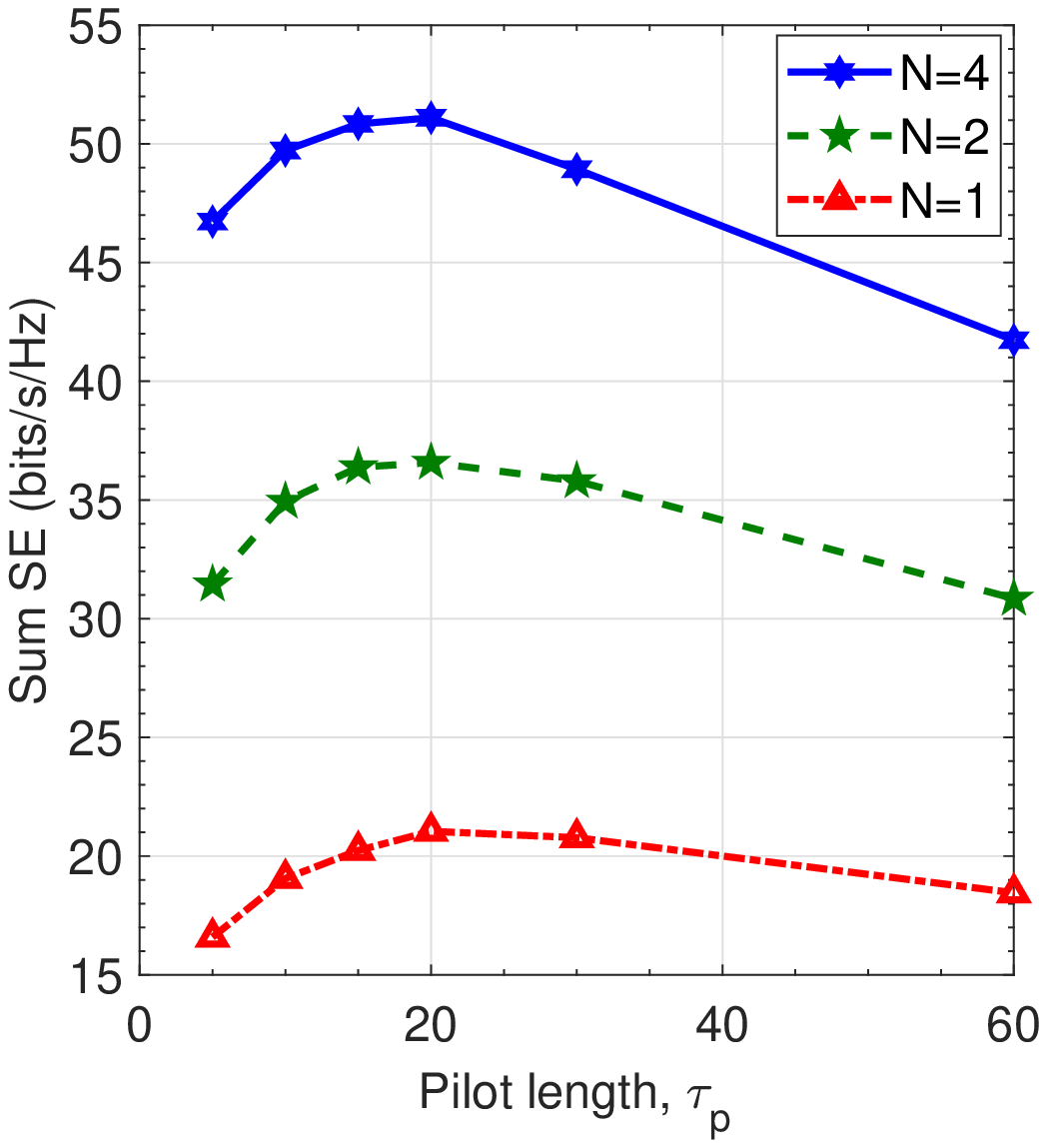}\vspace{-7pt}
		\caption{\small }
		\label{SE_vs_tauP}  
	\end{subfigure}
	\vspace{-12pt}
	\caption{a) Validation of closed-form SE; b) SE versus time instant $n$; c) SE versus the length of resource block $\tau_c$; and d) SE versus pilot length $\tau_p$. \vspace{-20pt}} 
	\label{fig:test2}
\end{figure*}

We first validate in Fig.~\ref{fig:SE_validation_ideal_dynamic_ADC} the derived closed-form SE expression in \eqref{eq_SINR_k} by comparing it with its {simulated} counterpart in~\eqref{eq_SE_expectations}, which numerically computes the expectations. For this study, we consider the following RF impairment and ADC/DAC combinations: {i) ideal RF with $\kappa_r = \kappa_t = \kappa = 0$, and ideal ADC/DAC with $b = \infty$-bit resolution; ii) non-ideal  RF with $\kappa=0.1$, and ideal ADC/DACs with $b = \infty$ bit; iii) non-ideal RF with $\kappa = 0.1$, and the following dynamic ADC architecture at the APs: $[b_1,b_2,b_3,b_4] = [1,2,4,6]$; and iv) ideal RF with  $\kappa = 0$ and $1$-bit ADC/DACs}. We see that for all of the  aforementioned RF and ADC/DAC impairments, the derived closed-form SE exactly matches with its simulated counterpart. This validates the derived analytical closed-form SE expression, which can thus be used for realistic evaluation of hardware-impaired CF mMIMO systems with channel aging. The dynamic ADC architecture with, varied ADC resolution at APs and non-ideal RF with $\kappa = 0.1$, is able to provide $85\%$ (resp. $75\%$) of the SE achieved by ideal ADCs (resp. ideal RF and ADC) case. \textit{The dynamic ADC architecture, with a careful bit resolution choice, is able to recover the spectral loss due to low-resolution ADCs and a non-ideal RF.}
%-------------------------------------------------------------------------
%\vspace{-0.8cm}
%-------------------------------------------------
\subsubsection{Impact of velocity on the length of the resource block $\tau_c$} 
We now investigate the combined effect of UE velocity and the resource block length $\tau_c$ on the sum SE. This will help in deciding the appropriate $\tau_c$ value for different UE velocities. We perform this study by plotting in Fig.~\ref{SE_vs_timeInstant} the SE versus the time instant $n$ in a resource block. For this study, we fix $\tau_c=500$, and consider three different UE velocities. We first see that for each UE velocity, the SE reduces with time instant $n$, and becomes zero. This is due the channel aging. The resource block length should be, thus, much lesser than the instant at which the SE becomes zero. This will avoid SE degradation towards the end of resource block. We also note that a small-length resource block will also increase the pilot overhead. 
We next investigate this trade-off in Fig.~\ref{SE_vs_resourceBlock} by plotting the SE versus the resource block length $\tau_c$ for different UE velocities. We see that, for all UE velocities,  the SE is relatively low for very small $\tau_c$ values. It then increases with $\tau_c$, and then reduces. This is because for very small $\tau_c$ values, the pilot overhead  is high, even though the channel remains fresh. The increased pilot overhead dominates for such low $\tau_c$ values, which leads to low SE values. The increase in SE with $\tau_c$ is due to the reduction in pilot overhead, even though the channel starts aging now. The reduced overhead dominates the degradation due to channel aging, which increases the SE. The final decrease in SE with increase in $\tau_c$ is because the channel ages too much for a long resource block. The degradation due to channel aging now dominates the reduced pilot overhead, which reduce the SE. We also note that the SE for a higher UE velocity, peaks for smaller $\tau_c$ value.  This is due to the increased channel aging. A system designer can thus, depending on the UE velocity, decide the $\tau_c$ value, where the SE peaks. 

\subsubsection{Impact of channel estimation on SE} 
{We now jointly investigate in Fig.~\ref{SE_vs_tauP} the impact of channel estimation on SE and resource allocation between channel training and data transmission, by plotting the sum SE versus the pilot sequence length  $\tau_p$ for different antenna values at the AP. We consider $M=100$~APs, $K=60$~UEs and velocity of UEs $v_k=54$~Km/hr.
We first observe that with increase in $\tau_p$, the SE first increases till a threshold value of $\tau_p=20$, and reduces after that. This is because the increase in $\tau_p$ increases the number of pilots. This reduces the number of UEs sharing the same pilot, which in turn, reduces the pilot contamination. This improves the channel estimation quality. For $\tau_p \le 20$, the increase in SINR due to improved channel estimates dominates the linear decrease in data transmission duration (resources), which increases the SE. For $\tau_p > 20$, the decrease in SE due to the reduced data transmission duration dominates the increased SE due to improved pilot channel estimation, which reduces the SE. 
This shows that $\tau_p=20$ is optimal for these system configurations.}

%-----------------------------------------------------------------
% We observe from Fig.~\ref{SE_vs_tauP} that with increase in $\tau_p$, the SE with LSFD initially increases till a threshold value of $\tau_p=20$, and reduces after that. This is because till $\tau_p= 20$, the increased pilot length increases the number of pilots. This reduces the number of UEs sharing the same pilots and consequently the pilot contamination. As a result the channel estimation quality improves, which the LSFD exploits to reduce the coherent interference due to pilot contamination, as shown in \eqref{LSFD_eqq} in the manuscript. We also note that increasing the pilot length reduces the symbols allocated to the uplink data transmission. For $\tau_p \le20$, the reduced coherent interference dominates the decreased uplink data transmission duration, which increases the SE. For $\tau_p > 20$, this does not happen, which reduces the SE. 
%
%\colb{We lastly note that for SLD, the threshold value of $\tau_p$ after which SE decreases is $\tau_p \le 10$, which is less than that of LSFD. This is because, even though the channel estimation quality improves, the SLD is not able to exploit it for canceling the coherent interference due to pilot contamination.} \colr{copy from the response sheet after modifying it there.}
%-----------------------------------------------------------------

\subsubsection{Impact of number of APs and UEs on SE}
{We numerically investigate in Fig.~\ref{SE_vs_APs} that the SE versus the number of APs $M$ for LSFD and SLD. We perform this study for $K=20$ and $K=40$ UEs, and consider a pilot length of $\tau_p=K/2$.  We see that the SE increases with  increase in $M$. This is  due to the increased array gain. We also note that the percentage of LSFD gain over SLD is higher  for $K=40$ UEs than $K=20$ UEs.
 {Specifically, for $K=40$~UEs and lower (resp. higher) number of APs, the LSFD offers $93\%$ (resp. $82\%$) SE gain over SLD, whereas these gains are $78\%$ (resp. $56\%$) for $K=20$ case.} This is because the increased number of UEs increases the interference experienced by the desired UE, which the LSFD can efficiently mitigate at the CPU by performing an extra level of decoding.}
%------------------------------------------------------------------------- 
\subsubsection{Effect of channel aging on the power of desired and interference terms}
\begin{figure*}[t] %[htbp]
	\centering\vspace{-15pt}
		\begin{subfigure}{.24\textwidth}
		\centering
		\includegraphics[width=\linewidth,height=\linewidth]{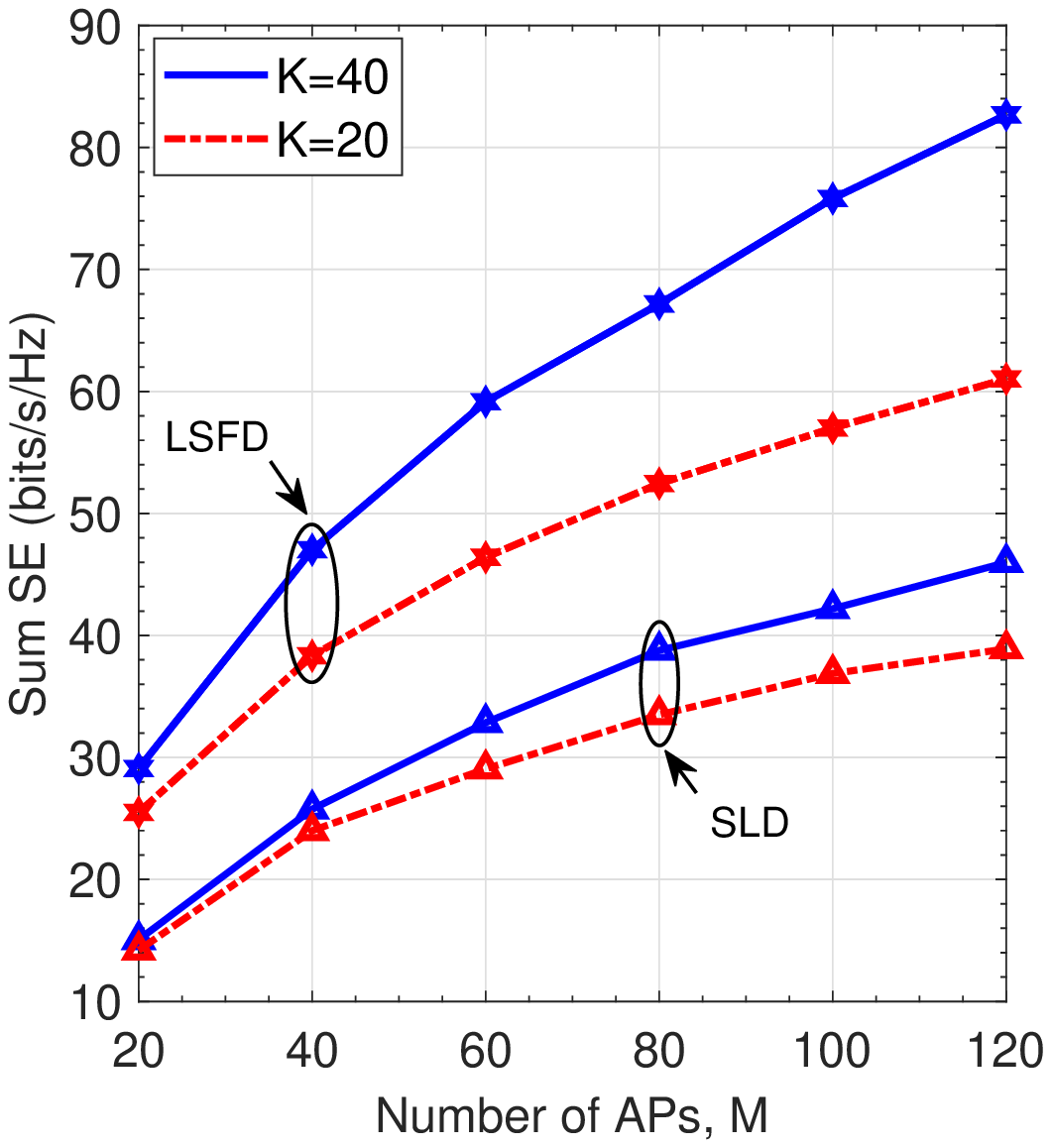}\vspace{-7pt}
		\caption{\small}
		\label{SE_vs_APs}
	\end{subfigure}
	\begin{subfigure}{.24\textwidth}
		\centering
		\includegraphics[width=\linewidth,height=\linewidth]{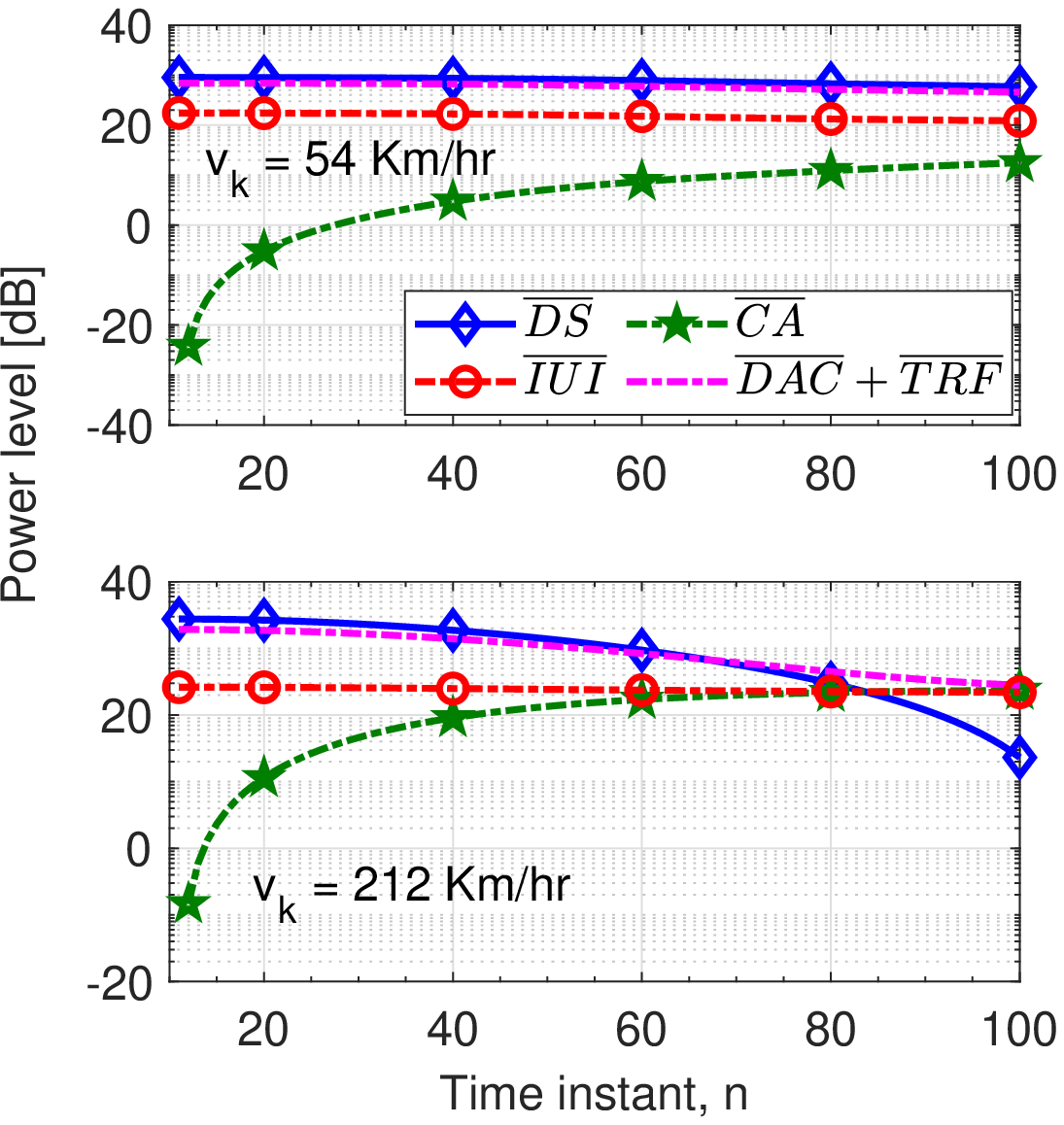}\vspace{-7pt}
		\caption{\small }
		\label{fig_CDF_LSFD_each_term_power} 
	\end{subfigure}
	\begin{subfigure}{.24\textwidth}
		\centering
		\includegraphics[width=\linewidth,height=\linewidth]{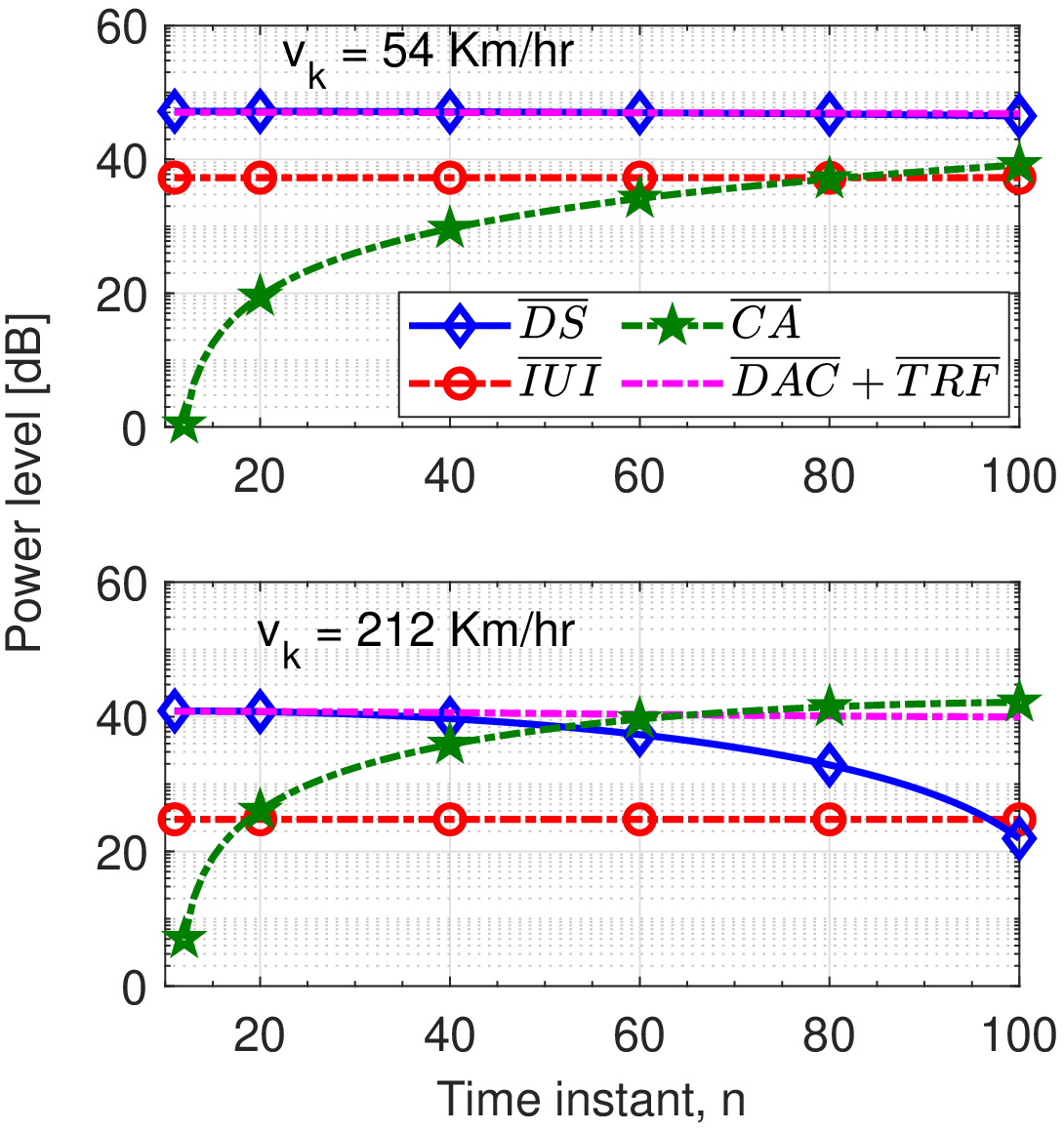}\vspace{-7pt}
		\caption{\small }
		\label{fig_CDF_SLD_each_term_power}
	\end{subfigure}
	\begin{subfigure}{.24\textwidth}
		\centering	
	\includegraphics[width=\linewidth,height=\linewidth]{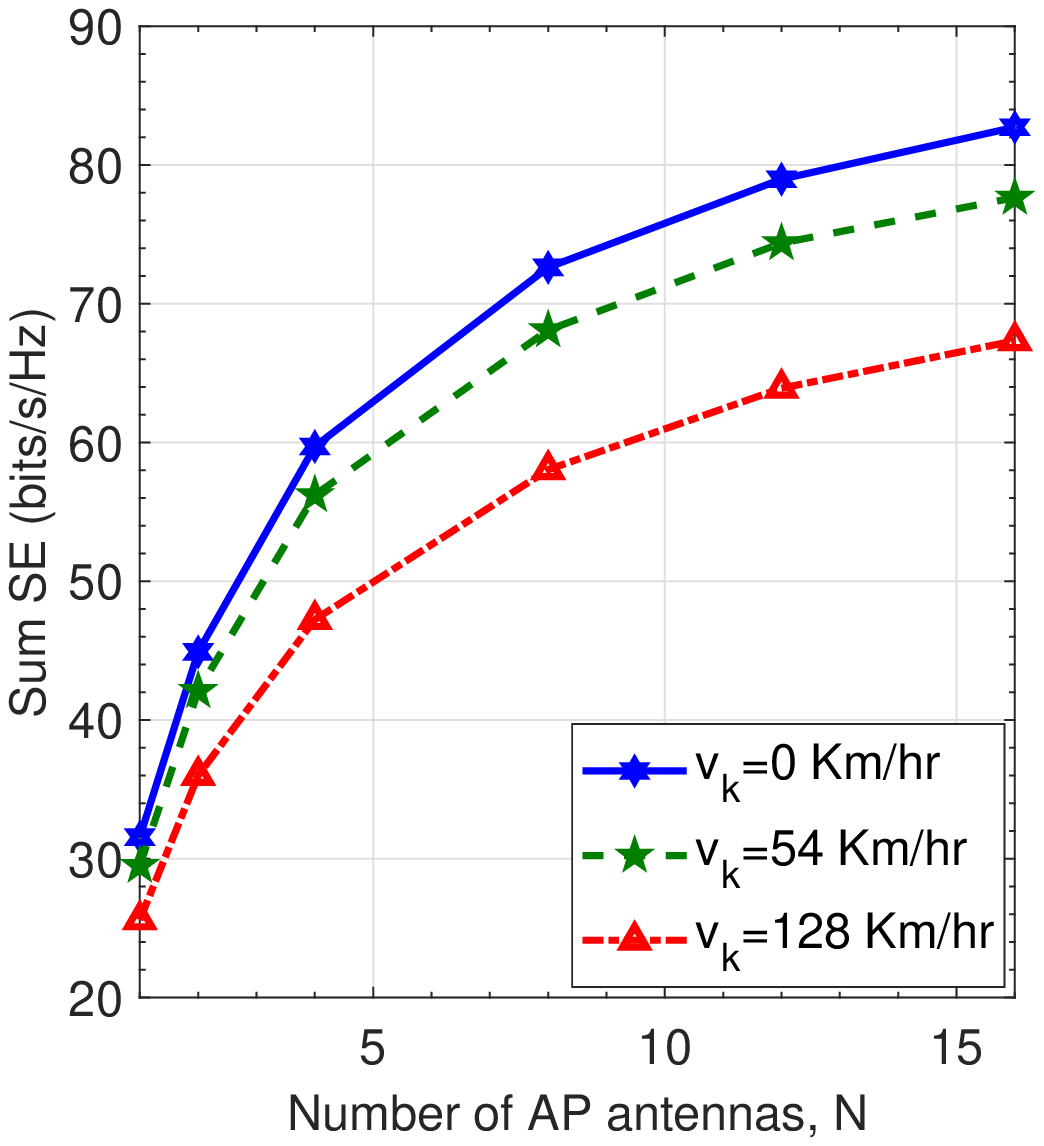}\vspace{-7pt}  
	\caption{\small }
	\label{SE_time_instant}		
		%			\includegraphics[width=\linewidth,height=\linewidth]{fig/combined_lsfdpercentage_gain_stp20_ALLk20.eps}\vspace{-7pt}
		%		\caption{\small}
		%		\label{fig_percentage_LSFD_gain_3D}
	\end{subfigure}
	\vspace{-15pt}
	\caption{a) Comparison of LSFD and SLD for different UE configuration; Effect of channel aging on the power of desired and interference terms for b) LSFD; and c) SLD; and d) SE versus number of antenna per AP for different UE velocities. \vspace{-20pt}} 
	\label{fig:test_individual_terms}
\end{figure*}

We now numerically investigate in Fig.~\ref{fig_CDF_LSFD_each_term_power}, the behavior of the desired signal  power $\overline{\text{DS}}_{k,n}$, channel aging power $\overline{\text{CA}}_{k,n}$, inter-user interference power $\sum_{i \neq k}^{K}\overline{\text{IUI}}_{ki,n}$, and UE RF and DAC distortion  power $\overline{\text{DAC}}_{k,n}+\overline{\text{TRF}}_{k,n}$ (labelled as $\overline{\text{DS}}$, $\overline{\text{CA}}$, $\overline{\text{IUI}}$ and $\overline{\text{DAC}}+ \overline{\text{TRF}}$, respectively), for LSFD and SLD schemes. For this study, we consider $K = 20$ UEs. The $K/2$ UEs have a velocity of $v_k=54$~Km/hr, and the remaining $K/2$ UEs have $v_k=212$~Km/hr. We plot in Fig.~\ref{fig_CDF_LSFD_each_term_power}, the power level of signal and interference terms of all UEs with LSFD. We observe the following:
	\begin{itemize}[leftmargin = *]
		\item For high-velocity UEs i.e., $v_k = 212$ Km/hr, the $\overline{\text{DS}}$ and $\overline{\text{DAC}}+\overline{\text{TRF}}$ powers reduce with time (see the bottom subplot in Fig.~\ref{fig_CDF_LSFD_each_term_power}). For example, at $n = 100$, they are roughly $20$~dB lesser than at $n = 10$. This is because both these powers are functions of the temporal correlation coefficient $\rho_{k}[n-\lambda]$, whose value, for a high UE velocity,  reduces greatly with time. For low-velocity UEs, as shown in the top subplot in Fig.~\ref{fig_CDF_LSFD_each_term_power}, these powers are almost time-invariant. This is because for such UEs, $\rho_{k}[n-\lambda]$ varies very slowly with $n$.
		\item For high-velocity UEs and $n \ge 60$, we see that $\overline{\text{DS}}$ reduces monotonically,  while $\overline{\text{DAC}}+\overline{\text{TRF}}$ power first tapers out, and then floors to a constant value. This is because, as also shown in Corollary~\ref{corollary_interference_order}, the $\overline{\text{DS}}$ power decreases as $\rho_k^2[n-\lambda]$, which decreases monotonically with $n$. The transmit RF + DAC impairments reduce as $\epsilon + \rho_k^2[n-\lambda]$, with $\epsilon > 1$ being a constant. For  $n\geq 60$, the constant $\epsilon $ term dominates the reduction due to $\rho_k^2[n-\lambda]$, which makes the  $\overline{\text{DAC}}+\overline{\text{TRF}}$ impairment power floor to a constant value.
		\item For both UE velocities, $\overline{\text{IUI}}$ power almost remains constant. This observation is in line with the result in Corollary~\ref{corollary_interference_order}, which showed that IUI power remains constant for all UE velocities.
	\end{itemize}
%-------------------------------------------------------------------------
%-------------------------------------------------------------------------
 %\vspace{-5pt}
%------------------------------------------------------------------------- 
%%-----------------------------------------------------
%\begin{figure*}[htpb] %\setcounter{figure}{1} %{0.5\textwidth}
%	\centering 
%	%\begin{subfigure}[b]{.45\linewidth}%\setcounter{subfigure}{3}
%	\includegraphics[width=8cm,height=8cm]{fig/vs_Antenna_M_64_K_20_tau_p_10_Dynamic_ADC_RF_0p01_Velocity_128_10setups.eps}  
%	\caption{\small SE versus number of antenna per AP for different UE velocities.}
%	\label{SE_time_instant}
%\end{figure*}
%-------------------------------------------------
We next compare Fig.~\ref{fig_CDF_LSFD_each_term_power} with Fig.~\ref{fig_CDF_SLD_each_term_power}, which plots the above power values for SLD. We see that for low-velocity UEs, LSFD has  a much lower $\overline{\text{IUI}}$, $\overline{\text{CA}}$  and $\overline{\text{DAC}}+\overline{\text{TRF}}$ power than SLD. This is because the channels of low-velocity UEs do not significantly age,  and consequently their channel estimate quality do not deteriorate. The LSFD can thus better suppress the IUI.  For high-velocity UEs, LSFD yields much lower $\overline{\text{CA}}$ values  than SLD, while both LSFD and SLD yield similar $\overline{\text{IUI}}$ values. This implies that LSFD can mitigate the effect of channel aging for high-velocity UEs, but not IUI. 
This study not only validates the interference-related insights in Corollary \ref{corollary_interference_order},  but also shows that LSFD can mitigate i) the effect of channel aging for low/high UE velocities;  and ii) IUI, but only for low-velocity UEs.% due to their better  channel estimates.

\subsubsection{Impact of AP antennas on the UE velocity}
{We now investigate in  Fig.~\ref{SE_time_instant} the SE versus number of antennas per AP ($N$) for different UE velocities: $v_k = \{0,54,128\}$~Km/hr. We see that the SE increasing with $N$. This is due to the increased array gain. We also infer that for different $N$ values,  the $\%$ of SE loss for $v_k = 54$~Km/hr (resp. $v_k = 128$~Km/hr) over $v_k = 0$~Km/hr is same i.e., $5\%$ (resp. $19\%$). \textit{This crucially informs  that the SE degradation due to channel aging is independent of the $N$ value, and depends only on the UE speed.}}
%------------------------------------------------------------------

%-------------------------------------------------------------------------
%\begin{figure*}[htbp]
%	\centering\vspace{-5pt}
%%	\begin{subfigure}{.45\textwidth}
%		\centering
%		\includegraphics[width=0.4\linewidth,height=0.4\linewidth]{fig/N_4_M_64_compare_Res.eps}
%%		\caption{\small}
%%	\end{subfigure}
%	\vspace{-5pt}
%	\caption{Comparison of different ADC architectures.} 
%			\label{SE_diff_ADC_ress}
%%	\label{fig:test}
%\end{figure*}
%-------------------------------------------------
\subsubsection{Dynamic ADC architecture across APs/antennas}
{To investigate the effect of dynamic ADC architecture, we plot in Fig.~\ref{SE_diff_ADC_ress} the SE versus transmit power and compare for the following ADC architectures: i) ideal ADCs; ii) \textit{Dynamic ADC - antennas}: $25\%$ of the antennas at each AP has $1$, $2$, $3$ and $4$ bit ADC resolutions; and iii) Dynamic ADC - APs: $25\%$ of the APs have $1$, $2$, $3$ and $4$ bit resolution. We see from Fig.~\ref{SE_diff_ADC_ress} that  \textit{Dynamic ADC $-$ antennas} architecture yields a similar SE as that of \textit{Dynamic ADC $-$ APs}  architecture. Also, both dynamic architectures provide nearly $84\%$ of the ideal ADCs SE. For this study, we assume that each AP has $4$ antennas.}

This shows the flexibility of dynamic ADC architecture, which allows the system designer to either have different ADC resolution across the antennas of each AP or across APs.

\subsubsection{SE versus UE transmit power}
\begin{figure*}[tbp]
	\centering\vspace{-15pt}
	\begin{subfigure}{.24\textwidth}
		\centering
		\includegraphics[width=\linewidth,height=\linewidth]{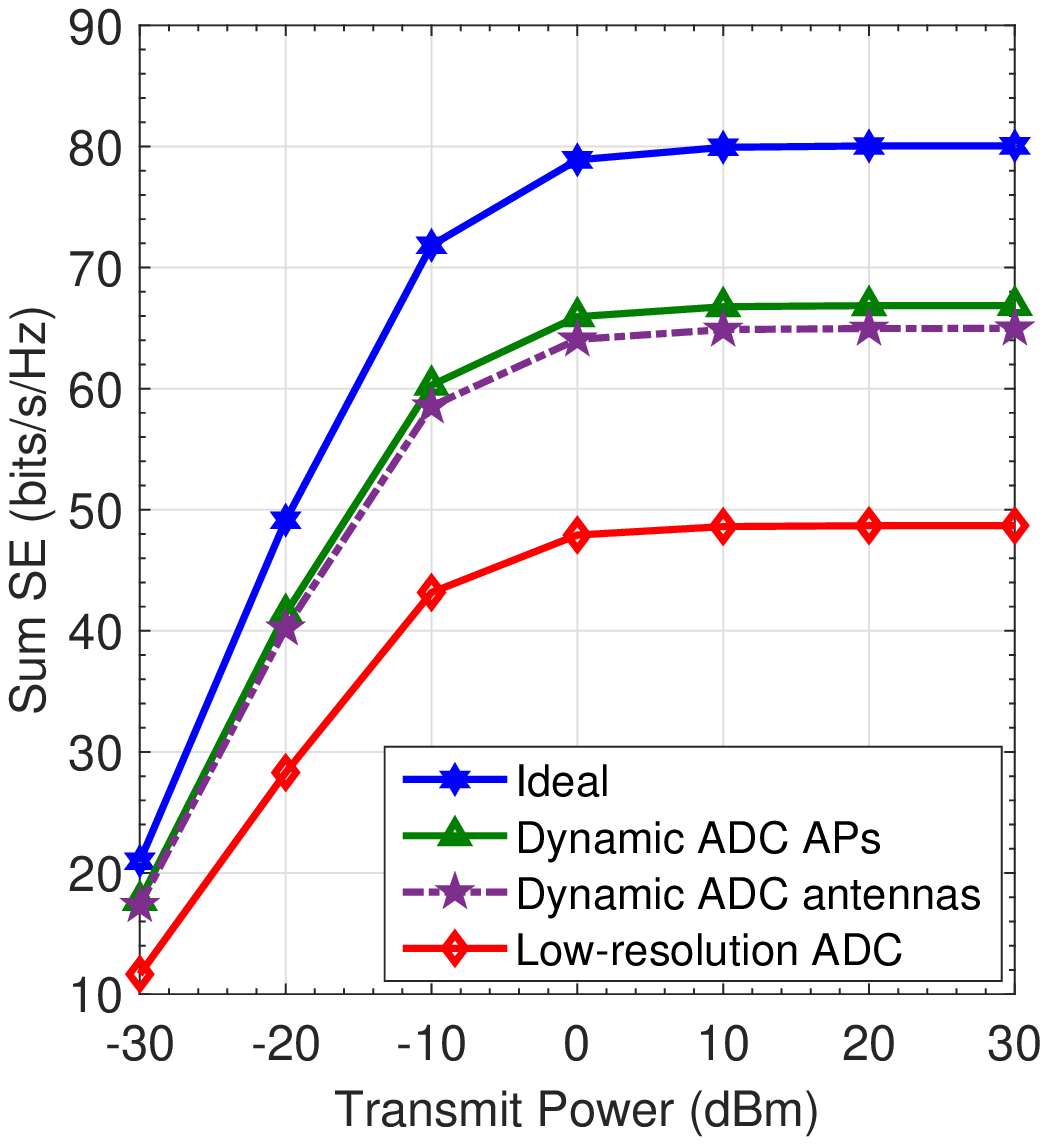}\vspace{-7pt}
		\caption{\small} 
			\label{SE_diff_ADC_ress}				
	\end{subfigure}
	\begin{subfigure}{.24\textwidth}
		\centering
		\includegraphics[width=\linewidth,height=\linewidth]{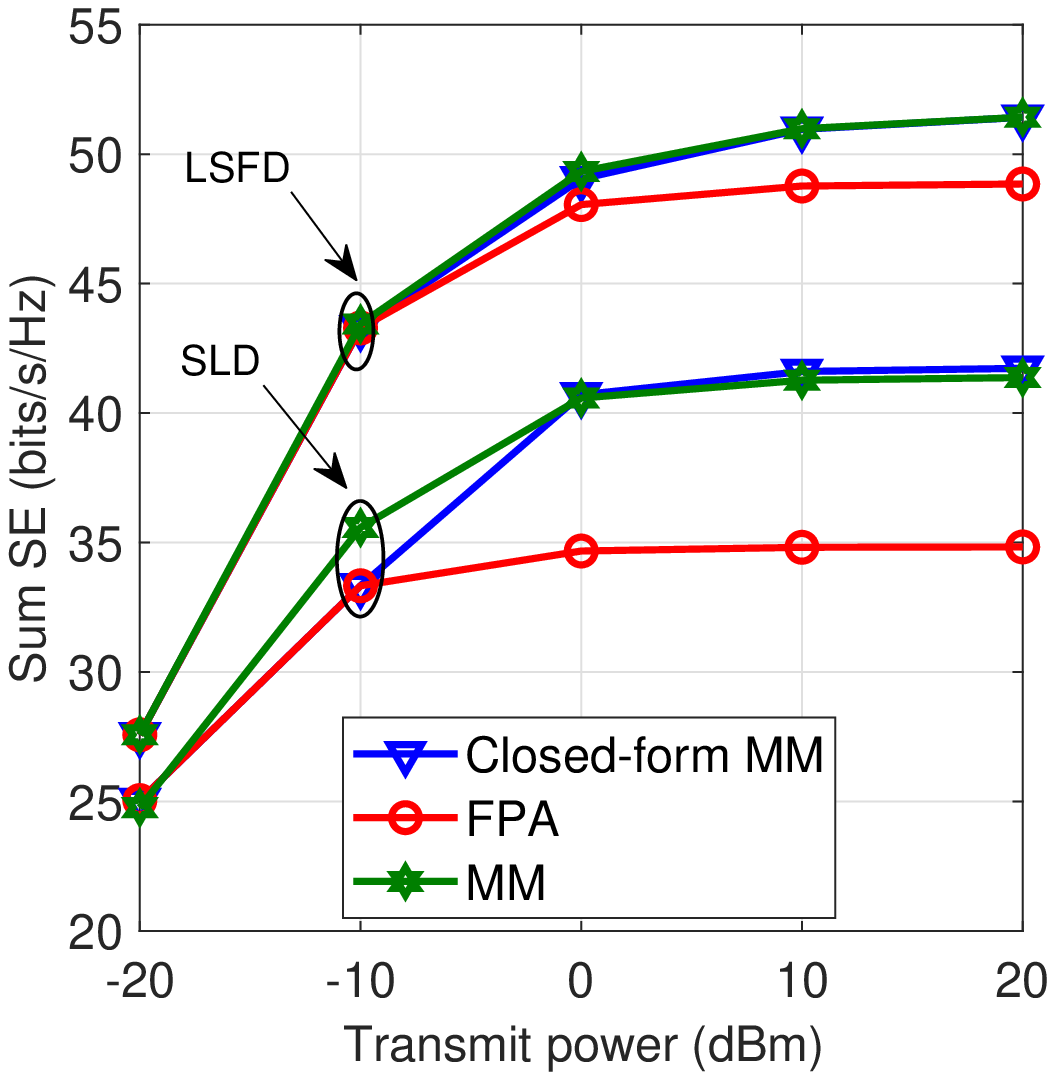}\vspace{-7pt}
		\caption{\small}
		\label{fig_SE_optimization_QT_CF_CVX}
	\end{subfigure}
	\begin{subfigure}{.24\textwidth}
		\centering
		\includegraphics[width=\linewidth,height=\linewidth]{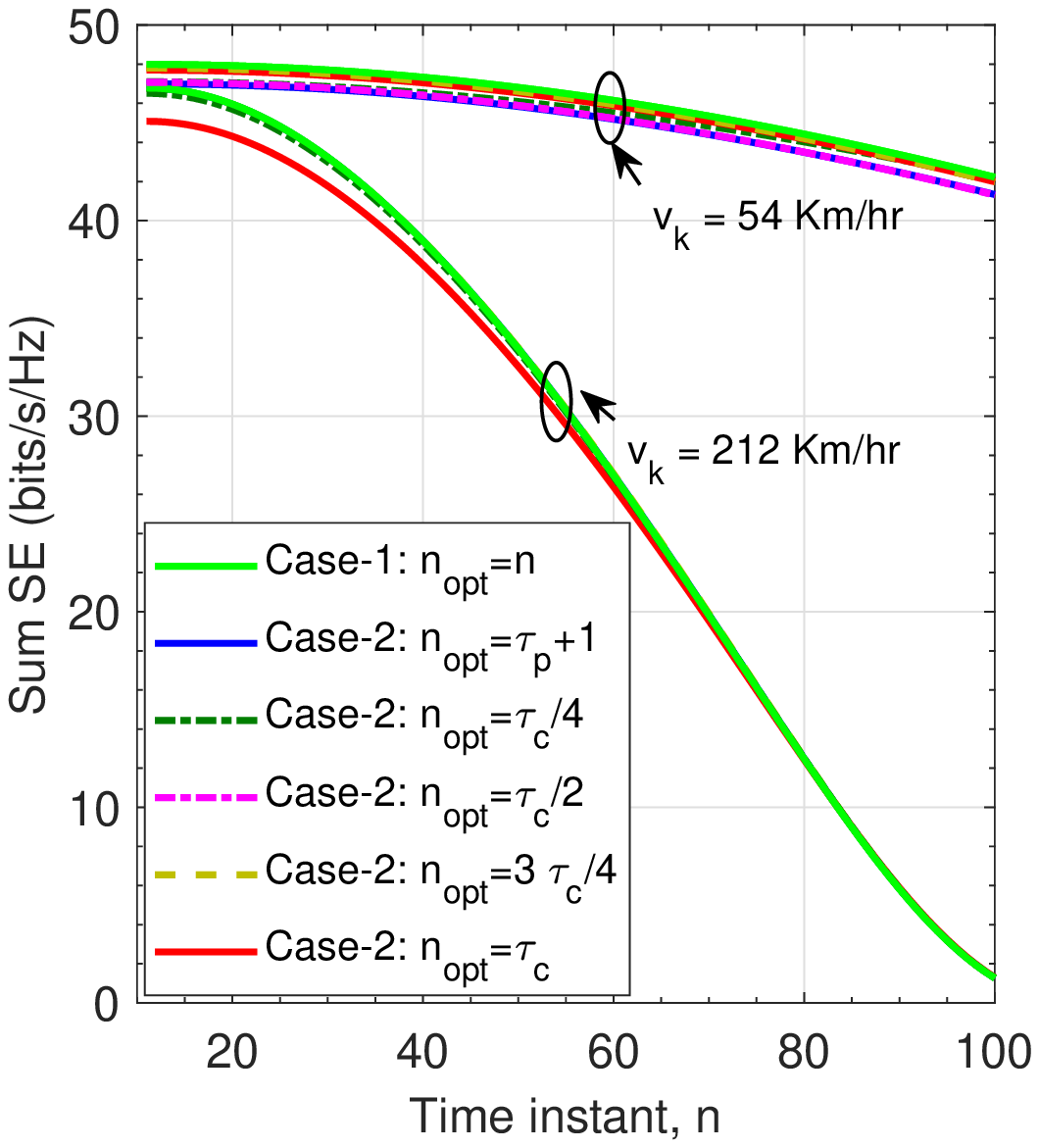}\vspace{-7pt}
		\caption{\small}
		\label{fig_optimization_vs_instant}
	\end{subfigure}
		\begin{subfigure}{.24\textwidth}
		\centering
		\includegraphics[width=\linewidth,height=\linewidth]{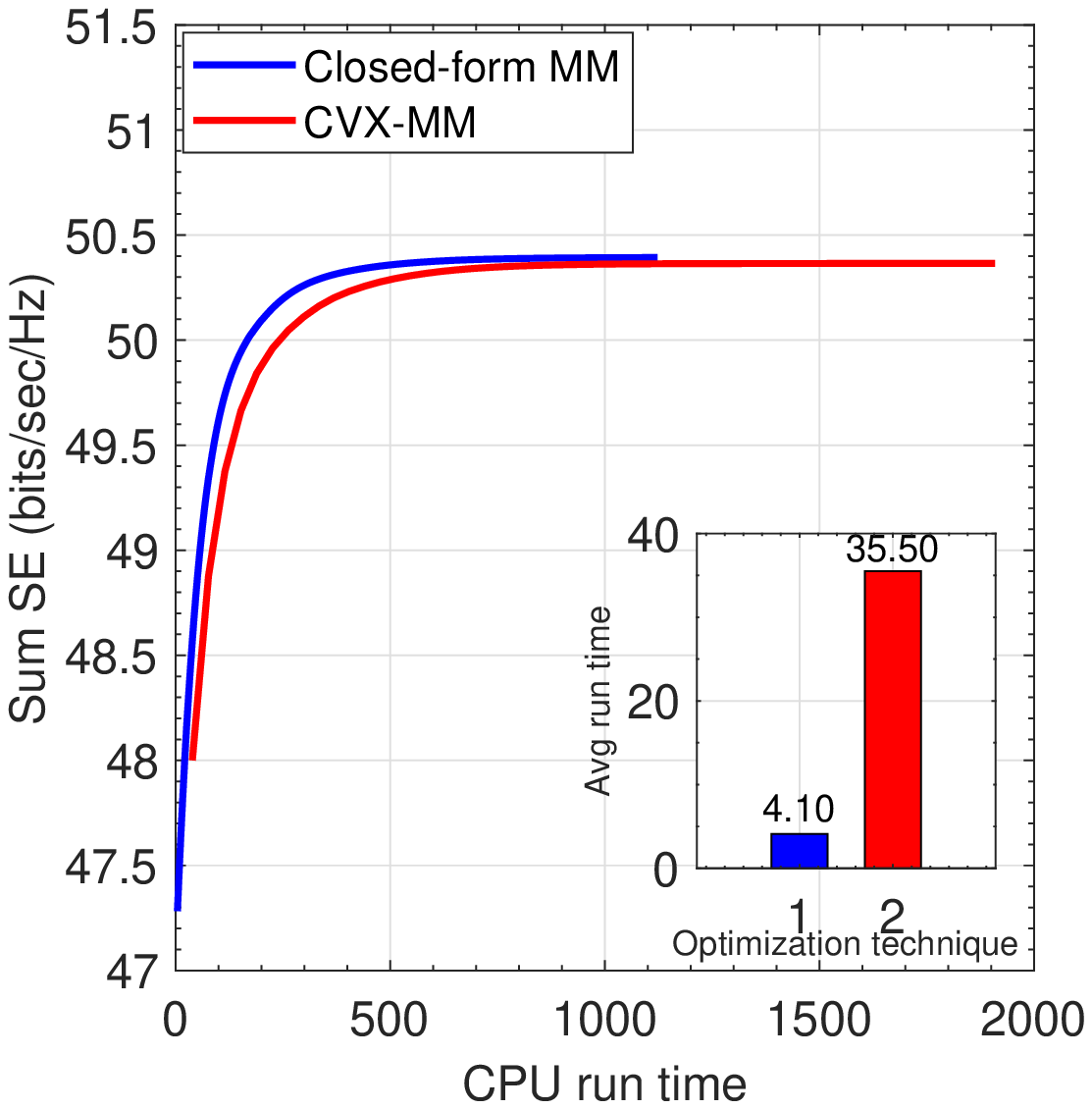}\vspace{-7pt}
		\caption{\small}
		\label{SE_vs_optimization_avgRuntime}
	\end{subfigure}
	\vspace{-15pt}
	\caption{a) Comparison of different ADC architectures; b) SE versus UE transmit power; c) SE versus time instant $n$;  d) CPU and average run time for different optimization techniques. 	\vspace{-20pt}} 
	\label{fig:test} 
\end{figure*}

We now plot in Fig.~\ref{fig_SE_optimization_QT_CF_CVX}, the SE obtained using the proposed optimization algorithms: CVX-based (labelled as MM) and Algorithm~\ref{algo_SE_closed-form_QT} (labelled as closed-form MM) for LSFD and SLD schemes. We also compare them with full power allocation (FPA) scheme, where each UE transmits at its maximum power. We fix $M=64$~APs, $K=20$ UEs, $N=4$ antennas, $\kappa = 0.1$ and $b=4$ bits at both APs and UEs.
We first see that both CVX-based and closed-form MM provide a similar SE, but the latter with its closed-form transmit power updates, has a trivial complexity. 
%Recall that Algorithm~\ref{algo_SE_closed-form_QT} provides closed form updates, and therefore has a trivial complexity.
It can, therefore, be easily implemented in commercial CF  systems. Also, both proposed optimizations improve the SE by $20\%$ in SLD, whereas they provide a reduced gain of $6.25\%$ in LSFD. This is because SLD is unable to effectively  mitigate the IUI,  which in turn is mitigated by using the optimal powers designed by the proposed algorithms. The LSFD, in contrast, mitigates the IUI by performing an extra level of decoding at the CPU, and optimizing transmit power, therefore, only provides marginal SE gain. 
%\textit{This study informs the designer that for SLD, the proposed optimization provides a significant gain, and its implementation is justified. For LSFD, its implementation, can be avoided.} 
%-------------------------------------------------------------------------
%\vspace{-20pt}
%-------------------------------------------------------------------------
%
% We observe that the SE with LSFD is higher than SLD.  This is because  LSFD can suppress the pilot contamination interference.
%\colr{We have not explained which plot consider the optimization, and which one do not.}
%------------------------------------------------------------------

\subsubsection{Effect of channel aging on SE optimization}
We now study in Fig.~\ref{fig_optimization_vs_instant}, the effect of optimization time instant on the sum SE for the following UE velocities $v_k=\{54,212\}$~Km/hr. We optimize the SE at a specific time-instant, denoted as $n_{\text{opt}}$, and use the obtained optimal transmit powers to calculate the SE for all time-instants. 
{ We compare in Fig.~\ref{fig_optimization_vs_instant},  the SE for:
	\begin{itemize}
		\item Case 1: SE is optimized at every data transmission instant  i.e., $n_{\text{opt}} = n$.
		\item Case 2: SE  is optimized at one of the following time instants $n_{\text{opt}} = \{\tau_p+1,\frac{\tau_c}{4},\frac{\tau_c}{2},\frac{3\tau_c}{4}\}$. 
	\end{itemize}} 
We perform this study for two different UE velocities of $v_k = 54$~Km/hr and $v_k =212$~Km/hr. We see that for a low UE velocity of $v_k = 54$~Km/hr,  Case 2, irrespective of the optimizing time-instant $n_{\text{opt}}$  yields a similar SE as that of Case 1 for all time instants. For a high UE velocity of $v_k = 212$~Km/hr, Case 1 and Case 2 with $n_{\text{opt}} \leq \frac{3\tau_c}{4}$ has similar SE. But, for $n_{\text{opt}} = \tau_c$, Case 2 yields only a minor $10\%$ lower SE than Case 1, and that too during initial time instants. The SE thus can be optimized in the beginning of the resource block and its solution could be reported only once, which makes the proposed optimization practical. 	

\subsubsection{Optimization complexity} We now analyze the time complexity of both the proposed iterative optimization MM techniques – CVX-based and closed-form -- by first plotting in Fig.~\ref{SE_vs_optimization_avgRuntime} their  total CPU run time required to converge. We see that the later approach requires much lesser time to converge. Also, both the techniques converge to the same SE value.  We next also investigate the average per-iteration CPU run time required by both algorithms (shown in the sub-plot). 
 We see that the closed-form technique has a much lesser run-time per iteration. This is due to its closed-form power updates with trivial complexity. 
%------------------------------------------------------------------
%We now analyze the time complexity of both the proposed iterative optimization MM techniques – CVX-based and closed-form -- by first plotting in Fig.~\ref{SE_vs_optimization_avgRuntime} their average per-iteration CPU run time. We see that the latter approach has a much lesser run-time per iteration. This is due to its closed-form power updates with trivial complexity.  We next investigate the total time required by both algorithms to converge by plotting the sum SE versus CPU run time (shown in the sub-plot). We again see that the closed-form technique requires much lesser time to converge. Also, both the techniques converge to the same SE value.
%------------------------------------------------------------------
\vspace{-0.3cm}
\section{Conclusion} \label{Conclusion7}\vspace{-0.1cm}
We derived a closed-form SE expression for a hardware-impaired spatially-correlated Rician-faded CF mMIMO system with channel aging and two-layer LSFD. We verified this expression for different transmit power and hardware impairment values.  We optimized the non-convex SE metric by proposing two novel optimization techniques. The first one has a high complexity,  while the second provides a closed-form solution with a trivial complexity.  We numerically showed that the two-layer LSFD effectively mitigates the interference due to channel aging for both low- and high-velocity UEs. It, however, mitigates inter-user interference only for the low-velocity UEs,  and not the fast ones. {We also showed that the SE loss due to hardware impairments can be compensated by the proposed dynamic ADC architecture.}

\appendices
\vspace{-0.5cm}
\section{} \label{appendix_LMMSE}  \vspace{-0.3cm}
The LMMSE estimate of the channel $\hv_{mk}$, based on received pilot signal in \eqref{eq:y_m[t_k]} is given by\vspace{-5pt}
\begin{align} 
    \hat{\hv}_{mk}[\lambda]=\Cmat_{\hv_{mk}[\lambda]\yv_{m}[t_k]}\Cmat_{\yv_{m}[t_k]\yv_{m}[t_k]}^{-1}\yv_{m}[t_k]. \label{eq_ch_estimate_appendix} \\[-33pt] \notag
\end{align}
Here $\Cmat_{\hv_{mk}[\lambda]\yv_{m}^p[t_k]}=\E\big\{\hv_{mk}[\lambda](\yv^p_{\text{ADC},m}[t_{k}])^H\big\}$ and $\Cmat_{\yv_{m}^p[t_k]\yv^p_{m}[t_k]}=\E\big\{\yv_{\text{ADC},m}^p[t_{k}](\yv^p_{\text{ADC},m}[t_{k}])^H   \big\}$.

We begin by simplifying the term $\Cmat_{\yv_{m}[t_k]\yv_{m}[t_k]}$ by substituting $\yv_{\text{ADC},m}[t_{k}]$ from \eqref{eq:y_m[t_k]} as\vspace{-4pt}
\begin{align}
   \Cmat_{\yv_{m}[t_k]\yv_{m}[t_k]} &\overset{(a)}{=} \sum_{j\in \mathcal{P}_k}\Amat_{m}\alpha_{d,j}^{2}\tilde{p}_{j}\bar{\Rmat}_{mj}\Amat_{m}^{H} + \sum_{j\in \mathcal{P}_k}\Amat_{m}(1-\alpha_{d,j}+\kappa_{t,j}^{2})\alpha_{d,j}\tilde{p}_{j}\bar{\Rmat}_{mj}\Amat_{m}^{H}  \nonumber\\[-4pt]
    &\qquad + \kappa^{2}_{r,m}\Jmat_k + \Bmat_{m}\big((1+\kappa_{r,m}^{2})\Jmat[t_k]+\sigma^{2}\Imat_{N}\big)  +\sigma^{2}\Amat_{m}\Amat_{m}^{H} \nonumber\\[-4pt]
    &= \sum_{j\in \mathcal{P}_k}\Amat_{m}\alpha_{d,j}(1+\kappa_{t,j}^{2})\tilde{p}_j\mathbf{\overline{R}}_{mj}\Amat_{m}^{H} 
    +\big(\mathbf{B}_{a}^{m}+\kappa^{2}_{r,m}\Amat_{m}\big)\Jmat+ \sigma^{2}\Amat_{m}.
    \label{eq_C_yy} \\[-33pt] \notag
\end{align}
Here $\Jmat_k = \sum_{j\in\mathcal{P}_{k}}(1+\kappa_{t,j}^{2})\alpha_{d,j}\tilde{p}_{j}\text{diag}\big(\bar{\Rmat}_{mj}\big)$.
Equality $(a)$ is obtained by noting that the i) distortion terms $\upsilon_{\text{DAC},k}^p[t_k],\xi_{\text{RF},k}^p[t_k],\boldsymbol{\eta}^p_{\text{RF},m}[t_k]$ and $\nv_{\text{ADC},m}^p[t_k]$ are independent of each other, and have a zero mean; and ii) variance of distortion/quantization noises are 
%-----------------
% $\E\{\upsilon_{\text{DAC},j}[t_k]\upsilon_{\text{DAC},j}^{H}[t_k]\}=\alpha_{d,j}(1\!-\!\alpha_{d,j})\tilde{p}_{j}$,   
%-----------------
$\E\{\xi_{\text{RF},j}^p[t_k](\xi^p_{\text{RF},j}[t_k])^{H}\}\!=\kappa_{t,j}^{2}\alpha_{d,j}\tilde{p}_{j}$,
$ \E\{\nv_{\text{ADC},m}^p[t_k](\nv_{\text{ADC},m}^p[t_k])^{H}\} =\Bmat_{m}\Smat^{m}[t_k]$, $\E\{\boldsymbol{\eta}_{\text{RF},m}^p[t_k](\boldsymbol{\eta}^p_{\text{RF},m}[t_k])^{H}\}=\kappa^{2}_{r,m}{\Wmat}^{m}[t_k]$; and  iii) impairment/quantization noises are independent of channel $\hv_{mk}$. Here the matrices $\Wmat^{m}[t_k] = \sum_{j\in\mathcal{P}_{k}}(1+\kappa_{t,j}^{2})\alpha_{d,j}\tilde{p}_{j}\text{diag}\big(\hv_{mj}[t_k]\hv_{mj}^{H}[t_k]\big)$ and  
 $ \Smat^{m}[n]  =(1+\kappa_{r,m}^{2})\Wmat^{m}[t_k]+\sigma^{2}\mathbf{I}_{N}$.
%--------------------------------
%To compute the expectations in \eqref{eq_C_yy}, we first compute $\E\Big\{\hv_{mk^{'}}[t_k]\hv_{mk^{'}}^{H}[t_k] \Big\}$ and $\E\Big\{\hv_{mk^{'}}[t_k]n_{DAC,k^{'}}[t_k]n_{DAC,k^{'}}^{H}[t_k]\hv_{mk^{'}}^{H}[t_k]\Big\}$ as below
%--------------------------------
We, similarly, have $\Cmat_{\hv_{mk}[\lambda]\yv_{m}^p[t_k]}$ as $ \Cmat_{\hv_{mk}[\lambda]\yv_{m}^p[t_k]} = \alpha_{d,k}\sqrt{\tilde{p}_{k}}\rho_{k}[\lambda\!-\!t_k]\bar{\Rmat}_{mk}\Amat_{m} $.
%Substituting \eqref{eq_C_yy} and $\Cmat_{\hv_{mk}[\lambda]\yv_{m}^p[t_k]}$ in \eqref{eq_ch_estimate_appendix}, we obtain the LMMSE estimate of $\hat{\hv}_{mk}[\lambda]$ in \colr{\eqref{eq:LMMSE_p_aware}}.
%-----------------------------------------------------
%\begin{align}
%    \hat{\hv}_{mk}[\lambda]=\alpha_{d,k}\sqrt{\tilde{p}_{k}}\rho_{k}[\lambda-t_k]\bar{\Rmat}_{mk}\Amat_{m}\boldsymbol{\Psi}_{mk}\yv_{,\text{ADC},m}[t_k].   
%\end{align} 
%-----------------------------------------------------
\vspace{-0.2cm}
%------------------------------------------------------APPENDIX C----------------------------
\section{} \label{appen_Lemma} \vspace{-0.1cm}
%\section*{Proof of Lemma 1} \vspace{-0.2cm}
To obtain the result (a) in Table~\ref{Lemma_Closed_form1}, we first express $\hv_{mi}[t_k]$ and $\hv_{mi}[n]$ in terms of $\hv_{mi}[\lambda]$, $\uv_{mi}[t_k]$ and $\uv_{mi}[n]$ as
\begin{align}
    &\E\left\{\hv_{mi}^{H}[t_k]\Amat_{a}^{m}\Pmat_{mk} \text{diag}(\hv_{mi}[n]\hv_{mi}^{H}[n])\Pmat_{mk}^{H}\Amat_{a}^{m}\hv_{mi}[t_k]   \right\} \nonumber\\
    & \quad  \qquad =\rho_{i}^{2}[\lambda-t_k]\E\big\{\hv_{mi}^{H}[\lambda]\Amat_{a}^{m}\Pmat_{mk} \text{diag}(\rho_{i}^{2}[n-\lambda]\hv_{mi}[\lambda]\hv_{mi}^{H}[\lambda])\Pmat_{mk}^{H}\Amat_{a}^{m}\hv_{mi}[\lambda]   \big\} \nonumber\\ 
    &\qquad\qquad + \big(\rho_{i}^{2}[\lambda-t_k]\Bar{\rho}_{i}^{2}[n-\lambda] + \Bar{\rho}_{i}^{2}[\lambda-t_k] \big)\text{tr}\left(   \bar{\Rmat}_{mi}  \Amat_{a}^{m}\Pmat_{mk} \text{diag}\big( \bar{\Rmat}_{mi} \big)\Pmat_{mk}^{H}\Amat_{a}^{m}    \right). \label{first_term_lemma1}
\end{align}
Equality $(a)$ is obtained by substituting $\E\{\uv_{mi}[t_k]\uv^H_{mi}[t_k]\}\!=\! \E\{\uv_{mi}[n]\uv^H_{mi}[n]\}\!= \!\Rmat_{mi}$ as~\cite{couillet_debbah_2011}.
%Equality $(b)$ is obtained by using the property that $\E\{\uv_{mi}[t_k]\uv^H_{mi}[t_k]\}\!=\! \E\{\uv_{mi}[n]\uv^H_{mi}[n]\}\!=\! \Rmat_{mi}$.
We now calculate the expectation in the first term of \eqref{first_term_lemma1} by substituting the value of $\hv_{mi}[\lambda]$ as 
%in \eqref{eq: h_mk[lambda]}
%$\hv_{mi}[\lambda] \!=\!\bar{\hv}_{mk}e^{j\phi_{mk}^{\lambda}}+ \Rmat_{mk}^{\frac{1}{2}}\tilde{\hv}_{mk}[\lambda]$~as 
\begin{align}
    &\E\big\{\hv_{mi}^{H}[\lambda]\Amat_{a}^{m}\Pmat_{mk}  \text{diag}(\hv_{mi}[\lambda]\hv_{mi}^{H}[\lambda])\Pmat_{mk}^{H}\Amat_{a}^{m}\hv_{mi}[\lambda]   \big\} 
    \nonumber\\
    &= \bar{\hv}_{mi}^{H}\Amat_{a}^{m}\Pmat_{mk}\text{diag}(\bar{\hv}_{mi}\bar{\hv}_{mi}^{H}+\Rmat_{mi})\Pmat_{mk}^{H}\Amat_{a}^{m}\bar{\hv}_{mi} +\text{Tr}\left(\Rmat_{mi}\Amat_{a}^{m}\Pmat_{mk}\text{diag}(\bar{\hv}_{mi}\bar{\hv}_{mi}^{H})\Pmat_{mk}^{H}\Amat_{a}^{m} \right) \nonumber\\
    &\quad  + 2 \text{real} \Big\{ \E\big\{\bar{\hv}_{mi}^{H}\Amat_{a}^{m}\Pmat_{mk}diag(\bar{\hv}_{mi}{\Tilde{\hv}}_{mi}^{H}[\lambda]\Rmat_{mi}^{1/2})\Pmat_{mk}^{H}\Amat_{a}^{m}\Rmat_{mi}^{1/2}{\Tilde{\hv}}_{mi}[\lambda]  \big\} \Big\} \nonumber\\
    &\quad+ \E\Big\{{\Tilde{\hv}}_{mi}^{H}[\lambda]\Rmat_{mi}^{1/2}\Amat_{a}^{m}\Pmat_{mk}diag(\Rmat_{mi}^{1/2}{\Tilde{\hv}}_{mi}[\lambda]{\Tilde{\hv}}_{mi}^{H}[\lambda]\Rmat_{mi}^{1/2})\Pmat_{mk}^{H}\Amat_{a}^{m}\Rmat_{mi}^{1/2}{\Tilde{\hv}}_{mi}[\lambda]  \Big\} \nonumber
        \end{align}
        %--------------------
    \begin{align}
    & = \bar{\hv}_{mi}^{H}\Amat_{a}^{m}\Pmat_{mk}\text{diag}(\bar{\hv}_{mi}\bar{\hv}_{mi}^{H}+\Rmat_{mi})\Pmat_{mk}^{H}\Amat_{a}^{m}\bar{\hv}_{mi} +\text{Tr}\left(\Rmat_{mi}\Amat_{a}^{m}\Pmat_{mk}\text{diag}(\bar{\hv}_{mi}\bar{\hv}_{mi}^{H})\Pmat_{mk}^{H}\Amat_{a}^{m} \right)
     \nonumber\\  
    & + \!2 \text{real}\Big\{\E\Big\{\!\sum_{j}\Big(\!\sum_{n_1}[\bar{\hv}_{mi}]_{n_1}^{*}[\bar{\hv}_{mi}]_{j}[\Amat_{m}\Pmat_{\!mk}]_{n_{1}j} \! \Big) \Big(\!\sum_{n_2}[\Pmat_{\!mk}^{H}\Amat_{m}]_{jn_{2}}[{\Tilde{\hv}}_{mi}^{H}[\lambda]\Rmat_{mi}^{1/2}]_{j}[\Rmat_{mi}^{1/2}\tilde{\hv}_{mi}[\lambda]]_{n_2} \! \Big)  \! \Big\} \Big\} \nonumber\\
    &  + \Big(\sum_{j}\sum_{n_1}\sum_{n_2}\sum_{n_3}\sum_{n_4} \E \Big\{[{\Tilde{\hv}}_{mi}^{*}[\lambda]]_{n_1}[{\Tilde{\hv}}_{mi}[\lambda]]_{n2}[{\Tilde{\hv}}_{mi}^{*}[\lambda]]_{n3}[{\Tilde{\hv}}_{mi}[\lambda]]_{n4} \Big\} \nonumber\\
    &\qquad\times [\Rmat_{mi}^{1/2}\Amat_{a}^{m}\Pmat_{mk}]_{n_{1}j}[\Rmat_{mi}^{1/2}]_{jn_{2}}[\Rmat_{mi}^{1/2}]_{n_{3}j}[\Pmat_{mk}^{H}\Amat_{a}^{m}\Rmat_{mi}^{1/2}]_{jn_{4}}\Big). \label{eq_t123}
\end{align}
%Equality $(c)$ is obtained by expressing each matrix and vector in terms of its elements.
Here, the terms $[\bar{\hv}_{mi}]_{n_1}$ and $[{\Tilde{\hv}}_{mi}^{*}[\lambda]]_{n_1}$ represent the  $n_1$th element of the vectors $\bar{\hv}_{mi}$ and ${\Tilde{\hv}}_{mi}^{*}[\lambda]$, respectively.
The terms $[\Amat_{a}^{m}\Pmat_{mk}]_{n_{1}j}$ and $[\Rmat_{mi}^{1/2}\Amat_{a}^{m}\Pmat_{mk}]_{n_{1}j}$ represent the  $(n_1,j)$th element of the matrices $\Amat_{a}^{m}\Pmat_{mk}$ and $\Rmat_{mi}^{1/2}\Amat_{a}^{m}\Pmat_{mk}$, respectively.
 We note that, in the last term, the expectation $\E\big\{[ {\Tilde{\hv}}_{mi}^{*}[\lambda]]_{n_1}[ {\Tilde{\hv}}_{mi}[\lambda]]_{n_2}[ {\Tilde{\hv}}_{mi}^{*}[\lambda]]_{n3}[ {\Tilde{\hv}}_{mi}[\lambda]]_{n4} \big\}$ is non-zero only when either $n_1\!=\!n_2\!=\!m_1;n_3\!=\!n_4\!=\!m_2$ or $n_1\!=\!n_4\!=\!m_1;n_2\!=n_3\!=m_2$. We can, therefore, calculate the expectations in \eqref{eq_t123} as \vspace{+0.1pt}
%--------------------------------------------
\begin{align}
 &   \!\!\!\! \E\Big\{\hv_{mi}^{H}[\lambda]\Amat_{m}\Pmat_{mk}  \text{diag}(\hv_{mi}[\lambda]\hv_{mi}^{H}[\lambda])\Pmat_{mk}^{H}\Amat_{m}\hv_{mi}[\lambda]   \Big\} 
    \nonumber \\
    &  \!\!\!\!\! = \bar{\hv}_{mi}^{H}\Amat_{m}\Pmat_{mk}\text{diag}(\bar{\hv}_{mi}\bar{\hv}_{mi}^{H}+\Rmat_{mi})\Pmat_{mk}^{H}\Amat_{a}^{m}\bar{\hv}_{mi} +\text{Tr}\left(\Rmat_{mi}\Amat_{m}\Pmat_{mk}\text{diag}(\bar{\hv}_{mi}\bar{\hv}_{mi}^{H})\Pmat_{mk}^{H}\Amat_{m} \right)
     \nonumber\\
     &  \!\!\!\!\!  \quad + 2 \text{real}\Big\{ \sum_{j}\Big(\sum_{n_1}[\bar{\hv}_{mi}\bar{\hv}_{mi}^{H}]_{jn_{1}}[\Amat_{m}\Pmat_{mk}]_{n_{1}j} \Big) \Big(\sum_{n_2}[\Pmat_{mk}^{H}\Amat_{m}]_{jn_{2}}[\Rmat_{mi}]_{n_2j} \Big) \Big\} 
    \nonumber%\\
        \end{align}
    \begin{align}
        &  \!\!\!\!\! \quad+ \sum_{j}\sum_{m_1}\sum_{m_2}[\Rmat_{mi}^{1/2}\Amat_{m}\Pmat_{mk}]_{m_{1}j} [\Rmat_{mi}^{1/2}]_{jm_{1}}[\Rmat_{mi}^{1/2}]_{m_{2}j}[\Pmat_{mk}^{H}\Amat_{m}\Rmat_{mi}^{1/2}]_{jm_{2}} \nonumber\\ &\quad+\sum_{j}\sum_{m_1}\sum_{m_2}[\Rmat_{mi}^{1/2}\Amat_{m}\Pmat_{mk}]_{m_{1}j}   [\Rmat_{mi}^{1/2}]_{jm_{2}}[\Rmat_{mi}^{1/2}]_{m_{2}j}[\Pmat_{mk}^{H}\Amat_{m}\Rmat_{mi}^{1/2}]_{jm_{1}}   \label{eq:Term_1a} 
    \\
        &  \!\!\!\!\! \overset{(d)}{=}\text{Tr}\!\left( \left(\bar{\hv}_{mi}\bar{\hv}_{mi}^{H} + \Rmat_{mi} \right) \Amat_{m}\Pmat_{mk} \text{diag}\left(\bar{\hv}_{mi}\bar{\hv}_{mi}^{H} + \Rmat_{mi} \right)\Pmat_{mk}^{H}\Amat_{a}^{m} \right) \nonumber\\
    &  \!\!\!\!\!  \quad +2\text{real}\left\{ \text{Tr}\!\left( \bar{\hv}_{mi}\bar{\hv}_{mi}^{H}\Amat_{m}\Pmat_{\!mk}\text{diag}\!\left( \Pmat_{\!mk}^{H}\Amat_{m}\Rmat_{mi} \right)      \right)\right\} \!+\!\text{Tr}\!\left( \Rmat_{mi}\Amat_{m}\Pmat_{\!mk}\text{diag}\!\left( \Pmat_{\!mk}^{H}\Amat_{m}\Rmat_{mi} \right)\right). \label{eq_first_term_CF_lemma1}
\end{align}
Equality $(d)$ is obtained by expressing the summation terms in matrix form and using algebraic operations. Substituting \eqref{eq_first_term_CF_lemma1} in \eqref{first_term_lemma1}, we obtain first result in Table~\ref{Lemma_Closed_form1}. 
Similarly, we can simplify the term $\E\{ \hv_{mi}^{H}[\lambda]\Amat_{m}\Pmat_{mk}\hv_{mi}[\lambda]\hv_{mi}^{H}[\lambda]\Pmat_{mk}^{H}\Amat_{m}\hv_{mi}[\lambda] \}$ to obtain the second result in Table~\ref{Lemma_Closed_form1}.

%Equality $(d)$ is obtained by i) calculating the expectation in $t_2$ and $t_3$ terms using the fact that $\E\{\tilde{\hv}_{mk}[\lambda]\tilde{\hv}_{mk}[\lambda]^H\} = \Imat_N$; and
% ii) expressing the summation terms in matrix form. 

%---------------------------------------------------------------
%------------------------------------------------------APPENDIX
\vspace{-0.4cm} %C----------------------------
\section{} \label{appen_SE_terms} \vspace{-0.2cm}
%\colr{We begin by proving the novel mathematical results which will be used in deriving Theorem~\ref{theorem_lower_bound}.}
We now derive Theorem~\ref{theorem_lower_bound} by computing the following expectations.
%-----------------------------------------------------
\newline
\underline{\textbf{Computation of $\overline{\text{DS}}_{k,n}$:}} The power of the desired signal can be calculated as
\begin{align}
    \overline{\text{DS}}_{k,n} = \Big|\sum_{m=1}^{M}\av_{mk}^{*}[n]\rho[n-\lambda]\left( \alpha_{d,k} \right)\sqrt{p_{k}}\E \left\{\hat{\hv}_{mk}^{H}[\lambda]\Amat_{m}\hv_{mk}[\lambda] \right\}  \Big|^{2}. \label{app_DS}
\end{align}
We now compute $ \E \big\{\hat{\hv}_{mk}^{H}[\lambda]\Amat_{m}\hv_{mk}[\lambda] \big\}$ by substituting $\hat{\hv}_{mk}$ using \eqref{eq: LMMSE} as
\begin{align}
& \E \left\{\hat{\hv}_{mk}^{H}[\lambda]\Amat_{m}\hv_{mk}[\lambda] \right\} =\alpha_{d,k}\sqrt{\tilde{p}_{k}}\rho_{k}[\lambda-t_k]\text{tr}\left( \Amat_{m}\E \left\{\hv_{mk}[\lambda]\yv_{\text{ADC},m}^{H}[t_k]\right \} \boldsymbol{\Psi}_{mk}\Amat_{m}\bar{\Rmat}_{mk}   \right) \nonumber\\[-5pt]
&\qquad =\alpha_{d,k}^{2}\tilde{p}_{k}\rho_{k}^{2}[\lambda-t_k]\text{tr}\left(\Amat_{m}\bar{\Rmat}_{mk}\Amat_{m}\boldsymbol{\Psi}_{mk}\Amat_{m}\bar{\Rmat}_{mk} \right) = \text{tr}\left(\Amat_{m} \boldsymbol{\overline{\Gamma}}_{mk} \right). \label{Appen_DS_m1}
\end{align} 
Substituting \eqref{Appen_DS_m1} in \eqref{app_DS}, we get $\overline{\text{DS}}_{k,n} \!= \!\big|\sum_{m=1}^{M} \!\av_{mk}^{*}[n]\rho[n\!-\!\lambda]\alpha_{d,k} \!\sqrt{p_{k}}\text{tr}\left(\Amat_{m} \boldsymbol{\overline{\Gamma}}_{mk} \right)  \big|^{2}$.
\newline
\underline{\textbf{Computation of $\overline{{\mbox{IUI}}}_{ki,n}$:}} The power of inter-user interference can be computed as 
\begin{align}
\overline{{\mbox{IUI}}}_{ki,n}\!& =\!\E\bigg\{ \!\Big| \sum\limits_{m=1}^{M}\!a_{mk}^{*}[n]\alpha_{d,i}\sqrt{p_{i}} \hat{\hv}_{mk}^{H}[\lambda]\Amat_{m}\hv_{mi}[n] \Big|^{2}  \bigg\} 
 \! = \! \alpha_{d,i}^2p_{i} \sum\limits_{m=1}^{M}\! \Big(|a_{mk}^{*}[n]|^2 \underbrace{ \E\Big\{ \big| \hat{\hv}_{mk}^{H}[\lambda]\Amat_{m}\hv_{mi}[n]\big|^2 \Big\}}_{c_{kin,m}} \nonumber \\[-9pt]
& \quad + \sum\limits_{m^{'}=1}^{M}a_{mk}^{*}[n]a_{m^{'}k}^{*}[n] \underbrace{ \E\Big\{ \hat{\hv}_{mk}^{H}[\lambda]\Amat_{m}\hv_{mi}[n]   \hat{\hv}_{m^{'}k}^{H}[\lambda]\Amat_{m^{'}}\hv_{m^{'}i}[n] \Big\}}_{c_{kin,mm^{'}}}  \Big). \label{eq_UI_expansion}
\end{align}
%-----------------------------------------------
%\begin{align}
%\overline{{\mbox{UI}}}_{kin}\!& =\!\E\bigg\{ \bigg| \sum\limits_{m=1}^{M}a_{mk}^{*}[n](\alpha_{d,i})\sqrt{p_{i}} \hat{\hv}_{mk}^{H}[\lambda]\Amat_{m}\hv_{mi}[n] \bigg|^{2}  \bigg\} \nonumber \\
% & = \! \alpha_{d,i}^2p_{i} \sum\limits_{m=1}^{M} \sum\limits_{m^{'}=1}^{M}a_{mk}^{*}[n]a_{m^{'}k}^{*}[n] \underbrace{ \E\Big\{ \hat{\hv}_{mk}^{H}[\lambda]\Amat_{m}\hv_{mi}[n]   \hat{\hv}_{m^{'}k}^{H}[\lambda]\Amat_{m}^{m^{'}}\hv_{m^{'}i}[n] \Big\}}_{c_{kin,mm^{'}}} \nonumber \\[-5pt]
% &\quad + \sum\limits_{m=1}^{M} |a_{mk}^{*}[n]|^2 (\alpha_{d,i})^2p_{i} \underbrace{ \E\Big\{ \big| \hat{\hv}_{mk}^{H}[\lambda]\Amat_{m}\hv_{mi}[n]\big|^2 \Big\}}_{c_{kin,m}}. \label{eq_UI_expansion}
%\end{align}
%-----------------------------------------------
%For the cases $i\in\mathcal{P}_{k}$ and $i\notin\mathcal{P}_{k}$, we represent $c_{kin,m}$ and $c_{kin,mm^{'}}$ as
%\begin{small}
%\begin{align}\label{Appendix_Interfe_eqs12}
%c_{kin,m} = \begin{cases} c_{kin,m}^{\in} &i\in\mathcal{P}_{k} \\
%c_{kin,m}^{\notin} & i\notin\mathcal{P}_{k},
%\end{cases} \quad
%c_{kin,mm^{'}} = \begin{cases} c_{kin,mm^{'}}^{\in} &i\in\mathcal{P}_{k} \\
%c_{kin,mm^{'}}^{\notin} & i\notin\mathcal{P}_{k}.
%\end{cases} 
%\end{align} 
%\end{small}
%-----------------------------------------------
%We now calculate the term $c_{kin}^{mm^{'}}\; \forall m,m^{'}$ for the cases  $i\in\mathcal{P}_{k}$ and  $i\notin\mathcal{P}_{k}$.
 We first compute the term $c_{kin,m}=\E\Big\{\Big|\hat{\hv}_{mk}^{H}[\lambda]\Amat_{m}\hv_{mi}[n] \Big|^{2}\Big\}$ for $i\in\mathcal{P}_{k}$ as
\begin{align}
c_{kin,m}^{\in}={\rho_{i}}^{2}[n-\lambda]\underbrace{\E\bigg\{\Big|\hat{\hv}_{mk}^{H}[\lambda]\Amat_{m}\hv_{mi}[\lambda] \Big|^{2}\bigg\}}_{\gamma_{1}}+\overline{\rho_{i}}^{2}[n-\lambda] \underbrace{\E\bigg\{\Big|\hat{\hv}_{mk}^{H}[\lambda]\Amat_{m}\uv_{mi}^{'}[n] \Big|^{2}\bigg\}}_{\gamma_2} \label{eq_c_kin_correlated}.
\end{align}
We now compute the term $\gamma_1$ by first substituting $\hat{\hv}_{mk}$ using \eqref{eq: LMMSE} as
\begin{align}
    \gamma_1 &=\E\Big\{\Big|\hat{\hv}_{mk}^{H}[\lambda]\Amat_{m}\hv_{mi}[\lambda] \Big|^{2}\Big\} =\E\Big\{\big|\alpha_{d,k}\sqrt{\tilde{p}_{k}}\rho_{k}[\lambda-t_k]\yv_{\text{ADC},m}^{H}[t_k]\boldsymbol{\Psi}_{mk}\Amat_{m}\bar{\Rmat}_{mk}\Amat_{m}\hv_{mi}[\lambda] \big|^{2}\Big\} \nonumber \\
   & \overset{(b)}{=}{\varsigma^{(1)}_{kin,m}}+{\varsigma^{(2)}_{kin,m}}+{\varsigma^{(3)}_{kin,m}}. \label{eq_gamma-1-expansion} 
\end{align} 
%--------------------------------------------------------
%To compute $\gamma_1$ in \eqref{eq_Gamma_1}, we first denote the received pilot signal $\yv_{\text{ADC},m}[t_k]$ in \eqref{eq_Rx_pilot_ADC} as
%\begin{align}
%    \yv_{\text{ADC},m}[t_k]&=\underbrace{\sum_{k^{'}\in \mathcal{P}_k}\Amat_{m}\hv_{mk^{'}}[t_k]\big(1-\rho_{d,k^{'}}\big)\sqrt{p_{k^{'}}}}_{x_{1}^{y_{m}[t_k]}} +\underbrace{\sum_{k^{'}\in \mathcal{P}_k}\Amat_{m}\hv_{mk^{'}}[t_k]\Big(n_{DAC,k^{'}}[t_k]+\eta_{t,k^{'}}^{\text{UE}}[t_k]\Big)}_{x_{2}^{y_{m}[t_k]}}  \nonumber\\
%    &\quad+\underbrace{\Amat_{m}\boldsymbol{\eta}_{\text{RF},m}[t_k]+\Amat_{m}\zv_{m}[t_k]+\nv_{ADC}^{m}[t_k]}_{x_{3}^{y_{m}[t_k]}}.      \label{eq: y_m[t_k] expansion1}
%\end{align}
%--------------------------------------------------------
Equality $(b)$ is obtained by substituting the received pilot signal $\yv^p_{\text{ADC},m}[t_k]$ in \eqref{eq:y_m[t_k]}. 
Here the terms ${\varsigma^{(1)}_{kin,m}} ={\E\{\xv_{1}^{H}\Pmat_{mk}\hv_{mi}[\lambda]\hv_{mi}^{H}[\lambda] \Pmat_{mk}^{H}\xv_{1} \}}$, ${\varsigma^{(2)}_{kin,m}} ={\E\{\xv_{2}^{H}\Pmat_{mk}\hv_{mi}[\lambda]\hv_{mi}^{H}[\lambda] \Pmat_{mk}^{H}\xv_{2} \}}$ and ${\varsigma^{(3)}_{kin,m}}={\E\{\xv_{3}^{H}\Pmat_{\!mk}\hv_{mi}[\lambda]\hv_{mi}^{H}[\lambda] \Pmat_{\!mk}^{H}\xv_{3} \}}$ 
with $\xv_{1} \!=\! \sum\limits_{j\in \mathcal{P}_k}\!\alpha_{d,j}\sqrt{\tilde{p}_{j}}\Amat_{m}\hv_{mj}[t_k]$, 
$\xv_{2} \!=\! \sum\limits_{j\in \mathcal{P}_k}\!\Amat_{m}\hv_{mj}[t_k](\upsilon^p_{\text{DAC},j}[t_k]+\xi^p_{\text{RF},j}[t_k])$,
 $\xv_{3} \!= \! \Amat_{m}\boldsymbol{\eta}_{\text{RF},m}^p[t_k]+\Amat_{m}\zv^p_{m}[t_k]+\nv^p_{\text{ADC},m}[t_k]$ and 
 $\Pmat_{\!mk}\!=\alpha_{d,k}\sqrt{\tilde{p}_{k}}\rho_{k}[\lambda-t_k]\boldsymbol{\Psi}_{mk}\Amat_{m}\bar{\Rmat}_{mk}\Amat_{m}$. 
We now simplify the term $\varsigma^{(1)}_{kin,m}=\E\{\xv_{1}^{H}\Pmat_{mk}\hv_{mi}[\lambda]\hv_{mi}^{H}[\lambda] \Pmat_{mk}^{H}\xv_{1} \}$ in \eqref{eq_gamma-1-expansion} as \vspace{+0.1pt}
\begin{align}
&  \varsigma^{(1)}_{kin,m}  
\overset{(b)}{=}\alpha_{d,i}^{2}\tilde{p}_{i}\rho_{i}^{2}[\lambda-t_k]\E\big\{ \hv_{mi}^{H}[\lambda]\Amat_{m}\Pmat_{mk}\hv_{mi}[\lambda]\hv_{mi}^{H}[\lambda]\Pmat_{mk}^{H}\Amat_{m}\hv_{mi}[\lambda] \big\}  \notag \\
    &\;\;+\alpha_{d,i}^{2}\tilde{p}_{i}\overline{\rho}_{i}^{2}[\lambda-t_k]\text{tr}\big(  \bar{\Rmat}_{mi}  \Amat_{m}\Pmat_{mk} \bar{\Rmat}_{mi}  \Pmat_{mk}^{H}\Amat_{m}\big) 
    +\sum_{j\in\mathcal{P}_{k}/i} \alpha_{d,j}^{2}\tilde{p}_{j} \text{tr}\big( \bar{\Rmat}_{mj}  \Amat_{m}\Pmat_{mk} \bar{\Rmat}_{mi}  \Pmat_{mk}^{H}\Amat_{m}\big). \notag \\
   & \quad \overset{(c)}{=}\sum_{j\in\mathcal{P}_{k}} \alpha_{d,j}^{2}\tilde{p}_{j} \text{tr}\left( \bar{\Rmat}_{mj} \Amat_{m}\Pmat_{mk}\bar{\Rmat}_{mi} \Pmat_{mk}^{H}\Amat_{m}\right) \label{eq_var_sigma_1_final} \\[-5pt] 
    & \qquad +\alpha_{d,i}^{2}\tilde{p}_{i}\rho_{i}^{2}[\lambda-t_k]\Big(  \left| \text{tr} \left( \Rmat_{mi}\Amat_{m}\Pmat_{mk}\right)    \right|^{2} 
     +2\text{real} \big\{ \overline{\hv}_{mi}^{H}\Amat_{m}\Pmat_{mk}{\overline{\hv}_{mi}} \text{tr} \left( \Rmat_{mi}\Pmat_{mk}^{H}\Amat_{m} \right)  \big\} \Big). \nonumber
\end{align}
{Equality $(b)$ is obtained by i) expressing $\hv_{mi}[t_k]$ in terms of $\hv_{mi}[\lambda]$ using \eqref{eq_channel_h[n]}; and ii) using the fact that innovation component  and channel $\hv_{mi}[\lambda]$ are independent, and have a zero mean. We can obtain equality $(c)$ by applying the second result in Lemma \ref{lemma_comp_term}.}

We can similarly calculate $\varsigma^{(2)}_{kin,m}$ in \eqref{eq_gamma-1-expansion} as follows:\vspace{+0.1pt}
\begin{align}
    &\varsigma^{(2)}_{kin,m}=\sum_{j\in\mathcal{P}_{k}} \alpha_{d,j}(1-\alpha_{d,j}+\kappa_{t,j}^{2})\tilde{p}_{j} \text{tr}\left( \bar{\Rmat}_{mj} \Amat_{m}\Pmat_{mk}\bar{\Rmat}_{mi} \Pmat_{mk}^{H}\Amat_{m}\right) \label{eq_var_sigma_2} \\[-5pt] 
    &+\alpha_{d,i}(1\!-\!\alpha_{d,i}+\kappa_{t,i}^{2})\tilde{p}_{i}\rho_{i}^{2}[\lambda \!-\! t_k]\Big(  \left| \text{tr} \left( \Rmat_{mi}\Amat_{m}\Pmat_{mk}\right)    \right|^{2} 
     \!+ 2\text{real}  \big\{ \bar{\hv}_{mi}^{H}\Amat_{m}\Pmat_{mk}{\overline{\hv}_{mi}} \text{tr} \left( \Rmat_{mi}\Pmat_{mk}^{H}\Amat_{m} \right) \! \big\} \Big). \nonumber
\end{align}
We now simplify $\varsigma^{(3)}_{kin,m}$ in  \eqref{eq_gamma-1-expansion} by substituting $\xv_{3}\! =\! \Amat_{m}\boldsymbol{\eta}^p_{\text{RF},m}[t_k]+\Amat_{m}\zv^p_{m}[t_k]+\nv^p_{\text{ADC},m}[t_k]$ as\vspace{+0.1pt} 
\begin{align}
  &  \varsigma^{(3)}_{kin,m}
    \!=  \text{tr}\left( \E\left\{\left(\Bmat_{m}\!+\kappa_{r,m}^{2}\Amat_{m} \right)\! \Wmat^{m}[t_k]\Pmat_{mk}\hv_{mi}[\lambda]\hv_{mi}^{H}[\lambda] \Pmat_{mk}^{H} \right\} \right)   \!+ \!\text{tr}\left( \E\left\{\sigma^{2}\Amat_{m}\Pmat_{mk}\hv_{mi}[\lambda]\hv_{mi}^{H}[\lambda] \Pmat_{mk}^{H} \right\} \right). \notag \\
    & \overset{(d)}{=} \!\!\sum_{j\in\mathcal{P}_{k}}\!(1\!+\! \kappa_{t,j}^{2})\alpha_{d,j}\tilde{p}_{j}\text{tr}\big(\!\! \left(\Bmat_{m}\!+\!\kappa_{r,m}^{2}\Amat_{m} \right) \text{diag}\left(  \bar{\Rmat}_{mj}  \!\right) \Pmat_{\!mk} \bar{\Rmat}_{mi}  \Pmat_{\!mk}^{H}  \big) \! +\! \text{tr}\left(\sigma^{2}\!\Amat_{m}\Pmat_{\!mk} \bar{\Rmat}_{mi}  \Pmat_{\!mk}^{H}  \right) \nonumber
    \\[-5pt]
    &\quad+(1+\kappa_{t,i}^{2})\alpha_{d,i}\tilde{p}_{i}\rho_{i}^{2}[\lambda-t_k]  \Big(2\text{real} \big\{ \text{tr}\big(\overline{\hv}_{mi}\overline{\hv}_{mi}^{H}\Pmat_{mk}^{H}  \left(\Bmat_{m}+\kappa_{r,m}^{2}\Amat_{m} \right) \text{diag}\left( \Pmat_{mk}\Rmat_{mi}\right) \big) \big\} \nonumber\\[-5pt]
     &\quad+ \text{tr}\left(\text{diag}\left(\Rmat_{mi}\Pmat_{mk}^{H} \right)(\Bmat_{m} +\kappa_{r,m}^{2}\Amat_{m}) \text{diag}\left(\Pmat_{mk}\Rmat_{mi} \right)  \right) \Big). \label{eq_var_sigma-3_final}  \\[-33pt] \notag
\end{align}
%--------------------------------------------------
%Here $\Bmat_{m} \!=\! \Amat_{m}\left(\Imat_{N}\!-\!\Amat_{m}\right)$.  We calculate the first expectation in \eqref{eq_var_sigma_3_expansion} by substituting $\Wmat^{m}[t_k]$ as
%---------------------------------------------
{Here $\Bmat_{m} \!=\! \Amat_{m}\left(\Imat_{N}\!-\!\Amat_{m}\right)$. Equality $(d)$ is obtained by i) substituting $\Wmat^{m}[t_k]$ in the first expectation; and ii) applying the first result from Lemma \ref{lemma_comp_term}.}
%---------------------------------------------
\begin{align}
    &\text{tr}\big( \E\left\{\left(\Bmat_{m}+\kappa_{r,m}^{2}\Amat_{m} \right) \Wmat^{m}[t_k]\Pmat_{mk}\hv_{mi}[\lambda]\hv_{mi}^{H}[\lambda] \Pmat_{mk}^{H} \right\} \big) \nonumber\\
    & = \text{tr}\big( \E\big\{\left(\Bmat_{m}+\kappa_{r,m}^{2}\Amat_{m} \right) \sum_{j\in\mathcal{P}_{k}}(1+\kappa_{t,j}^{2})\alpha_{d,j}\tilde{p}_{j}\text{diag}(\hv_{mj}[t_k]\hv_{mj}^{H}[t_k]) \Pmat_{mk}\hv_{mi}[\lambda]\hv_{mi}^{H}[\lambda] \Pmat_{mk}^{H} \big\} \big)  \nonumber
\end{align}

Using \eqref{eq_var_sigma_1_final}, \eqref{eq_var_sigma_2} and \eqref{eq_var_sigma-3_final} we can write $c_{kin,m}^{(\in)}$ in \eqref{eq_c_kin_correlated} as\vspace{-3pt}
\begin{align}
    c_{kin,m}^{(\in)}={\rho_{i}}^{2}[n-\lambda]\left(\varsigma_{kin,m}^{(1)} + \varsigma_{kin,m}^{(2)} + \varsigma_{kin,m}^{(3)} \right)+\bar{\rho}_{i}^{2}[n-\lambda]\text{tr}\left(\boldsymbol{\overline{\Gamma}}_{mk}\Amat_{m}\bar{\Rmat}_{mi}\Amat_{m}  \right). \\[-33pt] \notag
\end{align}

For $i\not\in\mathcal{P}_{k}$, the term $c_{kin,m}$ in \eqref{eq_c_kin_correlated} can be computed as\vspace{-3pt} 
\begin{align}
    c_{kin,m}^{(\notin)} \! =\!{\rho_{i}}^{2}[n\!-\! \lambda]\E\big\{\big|\hat{\hv}_{mk}^{H}[\lambda]\Amat_{m}\hv_{mi}[\lambda] \big|^{2}\big\}  \!+\!\bar{\rho}_{i}^{2}[n\!-\!\lambda] \E\big\{\big|\hat{\hv}_{mk}^{H}[\lambda]\Amat_{m}\uv_{mi}^{'}[n] \big|^{2}\big\} \! \overset{(e)}{=} \!\text{tr}(\boldsymbol{\overline{\Gamma}}_{\!mk}\Amat_{m}\bar{\Rmat}_{mi}\Amat_{m}).\notag \\[-33pt] \notag
\end{align}
Equality (e) follows from the fact that i) the channels $\hat{\hv}_{mk}[\lambda]$ and $\hv_{mi}[\lambda]$ are uncorrelated, ii) the variance of actual and estimated channel is $\bar{\Rmat}_{mi}$ and $\boldsymbol{\overline{\Gamma}}_{mk}$, respectively.
% covariance of $\hv_{mi}[\lambda]$ and $\uv_{mi}^{'}$ is $\bar{\Rmat}_{mi}$.

%--------------------------------------------------------------
We now compute the term $c_{kin,mm^{'}}$, for the case $m\neq m^{'}$ and $i\in\mathcal{P}_{k}$, we have%\vspace{+0.1pt}
\begin{align}
 c_{kin,mm^{'}}^{(\in)}& \overset{(d)}{=}\rho_{i}^{2}[n-\lambda]\text{tr}\big(\E\big\{\hv_{mi}[\lambda]\hat{\hv}_{mk}^{H}[\lambda]\Amat_{m} \big\} \big) \text{tr}\big(\E\big\{\hv_{m^{'}i}[\lambda]\hat{\hv}_{m^{'}k}^{H}[\lambda]\Amat_{m^{'}} \big\} \big) \label{eq_C_kin_m_neq_m_correlated} \\
&\overset{(e)}{=}\!\rho_{i}^{2}[n\!-\! \lambda] \alpha_{d,i}^{4}\rho_{i}^{2}[\lambda \!-\! t_k]\tilde{p}_{i}^{2}\text{tr}\left(\bar{\Rmat}_{mi}\Amat_{m}\boldsymbol{\Psi}_{mk}\Amat_{m}\bar{\Rmat}_{mk}\Amat_{m} \right) \text{tr}\big(\bar{\Rmat}_{m^{'}i}\Amat_{m^{'}}\boldsymbol{\Psi}_{m^{'}k}\Amat_{\!m^{'}}\bar{\Rmat}_{m^{'}k}\Amat_{m^{'}} \big).  \nonumber \\[-33pt]
    \nonumber
\end{align}
%-----------------------------------------------
%\begin{align}
%   & c_{kin,mm^{'}}^{(\in)} \overset{(d)}{=}\rho_{i}^{2}[n-\lambda]\text{tr}\big(\E\big\{\hv_{mi}[\lambda]\hat{\hv}_{mk}^{H}[\lambda]\Amat_{m} \big\} \big) \text{tr}\big(\E\big\{\hv_{m^{'}i}[\lambda]\hat{\hv}_{m^{'}k}^{H}[\lambda]\Amat_{m^{'}} \big\} \big) \label{eq_C_kin_m_neq_m_correlated} \\
%    &\overset{(e)}{=}\!\rho_{i}^{2}[n\!-\! \lambda](1\!-\! \rho_{d,i})^{4}\rho_{i}^{2}[\lambda \!-\! t_k]\tilde{p}_{i}^{2}\text{tr}\left(\bar{\Rmat}_{mi}\Amat_{m}\boldsymbol{\Psi}_{mk}\Amat_{m}\bar{\Rmat}_{mk}\Amat_{m} \right) \text{tr}\big(\bar{\Rmat}_{m^{'}i}\Amat_{m^{'}}\boldsymbol{\Psi}_{m^{'}k}\Amat_{\!m^{'}}\bar{\Rmat}_{m^{'}k}\Amat_{m^{'}} \big).  \nonumber \\[-33pt]
%    \nonumber
%\end{align}
%-----------------------------------------------
Equality (d) is due to fact that innovation component and the channel are uncorrelated at time instant $\lambda$ and has zero mean. Equality $(e)$ is obtained by substituting $\hat{\hv}_{mk}[\lambda]$ using \eqref{eq: LMMSE}.

 For the case $i\not\in\mathcal{P}_{k}$, the term $c_{kin,mm^{'}}\forall m \neq m^{'}$ can be calculated as \vspace{-5pt} 
 \begin{align}
c_{kin,mm^{'}}^{(\notin)} =\E\big\{(\hat{\hv}_{mk}^{H}[\lambda]\Amat_{m}\hv_{mi}[n])(\hat{\hv}_{m^{'}k}^{H}[\lambda]\Amat_{m}^{m^{'}}\hv_{m^{'}i}[n])^{*}  \big\} =0. \label{eq_C_kin_m_neq_m_uncorrelated} \\[-30pt]
    \nonumber
\end{align}
Using \eqref{eq_C_kin_m_neq_m_correlated} and \eqref{eq_C_kin_m_neq_m_uncorrelated}, we can write $\overline{\mbox{IUI}}_{kin}$ in \eqref{eq_UI_expansion}\vspace{-5pt}
\begin{align} \label{eq_interference_final} 
 \overline{\mbox{IUI}}_{ki,n}\! 
= \begin{cases} 
\alpha_{d,i}^2p_{i} \Big( \sum\limits_{m=1}^{M} \big( |a_{mk}^{*}[n]|^2 c_{kin,m}^{(\in)} +  \sum\limits_{m^{'}\neq m}^{M}a_{mk}^{*}[n]a_{m^{'}k}^{*}[n] c_{kin,mm^{'}}^{(\in)} \big) \Big), \; \; i\in\mathcal{P}_{k} \\[-1pt]
\alpha_{d,i}^2p_{i} \Big( \sum\limits_{m=1}^{M} \big( |a_{mk}^{*}[n]|^2 c_{kin,m}^{(\notin)} +  \sum\limits_{m^{'}\neq m}^{M}a_{mk}^{*}[n]a_{m^{'}k}^{*}[n] c_{kin,mm^{'}}^{(\notin)} \big) \Big). \; \; i\not\in\mathcal{P}_{k} .
 \end{cases}\!\! \!\!  \\[-32pt]
\nonumber
  \end{align}
%-------------------------------------------------------------
\underline{\textbf{Computation of $\overline{\text{BU}}_{k,n}$:}}
The power of the beamforming uncertainty is calculated as%\vspace{-8pt} 
\begin{align}
    \overline{\text{BU}}_{k,n}    &=\sum_{m=1}^{M}|\av_{mk}^{*}[n]|^{2}\alpha_{d,k}^{2}p_{k}\big(\E \big\{|\hat{\hv}_{mk}^{H}[\lambda]\Amat_{m}\hv_{mk}[\lambda]|^{2} \big\}-\big|\E \big\{\hat{\hv}_{mk}^{H}[\lambda]\Amat_{m}\hv_{mk}[\lambda] \big\}  \big|^{2}  \big)   \nonumber\\[-3pt]
    &\overset{(f)}{=} \sum_{m=1}^{M}|\av_{mk}^{*}[n]|^{2}\alpha_{d,k}^{2}p_{k}\big(\varsigma_{k,k,n,m}^{(1)} \!+ \varsigma_{k,k,n,m}^{(2)} \!+ \varsigma_{k,k,n,m}^{(3)} - \big| \text{tr}(\Amat_{m} \boldsymbol{\overline{\Gamma}}_{mk} )  \big|^{2}  \big)=  \av_{k}^{H}[n]\Bmat_{k,n}\av_{k}[n].  \notag \\[-33pt]
    \notag
\end{align}
Equality $(f)$ is obtained by following the similar steps as in \eqref{eq_c_kin_correlated}. We have $\E \big\{|\hat{\hv}_{mk}^{H}[\lambda]\Amat_{m}\hv_{mk}[\lambda]|^{2} \big\}$ as  $c_{kkn,m}^{\in}= \varsigma_{kkn,m}^{(1)} + \varsigma_{kkn,m}^{(2)} + \varsigma_{kkn,m}^{(3)}$. Here $\Bmat_{k,n} = \text{diag}\big(\alpha_{d,k}^{2}p_{k}\big(c_{kkn,m}^{\in} - \big| \text{tr}\left(\Amat_{m} \boldsymbol{\overline{\Gamma}}_{mk} \right)  \big|^{2}  \big) \big)$.
%--------------------------------
\newline
%-------------------------------------------------------------
\underline{\textbf{Computation of $\overline{\text{RRF}}_{k,n}$:}}
The power of the receiver RF impairments is calculated as%\vspace{-8pt} 
\begin{align}\label{app_RRF}
    \overline{\text{RRF}}_{k,n}  &= \sum\limits_{m=1}^{M}|a_{mk}^{*}[n]|^{2}\E\big\{\hat{\hv}_{mk}^{H}[\lambda]\Amat_{m}\boldsymbol{\eta_{r,m}^{\text{AP}}}[n]\boldsymbol{\eta}_{\text{RF},m}^{H}[n]\Amat_{m}\hat{\hv}_{mk}[\lambda] \big\} \\
%&= \sum\limits_{m=1}^{M}|a_{mk}^{*}[n]|^{2} \E\bigg\{\hat{\hv}_{mk}^{H}[\lambda]\Amat_{m}\E_{h}\Big\{\boldsymbol{\eta}_{r,m}^{\text{AP}}[n]\boldsymbol{\eta_{r,m}^{\text{AP},H}}[n] \Big\}\Amat_{m}\hat{\hv}_{mk}[\lambda]\bigg\} \nonumber  \\[-5pt]
& \overset{(g)}{=}\sum\limits_{m=1}^{M}|a_{mk}^{*}[n]|^{2} \sum_{i=1}^{K} \underbrace{\kappa_{r,m}^{2}(1+\kappa_{t,i}^{2})\alpha_{d,i}p_{i} \E\big\{\hat{\hv}_{mk}^{H}[\lambda]\Amat_{m}\text{diag}(\hv_{mi}[n]\hv_{mi}^{H}[n])\Amat_{m}\hat{\hv}_{mk}[\lambda]\big\}}_{d_{ki,n}^{m}}. \notag \\[-35pt]
\nonumber
\end{align}
Equality $(g)$ is obtained by i) using the that the receiver AP RF impairment $\boldsymbol{\eta_{r,m}^{\text{AP}}}[n] $ has pdf $\mathcal{CN}(0,\kappa_{r,m}^{2}\Wmat^{m}[n])$, and ii) substituting the expression of matrix $\Wmat^{m}[n]$ from Appendix~\ref{appendix_LMMSE}.

For $i \notin \mathcal{P}_{k}$, the term $d_{ki,n}^{m}$ is $d_{ki,n}^{m}  = \kappa_{r,m}^{2}(1+\kappa_{t,i}^{2})\alpha_{d,i}p_{i} \text{tr}\left( \boldsymbol{\overline{\Gamma}}_{mk}\Amat_{m}\text{diag}\left({\bar{\Rmat}_{mi}} \right)\Amat_{m} \right)$.
%\begin{align}
%    d_{k,i,n}^{m}  = \kappa_{r,m}^{2}(1+\kappa_{t,i}^{2})\alpha_{d,i}p_{i} \text{tr}\left( \boldsymbol{\overline{\Gamma}}_{mk}\Amat_{m}\text{diag}\left({\bar{\Rmat}_{mi}} \right)\Amat_{m} \right). \label{eq_d_kn_expect} \\[-33pt]
%\nonumber
%\end{align}
\newline
For $i \in \mathcal{P}_{k}$, the term $d_{ki,n}^{m}$ can be
{obtained as below by  i) following similar to steps given for i) $\gamma_1 $ in \eqref{eq_gamma-1-expansion}, $\varsigma_{kin,m}^{(1)}$, $\varsigma_{kin,m}^{(2)}$ and $\varsigma_{kin,m}^{(3)}$, respectively; and ii) applying the results from Lemma \ref{lemma_comp_term}. \vspace{-5pt} }
%--------------------------------------------
\begin{align}
   d_{ki,n}^{m} & =(1+\kappa_{t,i}^{2})\alpha_{d,i}p_{i} \Big( \sum_{j\in\mathcal{P}_{k}}(1+\kappa_{t,j}^{2})\alpha_{d,j}\tilde{p}_{j}\Big(\text{tr}\big(\bar{\Rmat}_{mj}\Amat_{m}\Pmat_{mk}  \text{diag}(\bar{\Rmat}_{mi})\Pmat_{mk}^{H}\Amat_{m}\big)   \nonumber\\[-4pt]
    &\quad+ \text{tr}\left(\left(\Bmat_{m}+\kappa_{r,m}^{2}\Amat_{m}\right)\text{diag}(\bar{\Rmat}_{mj}) \Pmat_{mk}\text{diag}(\bar{\Rmat}_{mi})\Pmat_{mk}^{H}  \right)  \Big) 
    + \sigma^{2}\text{tr}\left( \Amat_{m}\Pmat_{mk}  \text{diag}(\bar{\Rmat}_{mi})\Pmat_{mk}^{H}  \right) \nonumber \\[-4pt]
    &\quad+ (1+\kappa_{t,i}^{2})\alpha_{d,i}\tilde{p}_{i}\rho_{i}^{2}[n-\lambda]\rho_{i}^{2}[\lambda-t_k]\Big( \text{tr}\left(\left(\Bmat_{m}+\kappa_{r,m}^{2}\Amat_{m}\right)\text{diag}(\Rmat_{mi}\Pmat_{mk}^{H}) \Pmat_{mk}\Rmat_{mi} \right) \nonumber\\[-4pt]
        &\quad + 2\text{real}\big\{ \text{tr}\big( \bar{\hv}_{mi}\bar{\hv}_{mi}^{H}\Amat_{m}\Pmat_{mk}\text{diag}\left( \Pmat_{mk}^{H}\Amat_{m}\Rmat_{mi} \right)      \big)\big\} + \text{tr}\left( \Rmat_{mi}\Amat_{m}\Pmat_{mk}\text{diag}\left( \Pmat_{mk}^{H}\Amat_{m}\Rmat_{mi} \right)\right)\nonumber\\[-4pt]
    &\quad +2\text{real} \big\{\text{tr}\big( (\Bmat_{m}+\kappa_{r,m}^{2}\Amat_{m})\big( (\bar{\hv}_{mi}\bar{\hv}_{mi}^{H})\odot\Pmat_{mk}\big)\big( \Rmat_{mi}\odot\Pmat_{mk}^{H}\big)  \big)   \big\}  \Big) \Big). \label{eq_d_kn_m} \\[-33pt]\notag
    \end{align}
Substituting \eqref{eq_d_kn_m} in \eqref{app_RRF}, we have $ \overline{\text{RRF}}_{k,n} \!= \kappa_{r,m}^{2}\av_{k}^{H}[n]\sum\limits_{i=1}^{K}\mathbf{D}_{ki,n}\av_{k}[n]$, where $\mathbf{D}_{ki,n}\!=\text{diag}(d_{ki,n}^{1},\cdots,d_{ki,n}^{M})$. 
\newline
%\colr{$\Xi_{k,n}= diag (m,m)$th element- write in terms of $d_{k,n}^{m}$ and then give expression for $d_{k,n}^{m}$.}
%\colr{\begin{align}
%    \Xi_{k,n}&=\text{diag}\begin{bmatrix}\E\bigg\{\hat{\hv}_{mk}^{H}[\lambda]\Amat_{m}\E_{h}\Big\{\boldsymbol{\eta_{r,m}^{\text{AP}}}[n]\boldsymbol{\eta_{r,m}^{\text{AP},H}}[n] \Big\}\Amat_{m}\hat{\hv}_{mk}[\lambda]\bigg\} 
%    \end{bmatrix} 
%\end{align}}
%-------------------------------------------------------------
%--------------------------------------------------------
%-------------------------------------------------------------
\underline{\textbf{Computation of $\overline{\text{ADC}}_{k,n}$:}}
The power of the dynamic-ADC distortion at the $k$th UE is given~as\vspace{-3pt} 
\begin{align}
    \overline{\text{ADC}}_{k,n} = \E\big\{ \big| \sum\limits_{m=1}^{M}a_{mk}^{*}[n]\hat{\hv}_{mk}^{H}[\lambda]\nv_{\text{ADC},m}[n] \big|^{2} \big\} 
    \overset{(h)}{=} \av_{k}^{H}[n]\Big(\sum\limits_{i=1}^{K}\overline{\Dmat}_{ki,n} + \text{diag}\big(\sigma^{2}\text{tr}(\overline{\boldsymbol{\Gamma}}_{mk}\Bmat_{m}) \big) \Big) \av_{k}[n].\notag  \\[-33pt] \notag
\end{align}
Equality $(h)$ can be derived on lines similar to \eqref{app_RRF}.
For  $i\not\in\mathcal{P}_{k}$ the matrix $\Dmat_{ki,n}$ is
\begin{align*}
    \overline{\Dmat}_{ki,n} &= \text{diag}\big((1+\kappa_{t,i}^{2})(1+\kappa_{r,m}^{2})\alpha_{d,i}p_{i} \text{tr}\left(\boldsymbol{\overline{\Gamma}}_{mk}\Bmat_{m}\text{diag}\left(\bar{\Rmat}_{mi}\right)  \right)  +\sigma^{2}\text{tr}\left(\boldsymbol{\overline{\Gamma}}_{mk}\Bmat_{m}\right)\big). \\[-30pt]\notag
\end{align*}
For $i\in\mathcal{P}_{k}$ the matrix $\overline{\Dmat}_{ki,n}$ can be given as $\overline{\Dmat}_{ki,n} = (1\!+\!\kappa_{r,m}^{2}) \mathbf{D}_{ki,n}$.

Similarly, the closed-form expressions of the  transmit RF impairment $\overline{\mbox{TRF}}_{k,n}$, DAC impairment $\overline{\text{DAC}}_{k,n}$, channel aging $\overline{\text{CA}}_{k,n}$ and noise term $\overline{\text{NS}}_{k,n}$ in \eqref{eq_SE_expectations} are given respectively as: $    \overline{\mbox{TRF}}_{k,n} =\av_{k}^{H}[n] \sum_{i=1}^{K} \kappa_{t,i}^{2}\alpha_{d,i}p_{i}\Cmat_{kin}  \av_{k}[n]$, $ \overline{\text{DAC}}_{k,n} =  \av_{k}^{H}[n] \sum_{i=1}^{K} \rho_{d,i}\alpha_{d,i}p_{i}\Cmat_{kin}  \av_{k}[n]$, $\overline{\text{CA}}_{k,n} =  \av_{k}^{H}[n]\boldsymbol{\Lambda}_{k,n}\av_{k}[n]$ 
with $\boldsymbol{\Lambda}_{k,n} \!= \!\overline{\rho}_{k}^{2}[n-\lambda]\alpha_{d,k}^{2}p_k \text{diag}\big( \text{tr}\left(\boldsymbol{\overline{\Gamma}}_{1k}\Amat_{1} \bar{\Rmat}_{1k}  \Amat_{1}, \cdots,\boldsymbol{\overline{\Gamma}}_{Mk}\Amat_{M} \bar{\Rmat}_{Mk}  \Amat_{M} \right)\! \big)$ and $ \overline{\text{NS}}_{k,n} =\sigma^2\av_{k}^{H}[n]\Qmat_{k} \av_{k}[n]$ with $\Qmat_{k} \! = \!\text{diag} \big( \text{tr}(\overline{\boldsymbol{\Gamma}}_{1k}\Amat_{1}\Amat_{1}) ,\cdots , \text{tr}(\overline{\boldsymbol{\Gamma}}_{Mk}\Amat_{M}\Amat_{M}) \big)$.

\vspace{-0.4cm}
\section{}\label{appen_closed_form_qt_update}\vspace{-0.2cm}
We begin by evaluating the first order derivative of \textbf{P3} objective with respect to $p_k$ as\vspace{+0.1pt}
\begin{align}
     &\frac{\partial\sum_{k=1}^{K} \log_{2}(1+\gamma_{k,n}) -\frac{1}{\ln{2}}\left( \gamma_{k,n} - \left(2y_{k,n}\sqrt{\Delta_{k,n}(\pv)(1+\gamma_{k,n})} -y_{k,n}^{2}\left( \Delta_{k,n}(\pv)+\Omega_{k,n}(\pv) \right) \right)\right)}{\partial p_k} \nonumber \\[-4pt]
     & \quad=y_{k,n}\frac{\sqrt{(1+\gamma_{k,n})}}{\sqrt{\Delta_{k,n}(\pv)}}\frac{\partial \Delta_{k,n}(\pv)}{\partial p_k} -\bigg( y_{k,n}^{2}\frac{\partial \Delta_{k,n}(\pv)}{\partial p_k}+ \sum_{j=1}^{K} y_{j,n}^{2}\frac{\partial \Omega_{j,n}(\pv)}{\partial p_k}\bigg). \label{eq_power_Cform_update}
\end{align}
The partial derivative of desired signal $\Delta_{k,n}(\pv)$, using \eqref{eq_SINR_k} is given as $\frac{\partial \Delta_{k,n}(\pv)}{\partial p_k}= \alpha_{d,k}^{2}|\av_{k}^{H}[n]\boldsymbol{\delta}_{k,n}|^{2}$.
%\begin{align}
%    \frac{\partial \Delta_{k,n}(p)}{\partial p_k}&= \alpha_{d,k}^{2}|\av_{k}^{H}[n]\boldsymbol{\delta}_{k,n}|^{2}.  \label{eq_desired_derivative}
%\end{align}
The partial derivative of $\Omega_{j,n}(\pv) \forall j$ is calculated using \eqref{eq_SINR_k} as follows:
\begin{align}
    \sum_{i=1}^{K}y_{i,n}\frac{\partial \Omega_{i,n}(\pv)}{\partial p_k} =&  \sum\limits_{i=1}^{K}\av_{i}^{H}[n]\Big(y_{i,n}^{2}\alpha_{d,i}(1+\kappa_{t,i}^{2})\Cmat_{ik,n}  +\sum\limits_{i=1}^{K}y_{i,n}^{2}(\kappa^2_{r,m}\Dmat_{ik,n}+\overline{\Dmat}_{ik,n}) \Big)\av_{i}[n]  \nonumber\\[-4pt]
   &+ \av_{k}^{H}[n] y_{k,n}^{2}(\Bmat_{k,n} +\boldsymbol{\Lambda}_{k,n} - \alpha_{d,k}^{2}\Cmat_{kk,n})\av_{k}[n] = l_{k}^{d}. \label{eq_differentiation_l_kd}
\end{align}
The optimal $p_k$ can obtained by substituting  ${\partial \Delta_{k,n}(\pv)}/{\partial p_k}$ and \eqref{eq_differentiation_l_kd} in \eqref{eq_power_Cform_update}, and by equating it to zero. It is given as \vspace{-4pt}
%\begin{align}
$     p_{k} = \bigg( P_{max}, \frac{ y_{k,n}^{2}\left( 1 + \gamma_{k,n} \right) \alpha_{d,k}^{2}|\av_{k}^{H}[n]\boldsymbol{\delta}_{k,n}|^{2}} {\left(\alpha_{d,k}^{2}y_{k,n}^{2}|\av_{k}^{H}[n]\boldsymbol{\delta}_{k,n}|^{2} + l_{k}^{d}   \right)^{2} } \bigg)$.
\bibliographystyle{IEEEtran}
\bibliography{IEEEabrv,CF_aging_ref}
\end{document}